\documentclass[twocolumn]{aastex63}

\usepackage{gensymb}

\newcommand\Herschel{\textit{Herschel} }
\newcommand\Planck{\textit{Planck} }

\newcommand\HST{\textit{HST} }
\newcommand\WISE{\textit{WISE} }

\newcommand\LPs{\textit{LPs}}

\newcommand\Htwo{H$_{\rm 2}$}

\newcommand\kms{km s$^{-1}$}

\newcommand\logden{log($n({\rm H_{\rm 2}}) $)}

\newcommand\percc{cm$^{-3}$}

\newcommand\Lsun{L$_{\rm \odot}$}

\newcommand\Msun{M$_{\rm \odot}$}

\newcommand\z{$z$}

\newcommand\alphaunits{M$_{\rm \odot}$ (K km s$^{-1}$ pc$^{2}$)$^{-1}$}

\received{July 15, 2020}
\accepted{\today}
\submitjournal{ApJ}

\begin{document}

\title{Turbulent Gas in Lensed \textit{Planck}-selected Starbursts at $z \sim 1 - 3.5$ }

\correspondingauthor{Kevin C. Harrington\footnote{International Max Planck Research School of Astronomy and Astrophysics in Bonn and Cologne}}
\email{kharring@astro.uni-bonn.de}

\author{Kevin C. Harrington}
\affiliation{Argelander Institut f\"{u}r Astronomie, Auf dem H\"{u}gel 71, 53121 Bonn,  Germany}
\affiliation{European Southern Observatory, Alonso de C{\'o}rdova 3107, Vitacura, Casilla 19001, Santiago de Chile, Chile\\}

\author{Axel ~Wei{\ss}}
\affiliation{Max-Planck-Institut f\"{u}r Radioastronomie, Auf dem H\"{u}gel 69, 53121 Bonn, Germany\\}


\author{Min S. ~Yun}
\affiliation{Department of Astronomy, University of Massachusetts, 619E LGRT, 710 N. Pleasant Street, Amherst, MA 01003, USA\\}

\author{Benjamin ~Magnelli}
\affiliation{Argelander Institut f\"{u}r Astronomie, Auf dem H\"{u}gel 71, 53121 Bonn,  Germany}


\author{C. E. ~Sharon}
\affiliation{Yale-NUS College, 16 College Avenue West 01-220, 138527 Singapore \\}

\author{T. K. D. ~Leung}
\affiliation{Center for Computational Astrophysics, Flatiron Institute, 162 Fifth Avenue, New York, NY 10010, USA\\}


\author{A. ~Vishwas}
\affiliation{Department of Astronomy, Cornell University, Space Sciences Building, Ithaca, NY 14853, USA\\}

\author{Q. D. ~Wang}
\affil{Department of Astronomy, University of Massachusetts, 619E LGRT, 710 N. Pleasant Street, Amherst, MA 01003, USA\\}


\author{E. F. ~Jim{\'e}nez-Andrade}
\affiliation{Argelander Institut f\"{u}r Astronomie, Auf dem H\"{u}gel 71, 53121 Bonn,  Germany}
\affiliation{National Radio Astronomy Observatory, 520 Edgemont Road, Charlottesville, VA 22903, USA\\}

\author{D. T. ~Frayer}
\affiliation{Green Bank Observatory, 155 Observatory Rd., Green Bank, West Virginia 24944, USA\\}


\author{D. ~Liu}
\affiliation{Max-Planck-Institut f\"{u}r Astronomie, K\"{o}nigstuhl 17, D-69117 Heidelberg, Germany\\}

\author{P. ~Garc{\'i}a}
\affiliation{National Astronomical Observatory of China – NAOC, Camino el Observatorio 1515, Santiago, Chile \\}
\affiliation{Universidad Cat{\'o}lica del Norte, Av. Angamos 0610, Antofagasta, Chile\\}


\author{E. ~Romano-D{\'i}az}
\affiliation{Argelander Institut f\"{u}r Astronomie, Auf dem H\"{u}gel 71, 53121 Bonn,  Germany}

\author{B. L. ~Frye}
\affiliation{Department of Astronomy/Steward Observatory, 933 North Cherry Avenue, University of Arizona, Tucson, AZ 85721, USA\\}


\author{S. ~Jarugula}
\affiliation{Department of Astronomy, University of Illinois at Urbana-Champaign, 1002 West Green St., Urbana, IL 61801, USA\\}


\author{T. ~B\u{a}descu}
\affiliation{Argelander Institut f\"{u}r Astronomie, Auf dem H\"{u}gel 71, 53121 Bonn,  Germany}

\author{D. ~Berman}
\affil{Department of Astronomy, University of Massachusetts, 619E LGRT, 710 N. Pleasant Street, Amherst, MA 01003, USA\\}


\author{H. ~Dannerbauer}
\affiliation{Instituto de Astrof{\'i}sica de Canarias (IAC), E-38205 La Laguna, Tenerife, Spain\\}
\affiliation{Universidad de La Laguna Dpto. Astrof{\'i}sica, E-38206 La Laguna, Tenerife, Spain\\}

\author{A. ~D{\'i}az-S{\'a}nchez}
\affiliation{Departamento F{\'i}ısica Aplicada, Universidad Polit{\'e}cnica de Cartagena, Campus Muralla del Mar, 30202 Cartagena, Murcia, Spain\\}


\author{L. ~Grassitelli}
\affiliation{Argelander Institut f\"{u}r Astronomie, Auf dem H\"{u}gel 71, 53121 Bonn,  Germany}

\author{P. ~Kamieneski}
\affil{Department of Astronomy, University of Massachusetts, 619E LGRT, 710 N. Pleasant Street, Amherst, MA 01003, USA\\}


\author{W. J. ~Kim}
\affiliation{Instituto de Radioastronom{\'i}a Milim{\'e}trica (IRAM), Avda. Divina Pastora 7, Local 20, 18012 Granada, Spain\\}

\author{A. ~Kirkpatrick}
\affiliation{Department of Physics \& Astronomy, University of Kansas,Lawrence, KS 66045, USA\\}


\author{J. D. ~Lowenthal}
\affiliation{Department of Astronomy, Smith College, Northampton, MA 01063, USA \\}

\author{H. ~Messias}
\affiliation{Joint ALMA Observatory Alonso de C{\'o}rdova 3107, Vitacura 763-0355 Santiago, Chile\\}

\affiliation{European Southern Observatory, Alonso de C{\'o}rdova 3107, Vitacura, Casilla 19001, Santiago de Chile, Chile \\}


\author{J. ~Puschnig}
\affiliation{Argelander Institut f\"{u}r Astronomie, Auf dem H\"{u}gel 71, 53121 Bonn,  Germany}

\author{G. J. ~Stacey}
\affil{Department of Astronomy, Cornell University, Space Sciences Building, Ithaca, NY 14853, USA\\}


\author{P. ~Torne}
\affil{Instituto de Radioastronom{\'i}a Milim{\'e}trica (IRAM), Avda. Divina Pastora 7, Local 20, 18012 Granada, Spain\\}

\author{F. ~Bertoldi}
\affiliation{Argelander Institut f\"{u}r Astronomie, Auf dem H\"{u}gel 71, 53121 Bonn,  Germany}




\begin{abstract}
Dusty star-forming galaxies at high redshift ($1 < z < 3$) represent the most intense star-forming regions in the Universe. Key aspects to these processes are the gas heating and cooling mechanisms. Although it is well known that these galaxies are gas-rich, little is known about the gas excitation conditions, as only few detailed radiative transfer studies have been carried out due to a lack of line detections per galaxy. Here we examine these processes in a sample of 24 strongly lensed star-forming galaxies identified by the \textit{Planck} satellite (\LPs) at $z \sim 1.1 - 3.5$. We analyze 162 CO rotational transitions (ranging from J$_{\rm up} = 1 - 12$) and 37 atomic carbon fine-structure lines ([CI]) in order to characterize the physical conditions of the gas in sample of \LPs. We simultaneously fit the CO and [CI] lines, and the dust continuum emission, using two different non-LTE, radiative transfer models. The first model represents a two component gas density, while the second assumes a turbulence driven log-normal gas density distribution. These \LPs\ are among the most gas-rich, infrared (IR) luminous galaxies ever observed  ($\mu_{\rm L}$L$_{\rm IR(8-1000\mu m) } \sim 10^{13-14.6} $\Lsun;  $< \mu_{\rm L}$M$_{\rm ISM}> = 2.7 \pm 1.2 \times 10^{12}$ \Msun, with $\mu_{\rm L} \sim 10-30$ the average lens magnification factor). Our results suggest that the turbulent ISM present in the \LPs\ can be well-characterized by a high turbulent velocity dispersion ($<\Delta V_{\rm turb}> \sim 100 $ \kms) and gas kinetic temperature to dust temperature ratios $<T_{\rm kin}$/$T_{\rm d}> \sim 2.5$, sustained on scales larger than a few kpc. We speculate that the average surface density of the molecular gas mass and IR luminosity $\Sigma_{\rm M_{\rm ISM}}$ $\sim 10^{3 - 4}$ \Msun pc$^{-2}$ and $\Sigma_{\rm L_{\rm IR}}$ $\sim 10^{11 - 12}$ \Lsun  kpc$^{-2}$, arise from both stellar mechanical feedback and a steady momentum injection from the accretion of intergalactic gas.

\end{abstract}


\section{Introduction} \label{sec:intro}

Star-forming galaxies at redshifts $z \sim 1 - 3$ probe the cosmic epoch when most of the stellar mass assembly in the Universe took place \citep[][and references therein]{madau2014}. A better understanding of star formation (SF) during this epoch is therefore imperative to understand SF across cosmic time. Locally, less than 5\% of the galaxy population has a star formation rate (SFR) that is significantly higher than the empirical main-sequence for star-forming galaxies, i.e. the tight correlation ($\sim$0.3 dex) between the SFR and stellar mass, M$_{\star}$ \citep{brinchmann2004, Noeske2007a, elbaz2007, elbaz2011, rodighiero2011, goto2011, sargent2012, whitaker2012a, whitaker2014, salmon2015}. These often-called starburst galaxies, with an infrared (IR) luminosity $L_{\rm IR} \sim 0.1 - 5 \times 10^{12}$ \Lsun  \citep[e.g.][]{sanders1996, Downes1998}, become increasingly more common at high-\z. In fact, (sub)mm number counts reveal that galaxies with $L_{\rm IR} > $ 10$^{12-13}$ \Lsun, at $z >$ 0.5, are many hundreds of times more likely to exist than in the local Universe \citep{blain2002, chapman2005, berta2011, magnelli2011, bethermin2012a, magnelli2013, casey2013, geach2013, simpson2014, strandet2016, casey2014, brisbin2017}. Meanwhile, the cosmic molecular gas density also peaks at $z \sim 1 - 3$ \citep{decarli2014, walter2014, lentati2015, decarli2016a, decarli2016b, pavesi2018, riechers2019, decarli2019b, liu2019}. This suggests a strong link between molecular gas and SF. Rest-frame far-infrared (FIR) measurements of spectral lines and thermal dust continuum emission have been used to investigate the cooling and heating processes of the interstellar medium (ISM) in star-forming galaxies, however the physical conditions at high-\z\ is still, in general, poorly investigated \citep{popesso2012, bothwell2013, carilli2013, genzel2013, yang2017, tacconi2018, tacconi2020, lenkic2020, aravena2020, birkin2020, boogaard2020}. \\

Turbulence regulates SF within cold and dense molecular clouds in most star-forming regions, and turbulence-regulated feedback seems to properly describe the main characteristics of the star-forming ISM \citep{shu1987, elmegreen2004, krumholz2005, mckee2007, krumholz2014b}. A log-normal probability distribution function (PDF) is often used to describe both the molecular gas velocity dispersion and volume density \citep{Vazquez-Semadeni1994, padoan1997, ostriker1998, klessen2000, wada2001, kowal2007, narayanan2008a, narayanan2008b, krumholz2009b, molina2012, hopkins2012a, hopkins2012b, hopkins2013a}. This is because the turbulent activity sets the local gas density as a consequence of randomly distributed shocks that compress the gas. This processes eventually converge towards a log-normal distribution of density due to the central-limit theorem \citep{Vazquez-Semadeni1994, Kevlahan2009, krumholz2014b}. Such turbulent models are supported by observational evidence using optically thin, diffuse \textit{and} dense molecular gas tracers of clouds within the Milky Way \citep[e.g. ][]{ginsburg2013}. A commonly used simplification to deal with these complex models is to adopt the Large Velocity Gradient (LVG) approximation \citep{goldreich1974, scoville1974} to model the photon escape probabilities within large-scale velocity flows. This assumption is applicable for clouds in the Milky Way, where the local thermal motions are much smaller than flow velocities, although non-local thermodynamic equilibrium (non-LTE) gas conditions may be present. The radial motion assumed in early applications of non-LTE LVG models would lead to higher SF efficiencies than observed, leading to the conclusion that a form of turbulent feedback must be present in the ISM to regulate SF \citep{zuckerman1974,zuckerman1975}. In addition, turbulent motion is set at the `driving scale' \citep[e.g. ][]{elmegreen2004, scalo2004}, determined by the largest physical size of the system. Diverse studies have found that gas turbulence increases as a function of $z$, suggesting such star-formation processes cannot be maintained for long periods of time (several orbital times) -- particularly in the most extreme star-forming galaxies \citep{kassin2012, wisnioski2015, johnson2018, ubler2019}. How the turbulent ISM behaves at high-\z, given the strong cosmic evolution of star-forming gas, is still an open question. \\

Star-forming galaxies at high-\z\ may have turbulent SF extending many kiloparsecs beyond their center, with total molecular gas masses up to an order of magnitude larger than local starbursts \citep{tacconi2006, hodge2010, ivison2010, hodge2016, kirkpatrick2017}. Therefore, it is crucial to derive the bulk molecular gas mass content in these host galaxies in order to properly quantify the SF activity as a function of the molecular gas properties. The main challenge in studying the star-forming gas is that cloud collapse requires cold environments, with T $\le 100$ K, yet the lowest energy transitions of \Htwo\ are much higher than these temperatures. Therefore \Htwo\ is unable\footnote{\Htwo\ is a low-mass quantum rotor with the first quadropole transition, J = 2$\rightarrow$ 0 at 28${\rm \mu m}$, requires $h\nu/k_{\rm B} \sim$  500 K to excite the lowest transition \citep{dabrowski1984}.} to trace the total column density (i.e. total mass) of gas \citep[e.g. 1-30\% of the gas column density;][]{roussel2007}. Carbon Monoxide (CO) is the second most abundant molecule in the ISM and is almost exclusively excited by collisions with \Htwo. Since it is one of the primary tracers of the \Htwo\ gas column density, it offers a unique opportunity to study these cold gas properties in star-forming galaxies. \\

The CO line luminosity to molecular gas mass conversion factor, $\alpha_{\rm CO}$, has been reviewed by several detailed studies \citep{magdis2011, genzel2012, narayanan2012,  schruba2012, bolatto2013, hunt2015, narayanan2015, amorin2016, accurso2017}. The CO(1-0) and CO(2-1) rotational transitions, and their less optically thick isotopologues ($^{13}$CO, C$^{18}$O), have been vital in determining the $L'_{\rm CO}$-to-$M_{\rm H_{\rm 2}}$ conversion factor, $\alpha_{\rm CO}$. The Galactic value is $\alpha_{\rm CO} \sim 4$ \alphaunits, whereas the canonical value for local starburst galaxies is $\alpha_{\rm CO} \sim $0.8 \alphaunits \citep{Downes1998}. CO traces diffuse and dense gas, however the atomic carbon, fine-structure line transitions ([CI] lines) are able to trace mostly diffuse gas \citep{glover2012, israel2015}. [CI] is an additional tracer capable of determining the  molecular \Htwo\  gas mass \citep{weiss2003, weiss2005carbon}. Efforts to calibrate the [CI] transitions as a tracer of the molecular gas mass, and attempts to constrain the gas-phase carbon abundance, have grown significantly, including high-\z\ massive star-forming galaxies on the main-sequence and bright quasars \citep{walter2011, Alaghband-Zadeh2013, bothwell2017, valentino2018, dannerbauer2019, valentino2020}. This is mostly based on the relative increase in detection efficiency of the [CI] ground transition as it gets redshifted at $z > 1$ into mm wavelengths, due to the higher photon energy in the [CI] lines. This makes it easier to detect than the faint ground-state CO(1-0) line. For the cold (T $\sim$ 20 K) and low density (\logden\ $\sim$ 2 \percc) ISM, dominating the emission in the Milky Way \citep{dame1986, bronfman1988, fixsen1999, garcia2014}, the CO(1-0) line luminosity has traditionally been used as a tracer of the total molecular gas content \citep{bolatto2013}, as higher molecular rotational levels are poorly populated under these conditions. Following early studies in the Milky Way, this approach has been widely applied to determine the molecular gas content in nearby star-forming galaxies. The general scenario may differ for higher excitation gas (or increased SF activity), as the higher-J level populations can contribute a more significant fractional contribution to the CO partition function. For intense star-forming environments, where the mean gas density is larger than 10$^{3-4}$ \percc\ and/or the gas kinetic temperature is higher than 20 K, the CO(2-1) and CO(3-2) and even higher rotational transitions begin to contribute a higher fraction to the partition function, as less molecules sit at the J$_{\rm up}$ = 1 state. Thus, these low-J lines can trace comparable, if not larger, fractions of the total CO column densities (and thus more molecular gas) as the CO(1-0) line. This highlights the need to measure multiple CO transitions and conduct a proper modelling of the line intensities to obtain meaningful conversion factors for star-forming galaxies at $z > 1$.\\

Local measurements of the CO ladder in large samples of star-forming and starburst systems have been conducted using the \Herschel\-SPIRE \citep[and HIFI; ][]{rangwala2011, liu2015, kamenetzky2016, rosenberg2015, lu2017}. On average, the majority of the SF in local starburst galaxies is confined to the central few hundred parsecs of the Galactic nucleus. Most extreme IR luminosities in the local Universe are induced by merger-driven processes, although, in general, there is a strong presence of warm and diffuse molecular gas \citep[see e.g.][]{Downes1998}, well-traced by the mid-to-high-J CO lines \citep{rosenberg2015, kamenetzky2016}. Constraints on such galaxy-wide molecular ISM properties at $z > 1$ have been limited to large integration times required to sample the low-, mid-, and high-J CO lines. Less than twenty years ago, only $\sim$40 galaxies at $z > 1$ had been detected in CO emission \citep{solomon2005, omont2007}. \citet{carilli2013} reviewed $\sim$ 200 galaxies, most with a single line detection (J$_{\rm up}$ = 2-5). At the time, only eleven high-\z\ galaxies had one (or both) [CI] line detection(s) \citep{weiss2005carbon, walter2011, carilli2013}.\\ 

Strong gravitational lensing of high-\z\ star forming galaxies offers a unique way to examine highly magnified molecular gas. The method for selecting strongly lensed dusty galaxy candidates, at $z > 1$, is primarily based on unusually bright (sub)mm fluxes compared to the expected steep drop-off in (sub)mm number counts \citep[e.g.][]{negrello2007,negrello2010}. This method has since identified a large number across the extragalactic sky, i.e. more than 100 lensed candidates at $z > 1$ \citep{ivison2010, wardlow2013, negrello2017, bussmann2013, bussmann2015, vieira2010, vieira2013, weiss2013, strandet2016, canameras2015, harrington2016, diaz-sanchez2017, bakx2018}. The lensed population of dusty star-forming galaxies selected by the South Pole Telescope \textit{SPT}, \textit{Herschel} Space Observatory and \textit{Planck} have now been detected in more than two CO transitions \citep[e.g. ][and this work]{spilker2016, strandet2017, yang2017, bakx2020}. The \textit{Herschel}-selected, strongly lensed galaxy sample \citep{bussmann2013} offered the first systematic approach to producing a statistically significant sample of CO/[CI] lines \citep{yang2017}, followed by a compilation in 11 \Planck\ and \Herschel\ selected lensed galaxies \citep{canameras2018b}, including four galaxies with both [CI] lines detected \citep{nesvadba2019}. The IR to CO luminosity relations of local starbursts and high-\z\ star-forming galaxies explored by \citet{greve2014} indicate that the ISM radiation field is an important component to consider when understanding CO line excitation, yet this investigation was limited to 23 unlensed and 21 lensed dusty star-forming systems -- all with more than three frequency measurements of the dust continuum and usually a single CO line detection. Most previous studies used only single and/or double component gas emitting regions to reproduce the observed CO emission, excluding the simultaneous modeling of the available [CI] emission, but also ignoring the role of the dust continuum emission as a heating source of the gas.\\

In this work, we apply state-of-the-art non-LTE models to $\sim$ 200 CO and [CI] emission lines, from single-dish line measurements, for a flux-limited sample of 24 lensed galaxies identified by the \textit{Planck} satellite. This sample builds off of our pilot \Planck and \Herschel selection in \citet[][]{harrington2016}, expanded since then (Berman et al. in prep.). We have selected 24 of these galaxies to investigate the physical gas conditions responsible for driving such bright apparent FIR luminosities ($\mu_{\rm L} L_{\rm IR} > 10^{14}$ \Lsun). We have a systematic focus on detecting the rise, peak and turnover in the CO excitation ladder in order to investigate the gas volume densities and turbulent properties, the relationship between the gas kinetic temperature and dust temperature, and the derivation of $\alpha_{\rm CO}$. We follow a novel approach when modelling all emission lines detected based on a turbulence-driven gas density PDF. Unlike most high-\z\ studies, we have simultaneously modelled these lines in the presence of both dust continuum radiation field and CMB radiation as background excitation sources. This work is organized as follows. In \S 2 we describe the sample selection and ancillary dust photometry of the 24 strongly lensed galaxies in our sample. In \S 3 we provide the details of the novel GBT, IRAM 30m, and APEX single-dish measurements of the CO ladder, ranging from J$_{\rm up} = 1 - 12$, and both [CI] lines. In \S 4 we provide a summary of the emission line profiles. We summarize the model and model assumptions we applied in \S 5. In \S 6 we discuss our main results, and in \S 7 we provide an interpretation of the physical gas conditions of these extreme starburst galaxies. Our conclusions are summarized in \S 8. We adopt a fiducial $\Lambda$CDM cosmology with ${\rm H_{0} = 69.6 \, km s^{-1} Mpc^{-1} }$ with ${\rm \Omega_{m} = 0.286}$, and ${\rm \Omega_{\Lambda} = 1 - \Omega_{m}}$ throughout this paper \citep{bennett2014}\footnote{We have used \textit{astropy.cosmology} \citep{AstropyCollaboration2018}. }.\\

		
								
\section{Sample}
\label{sec:thesample}

\subsection{Selection}
Here we outline our sample of strongly lensed \textit{Planck} selected, dusty star-forming galaxies, hereafter ``\LPs'' (Table \ref{tab:summaryLPs}). Our sample  of 24 \LPs\ began with a \Planck \&  \Herschel cross-match identification of eight objects (8/24) with continuum detections at 857 GHz \citep{harrington2016} greater than 100 mJy. The remaining 16/24 \LPs\ were selected based on continuum detections by \textit{Planck}, at 857, 545 and/or 353 GHz in the maps of all the available, clean extragalactic sky. These bright \textit{Planck} point sources were then analyzed through a filtering process using a WISE color selection for the four WISE bands \citep[3.4$\mu$m, 4.6$\mu$m, 12$\mu$m, 22$\mu$m][Berman et al in prep.]{yun2008}. Other methods to identify strong gravitational lenses using (sub)mm data were independently verified by other teams using \Planck \&  \Herschel color criteria \citep{canameras2015}. The 24 \LPs\ presented in these analyses include eight systems identified by \citet[][]{canameras2015}. The use of \Planck and WISE data resulted in the discovery of the brightest known, dusty starburst galaxy at $z > 1$, the 'Cosmic Eyebrow' \citep{diaz-sanchez2017, dannerbauer2019}, which has also been independently recovered as one of the \LPs\ presented in this survey work. Note that LPs-J1329 corresponds to the location on the sky associated with the \textit{Cosmic Eyebrow-A} lens component \citep{dannerbauer2019}. Table \ref{tab:summaryLPs} shows the size of the lensed emission for each of the \LPs, in which there are 21/24 with lens sizes $\leq 10^{\prime \prime}$. Half of the \LPs\ are galaxy-galaxy lenses, while the other half are a mix of cluster or group lensing. The foreground lens galaxies have a negligible contribution to the observed far-IR emission of the lensed galaxy \citep{harrington2016}. The \LPs\ have CO-based spectroscopic redshifts ranging from $z_{\rm CO} \sim$ 1.1 - 3.6 \citep[][and this work]{harrington2016, harrington2018, canameras2018}. They are comparable or brighter in CO and FIR luminosity than other strongly lensed SPT \citep{strandet2016, strandet2017, weiss2013} or \textit{Herschel}-selected dusty star forming galaxies \citep{harris2012, bussmann2013, bussmann2015, yang2017}. The \Planck \& \Herschel wavelength selections preferentially target $z \sim 2 - 3$ galaxies, versus the mm-selected SPT sources with a median closer to $z \sim 4$, although with a wide range between $z \sim 2 - 7$ \citep{weiss2013, strandet2016, spilker2016, reuter2020}. \\

Our selection method only picks out sub-mm bright point sources, and the WISE data assists us in interpreting that these systems do not have the same mid-IR characteristics as the luminous WISE-selected, dust-obscured QSOs \citep{tsai2015}. The prevalence of a dust-obscured AGN within the \LPs \, is uncertain, however \citet{harrington2016} and Berman et al. (in prep.) have shown, using \WISE and \Herschel data \citep[see methods in, e.g. ][]{kirkpatrick2015}, that the majority of the \LPs \, have a substantial contribution to the total IR luminosity from SF activity instead of AGN activity \citep[see also][]{canameras2015}. The dusty nature of the \LPs\ has thus far resulted in the absence of stellar mass estimates, yet the extreme nature of their IR luminosities suggests that it would be reasonable to assume they would lie above the main-sequence for star-forming galaxies at these redshifts. We therefore consider them starburst galaxies, without alluding to an assumed SF history.\\

\startlongtable
\begin{deluxetable*}{ccccccccc}




\tablecaption{Sample summary}


\tablehead{\colhead{ID} & \colhead{RA} & \colhead{DEC} & \colhead{$z_{\rm fg}$} & \colhead{$z_{}$} & \colhead{$\mu_{\rm L}$} & \colhead{$\mu_{\rm L}^{\dagger}$} & \colhead{Lens size} & \colhead{Reference} \\ 
\colhead{(---)} & \colhead{(h \, m \, s)} & \colhead{(\degree \, $^{\prime}$ \, $^{\prime \, \prime}$)} & \colhead{(---)} & \colhead{(---)} & \colhead{(---)} & \colhead{(---)} & \colhead{(\arcsec)} & \colhead{(---)} } 

\startdata
LPsJ0116 & 01:16:46.77 & -24:37:01.90 & 0.4 & 2.12453 & - & 23 & $\sim$4.5$^{GG}$ & 1 \\
LPsJ0209 & 02:09:41.3 & 00:15:59.00 & 0.202 & 2.55274 & 7 -- 22 & 58 & $\sim$3$^{GG}$ & 2,3,4,5,6,7,8 \\
LPsJ0226 & 02:26:33.98 & 23:45:28.3 & 0.34 & 3.11896 & - & 40 & $\sim$3.5$^{GG}$ & 1,8 \\
LPsJ0305 & 03:05:10.62 & -30:36:30.30 & 0.1-0.5 & 2.26239 & - & 18 & $\sim$2$^{GG}$ & 1 \\
LPsJ0748 & 07:48:51.72 & 59:41:53.5 & 0.402 & 2.75440 & - & 21 & $\sim$13$^{GC}$ & 1,9,10 \\
LPsJ0846 & 08:46:50.16 & 15:05:47.30 & 0.1 & 2.66151 & - & 27 & $\sim$10$^{GC}$ & 1 \\
LPsJ105322 & 10:53:22.60 & 60:51:47.00 & 0.837 & 3.54936 & 5 -- 12 & 41 & $\sim$6$^{GG}$ & 1,11,12,13 \\
LPsJ105353 & 10:53:53.00 & 05:56:21.00 & 1.525 & 3.00551 & 9 -- 48 & 20 & $\sim$1.5$^{GG}$ & 2,11,12,13,14,15,16 \\
LPsJ112714 & 11:27:14.50 & 42:28:25.00 & 0.33-0.35 & 2.23639 & 20 -- 35 & 25 & $\sim$13$^{GC}$ & 2,11,12,13,17 \\
LPsJ112713 & 11:27:13.44 & 46:09:24.10 & 0.415 & 1.30365 & - & 21 & $\sim$1.5$^{GG}$ & 1 \\
LPsJ1138 & 11:38:05.53 & 32:57:56.90 & 0.6 & 2.01833 & - & 10 & $\sim$1$^{GG}$ & 1 \\
LPsJ1139 & 11:39:21.74 & 20:24:50.90 & 0.57 & 2.85837 & 6 -- 8 & 19 & $\sim$1$^{GG}$ & 1,11,12 \\
LPsJ1202 & 12:02:07.60 & 53:34:39.00 & 0.212 & 2.44160 & - & 25 & $\sim$5-10$^{GC}$ & 2,11,12,16 \\
LPsJ1322 & 13:22:17.52 & 09:23:26.40 & - & 2.06762 & - & 20 & $\sim$10$^{GC}$ & 1 \\
LPsJ1323 & 13:23:02.90 & 55:36:01.00 & 0.47 & 2.41671 & 9 -- 12 & 25 & $\sim$10$^{GC}$ & 2,11,12,16 \\
LPsJ1326 & 13:26:30.25 & 33:44:07.40 & 0.64 & 2.95072 & 4 -- 5 & 33 & $\sim$1.5$^{GG}$ & 1,18,19 \\
LPsJ1329 & 13:29:34.18 & 22:43:27.30 & 0.443 & 2.04008 & 9 -- 13 & 31 & $\sim$11$^{GC}$ & 1,20,21,22 \\
LPsJ1336 & 13:36:34.94 & 49:13:13.60 & 0.28 & 3.25477 & - & 24 & $\sim$1.5$^{GG}$ & 1 \\
LPsJ1428 & 14:28:23.90 & 35:26:20.00 & - & 1.32567 & - & 4 & $\sim$1$^{GG}$ & 2,23,24,25,26,27 \\
LPsJ1449 & 14:49:58.59 & 22:38:36.80 & - & 2.15360 & - & 8 & $\sim$10$^{GC}$ & 1 \\
LPsJ1544 & 15:44:32.35 & 50:23:43.70 & 0.673 & 2.59884 & 10-17 & 10 & $\sim$7$^{GC}$ & 1,11,12,13,27 \\
LPsJ1607 & 16:07:22.6 & 73:47:03 & 0.65 & 1.48390 & - & 4 & $\sim$1$^{GG}$ & 2,16 \\
LPsJ1609 & 16:09:17.80 & 60:45:20.00 & 0.45 & 3.25550 & 12 -- 16 & 44 & $\sim$7$^{GC}$ & 2,11,12,13,16 \\
LPsJ2313 & 23:13:56.64 & 01:09:17.70 & 0.56 & 2.21661 & - & 57 & $\sim$3$^{GG}$ & 1 \\
\enddata


\tablecomments{Foreground lens redshfits, $z_{\rm fg}$, are reported in references. $z_{}$ is the average redshift of the \LPs\ based on all CO/[CI] line detections. $\mu_{\rm L}$ Lens magnfication factor range. Measured with single line / single-band CO / dust emission or HST near-IR imaging. $\mu_{\rm L}^{\dagger}$ = Estimated using Tully-Fisher'' method (see Appendix \ref{difflens}). $^{GG}$ = Galaxy-Galaxy lens. The lens arc size corresponds to the effective Einstein radius, or the inferred circular radius. $^{GC}$ = Galaxy-Galaxy Cluster (or group) lens. The lens arc size corresponds to the largest lens arclet or effective Einstein radius.}

\tablerefs{ (1) Berman et al. (in prep), (2) \citep{harrington2016}, (3) \citep{harrington2019}, (4) \citep{geach2015}, (5) \citep{su2017}, (6) \citep{geach2018}, (7) \citep{rivera2019}, (8) Kamieneski et al. (in prep), (9) \citep{khatri2016}, (10) \citep{amodeo2018}, (11) \citep{canameras2015},  (12) \citep{canameras2018b}, (13) \citep{frye2019},  (14) \citep{canameras2017a}, (15) \citep{canameras2017b},  (16) \citep{harrington2018}, (17) \citep{canameras2018a},  (17) \citep{bussmann2013}, (18) \citep{yang2017},  (19) \citep{diaz-sanchez2017}, (20) \citep{dannerbauer2019},  (21) \citep{iglesias-groth2017},  (22) \citep{borys2006}, (23)  \citep{iono2009}, (24) \citep{sturm2010}, (25) \citep{stacey2010}, (26) \citep{hailey-dunsheath2012}, (27) \citep{nesvadba2019}}

\label{tab:summaryLPs}
\end{deluxetable*}

\subsection{Continuum data}

The observed dust continuum and spectral energy distribution are used to constrain the excitation conditions, and a database of continuum measurements between 2mm and 250 \textnormal{$\mu$m} is compiled from new and archival photometry by \textit{Planck}, \textit{Herschel}, ALMA, LMT, JCMT, and IRAM 30-m telescopes. All of the ancillary (sub)mm photometric data used in this work can be accessed online (see abridged version in Table~\ref{tab:continuumdata}). We also provide the modelled continuum data from Berman et al. (in prep.), and we refer the reader to more detailed information reported in the literature for previous (sub)mm observations with the SMA, NOEMA and ALMA for a sub-set of the \LPs\ \citep{bussmann2013, canameras2015, harrington2016, su2017, geach2018, rivera2019, diaz-sanchez2017, dannerbauer2019}. With the exception of the sources with ALMA 1 mm imaging data that fully resolves the continuum structure with better than 1\arcsec\ angular resolution, all other photometry come from low resolution observations that do not resolve the dust emission. There are ten \LPs\ with ALMA 1 mm continuum measurements (Berman et al. in prep.). Six of these ten also have LMT-AzTEC measurements, which agree well with the comparable ALMA detection (see Table~\ref{tab:continuumdata}). The continuum measurements at $\lambda_{\rm obs} =$ 1-2 mm come from LMT-AzTEC (1.1mm) and/or IRAM 30m-GISMO2 (2mm) observations \citep[][Berman in prep]{canameras2015,harrington2016}, and in some cases archival SCUBA-2 850$\mu$m data was available \citep[][Berman et al. in prep.]{diaz-sanchez2017}. \\
The measured (sub)mm flux densities of 10s to 100s of mJy are so large that source confusion is not relevant. An exception is the \textit{Planck} data with effective resolution of 5$\arcmin$. Here we adopt the photometry and uncertainty which fully incorporates the confusion noise based on the measured local foreground, leading to the conservative photometric uncertainties from point sources identified in the \Planck\ maps. The majority of the \LPs\ have ancillary \textit{Herschel}-SPIRE (250$\mu$m, 350$\mu$m, 500$\mu$m) and/or mm-wavelength measurements from both wide-field maps and pointed observations, which are useful to constrain the peak wavelength and the long-wavelength tail of the thermal dust emission. Additionally, previous work by \citet{harrington2016} has shown that a minimal fraction of the far-IR emission is expected to come from the foreground lens for these systems.  \\

		

\section{Spectral Line Observations}
\label{theobs}
The number of new line measurements we present in this work is $\sim$70\% of the following: 20 CO(1-0), 6 CO(2-1), 24 CO(3-2), 15 CO(4-3), 16 CO(5-4), 15 CO(6-5), 18 CO(7-6), 17 CO(8-7), 16 CO(9-8), 6 CO(10-9), 8 CO(11-10), 1 CO(12-11),  19 [CI](1-0) and 18 [CI](2-1).  For a thorough analysis, we complemented our line observations with nearly 50 line measurements previously reported in the literature \citep[][Berman et al. in prep.]{canameras2015,harrington2016, canameras2017b, canameras2018, harrington2018, dannerbauer2019, nesvadba2019}, for the \LPs\ in our catalogue. In Table \ref{tab:telescopes}, we summarize the astronomical facilities, receiver names, observed bandwidths, and telescope's beam sizes involved in the data acquisition for this work. Table \ref{tab:telescopes} also includes the beam size for the LMT and ALMA/Band 3 spectral line measurements to be presented in Berman et al. (in prep.). Both the LMT and ALMA/Band 3 observations had targeted the same CO transition, CO(2-1) or CO(3-2), with comparable line fluxes.\\

\subsection{GBT, IRAM 30m and APEX Observations}

\begin{deluxetable}{cccc}




\tablecaption{Observational Facilities}


\tablehead{\colhead{Telescope} & \colhead{Receiver} & \colhead{FrequencyCoverage} & \colhead{BeamSize} \\ 
\colhead{(---)} & \colhead{(---)} & \colhead{(GHz)} & \colhead{(arcsec)} } 

\startdata
GBT 100m & K$_{a}$ band & 26 - 40 & 19 - 29 \\
LMT 32m & RSR & 75 - 115 & 21 - 31 \\
ALMA & Band 3 & 85 - 116 & 0.4 - 0.6 \\
IRAM 30m & E150 & 125 - 175 & 14-20 \\
IRAM 30m & E230 & 202 - 274 & 9 - 12 \\
IRAM 30m & E330 & 277 - 350 & 7 - 9 \\
APEX 12m & PI230 & 200 - 270 & 23 - 31 \\
APEX 12m & FLASH345/460 & 268 - 516 & 12 - 17 \\
\enddata


\tablecomments{The largest angular scale for the ALMA spectral line observations is 4\arcsec. See \citep[][and Berman et al. in prep.]{harrington2016} for details about the LMT and ALMA observations. }


\label{tab:telescopes}

\end{deluxetable}

We observed the CO(1-0) line with the Ka-band receiver on the Green Bank Telescope GBT (Pr. ID: 17B-305; PI: K. Harrington) between October 7 - 31, 2017, in Green Bank, West Virginia, U.S.A. under stable atmospheric conditions during both night and day hours. The observing procedure and data reduction is identical to that presented in \citet{harrington2018, dannerbauer2019}, and we briefly describe the procedure below. We executed a SubBeamNod observing mode, with 4 min integration per scan. Each session started with a pointing and focus check, followed by a pointing every 1-1.5hr. Focus measurements were conducted every 3hr for longer observing sessions. We tuned the backend spectrometer, VEGAS, to its low-resolution, 1.5 GHz bandwidth mode. Using GBTIDL \citep{marganian2013} we computed all On-Off measurements and corrected for atmospheric attenuation. Each spectrum was inspected by eye after baseline subtraction \citep[see][]{harrington2018}, and roughly 10-15\% of scans were dropped. After subtracting a baseline, and then averaging, we smoothed the spectra to $\sim$ 100 \kms channel resolution.\\

We observed mid-high-J CO and [CI] emission lines in the sources available in the Southern hemisphere using both the PI230 and dual-frequency FLASH+ 345/460L receivers on the APEX telescope in San Pedro de Atacama, Chile \citep{guesten2006}. We used Max Planck Society observing time between 22 May and 28 September, 2018 (Pr. M-0101.F-9503A-2018; PI: Harrington), soon after the new telescope surface was installed and commissioned. Observations took place in a range of very good to reasonable weather conditions, i.e. precipitable water vapor (PWV) $\sim$ 2-3mm for PI230 and PWV $<$2mm for FLASH+. FLASH \citep{heyminck2006} is a 2 side-band (SB) dual-frequency heterodyne receiver with a single orthogonal linear polarization for each of the 345 GHz and 460 GHz atmospheric windows. Both the FLASH 345/460 channels have an upper and lower side-band with 4 GHz bandwidth. The PI230 receiver is a 2-sideband heterodyne receiver with a dual-polarization capability and 2$\times$8 GHz bandwidth. We used a standard wobbler switching with a chopping rate of 1.5 Hz, and an azimuthal throw offset of 30''. Each scan consisted of a hot/sky/cold calibration 600'' off-source, followed by 12 subscans of 20s per on-source integration time. Focus checks were performed roughly every 3-4hr, whereas pointing checks on Jupiter or nearby star were performed every 1-2h (pointing accuracy within 2-3\arcsec). All data was recorded using the MPIfR eXtended bandwidth Fast Fourier Transform spectrometers \citep[FFTS; ][]{klein2006}, and each of the scans were reduced and analyzed using the CLASS and GREG packages within the GILDAS\footnote{Software information can be found at: http://www.iram.fr/IRAMFR/GILDAS. } software \citep{pety2005_gildas}. The spectrum from each scan was smoothed to $\sim$ 100 \kms channel resolution and assessed after subtracting a first-order baseline from the emission line-free channels. The baseline stability depends strongly on the observed frequency and/or weather conditions, therefore we dropped 10-25\% of the scans before co-adding the base-line subtracted, rms-weighted spectrum. \\

We observed low-to-high-J CO and [CI] emission lines with the IRAM 30m telescope during three observing semesters (Pr. 187-16, 170-17, 201-18; PI: K. Harrington), between January 29th, 2017 and April 24th, 2019. Overall, weather conditions varied from excellent to poor, with the reference zenith opacity at 225 GHz, $\tau_{\rm \nu 225 GHz} \sim 0.05-0.8$. We utilized all four of the EMIR receivers \citep{carter2012}, E090, E150, E230 and E330, often with dual tuning modes to target more than one CO/[CI] emission line. In total, the EMIR receiver has a dual polarization, with a 16 GHz bandwidth backend spectrometer, the fast Fourier Transform Spectrometre (FTS200), and an 8 GHz bandwidth spectrometer, the WIde-band Line Multiple Auto-correlator (WILMA). The FTS200 has a finer channel resolution, however is subject to baseline instabilities such as platforming features in the bandpass. The WILMA has a lower native channel resolution and was used almost always alongside the FTS200 to verify observed line features. We carried out a standard wobbler switching observing mode with offset throws of 40\arcsec every second.  Each wobbler switching mode procedure includes three, 5 minute integrations (i.e. twelve 25-s subscans). Pointing corrections were performed (e.g. Uranus, Venus, J1226+023, J1418+546) every 1-2hr, with azimuth and elevation pointing offsets typically within 1-3\arcsec. Focus measurements were repeated roughly 1.5hr after sunset/sunrise and every 3-4hr to correct for thermal deformations of the primary dish and/or secondary mirror.  In the same manner presented in \citet{harrington2019}, all scans were reduced using GILDAS package\footnote{\url{http://www.iram.fr/IRAMFR/GILDAS}.}, smoothed to $\sim 50-150$ \, \kms channel resolution, followed by a visual inspection of a baseline-subtracted spectrum and subsequent averaging of the rms-weighted spectrum. We dropped 5-20\% of the scans per line due to unstable baselines or noise spikes, which may strongly depend on the specific tuning setup and weather. \\

\subsection{Absolute Calibration Errors}
\label{sec:adopteduncertainties}
In the following analyses we model the \textit{apparent} (not corrected for lens magnification) velocity-integrated flux density, integrated across the entire line profile. We apply a total error based on the typical systematic uncertainties associated with pointed single-dish spectroscopic observations. These include: atmospheric instabilities (transmission varying on the order of seconds/minutes), pointing/focus corrections, baseline subtraction procedures, the calibration of the Jy/K gain conversion, receiver stability across the entire bandpass. For all CO(1-0) lines, we adopt a 25\% uncertainty for systematic effects with the GBT \citep[see][]{harrington2018,frayer2018}. We adopt a 20\% uncertainty for all APEX and IRAM 30m measurements less than $\sim$ 240 GHz and a 35\% uncertainty for lines observed at higher frequencies. We add an additional 5-10\% total uncertainty to those emission lines which were detected at the edge of the EMIR receiver capabilities and at lower atmospheric transmission. Despite careful pointing/focus/calibration measurements, we add an additional 5\% total uncertainty to all integrated fluxes used in this study due to the heterogeneous observing conditions among all of the emission lines observed or reported in other studies. Sources LPs-J1322, LPs-J0846, LPs-J0748 have extended emission as detected by AzTEC 1.1mm continuum (Berman et al in prep.). Therefore, we measured the emission surrounding the reported RA/DEC, which we consider to be representative of the entire galaxy. As noted in Berman et al. (in prep.) for LPs-J1322, the ALMA measurements did not account for $\sim$ 35\% of the LMT/AzTEC 1.1mm continuum flux due to its large Einstein Ring (also see Table \ref{tab:summaryLPs} and Table \ref{tab:telescopes}). Therefore we have adopted an additional 35\% total error for the lines in this source. High-frequency measurements may underestimate the total flux for the most extended \LPs\ due to smaller beam-sizes and pointing errors. The conservative total uncertainties we adopt for these single-dish measurements thereby include a wide variation in the value of the flux density in an attempt to constrain the average global ISM properties. \\

\section{Emission Line Profiles}
\label{thelines}
\begin{figure*}
\centering
 \includegraphics[scale = 0.25]{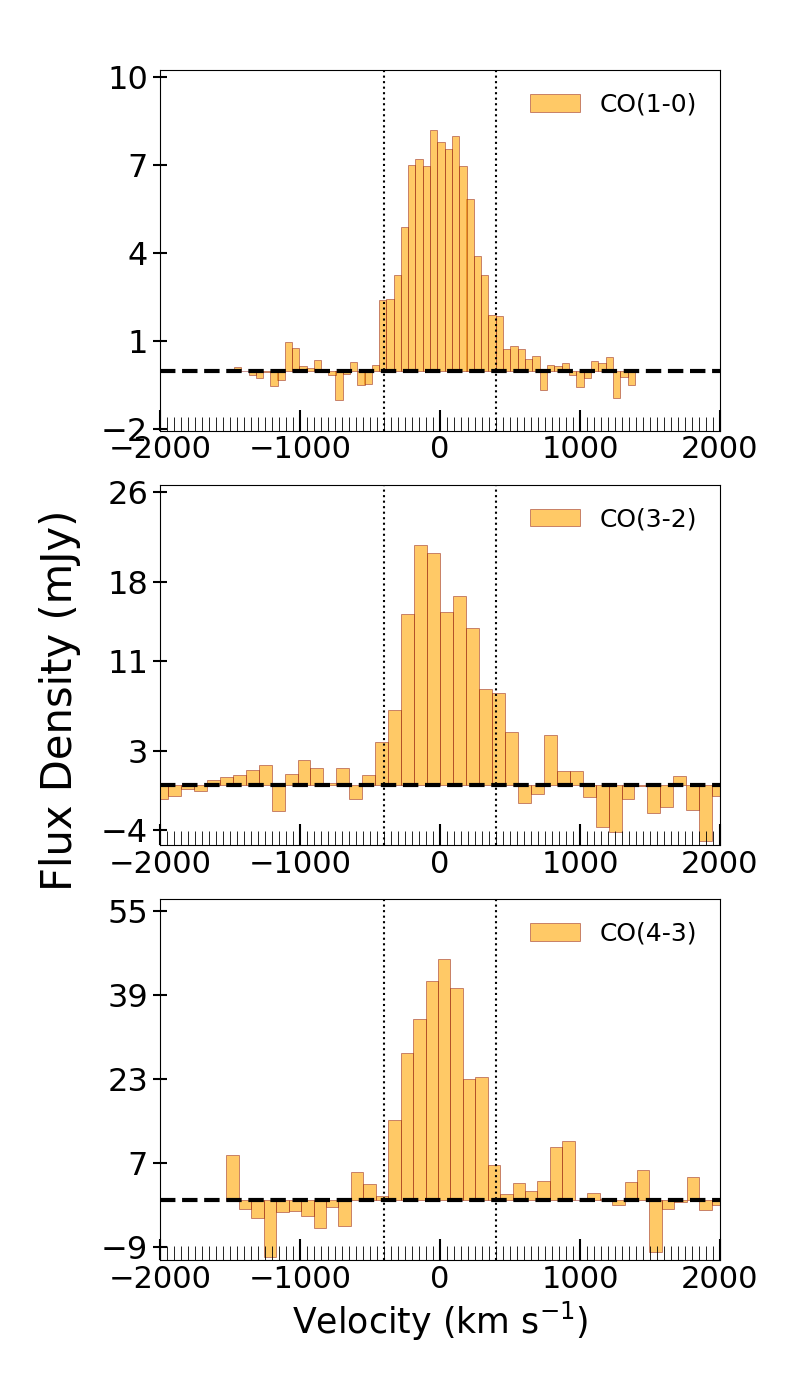}
 \includegraphics[scale = 0.25]{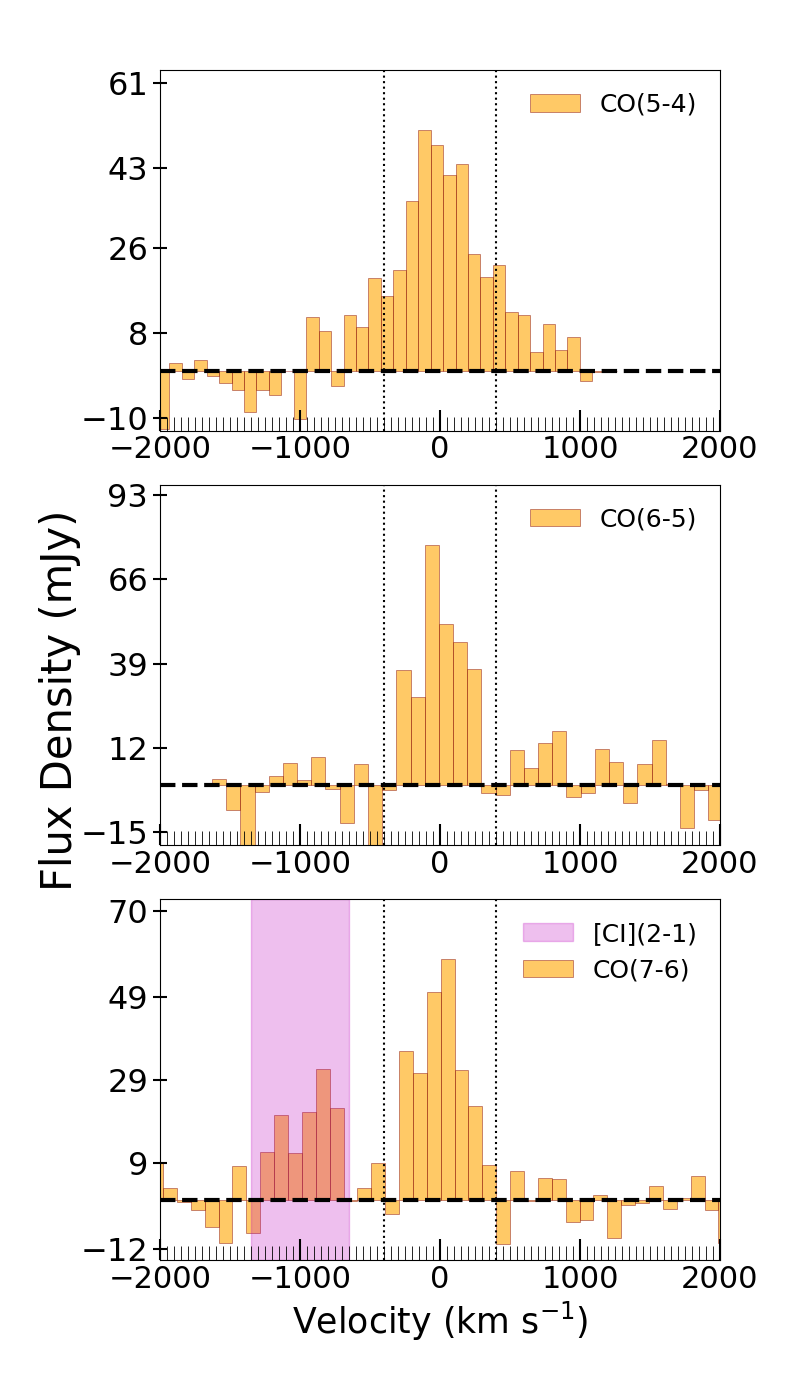}
  \includegraphics[scale = 0.25]{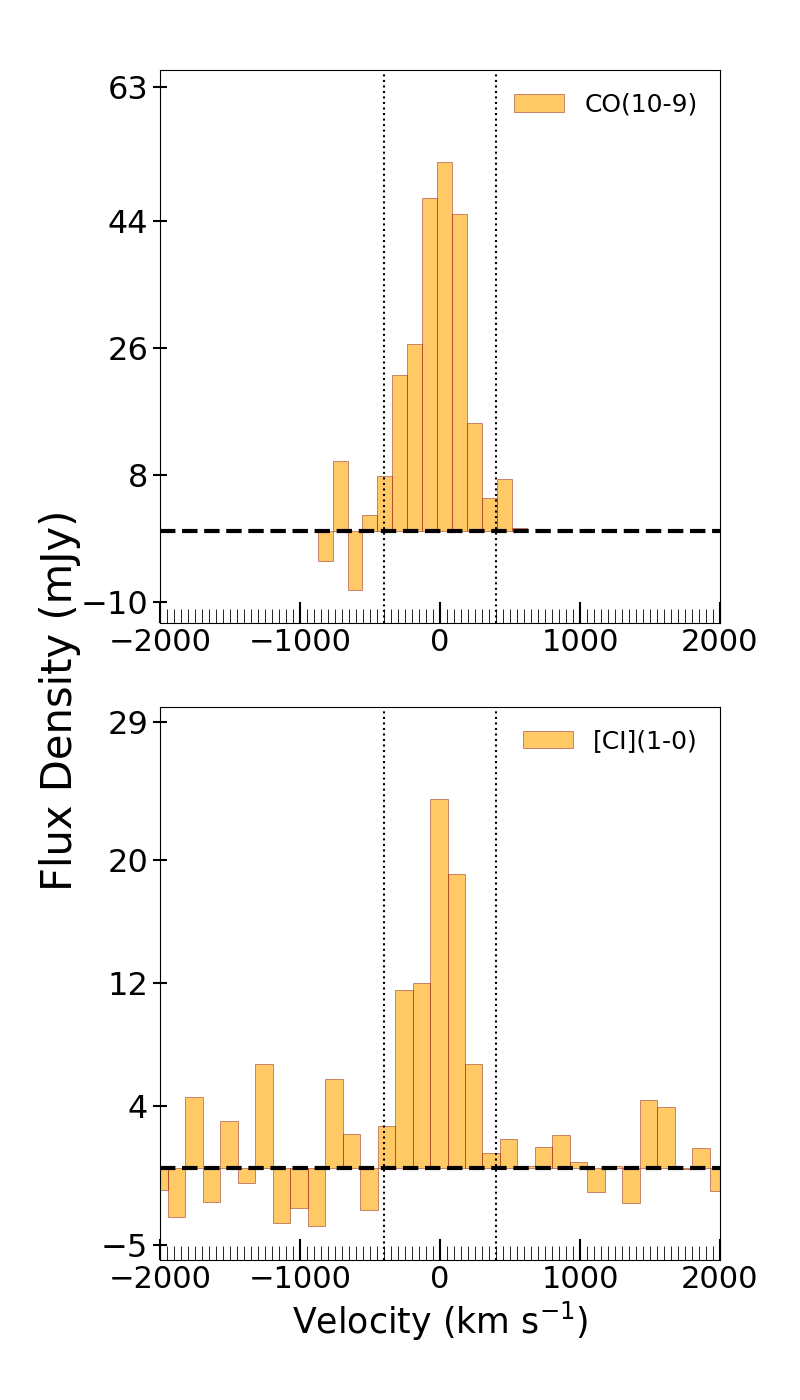}
 \caption{Apparent flux density versus velocity for the CO and [CI] line detections. The best fit models of all line and continuum data are shown for LPs-J1323 in Fig. \ref{fig:bfJ1323}. The CO(1-0) line was previously presented in \citet{harrington2018}. Spectra and best-fit models for the other \LPs\ can be accessed online. }
 \label{fig:examplespectra}
\end{figure*}

Figure \ref{fig:examplespectra} shows an example set of low-to-high-J CO and [CI] line detections for LPs-J1323. The remaining figures for the line spectra for the \LPs\ can be accessed online in the supplemental journal. In all of the 21/24 \LPs\ with a [CI] line detection, the emission line profile matches that of the spectrally adjacent CO emission. Specific examples of this can be seen in the [CI](2-1)/CO(7-6) spectrally adjacent pair (see e.g. LPs-J0116, LPs-J0209, LPs-J0305, LPs-J0748, LPs-J1326). Most of the CO and [CI] lines have similar line-widths and shape. These spectrally resolved measurements indicate that the emitting regions follow the same large-scale dynamics, based on these spatially unresolved measurements. \\

\begin{figure}
\resizebox{0.49\textwidth}{!}{%
 \includegraphics[width=0.98\textwidth]{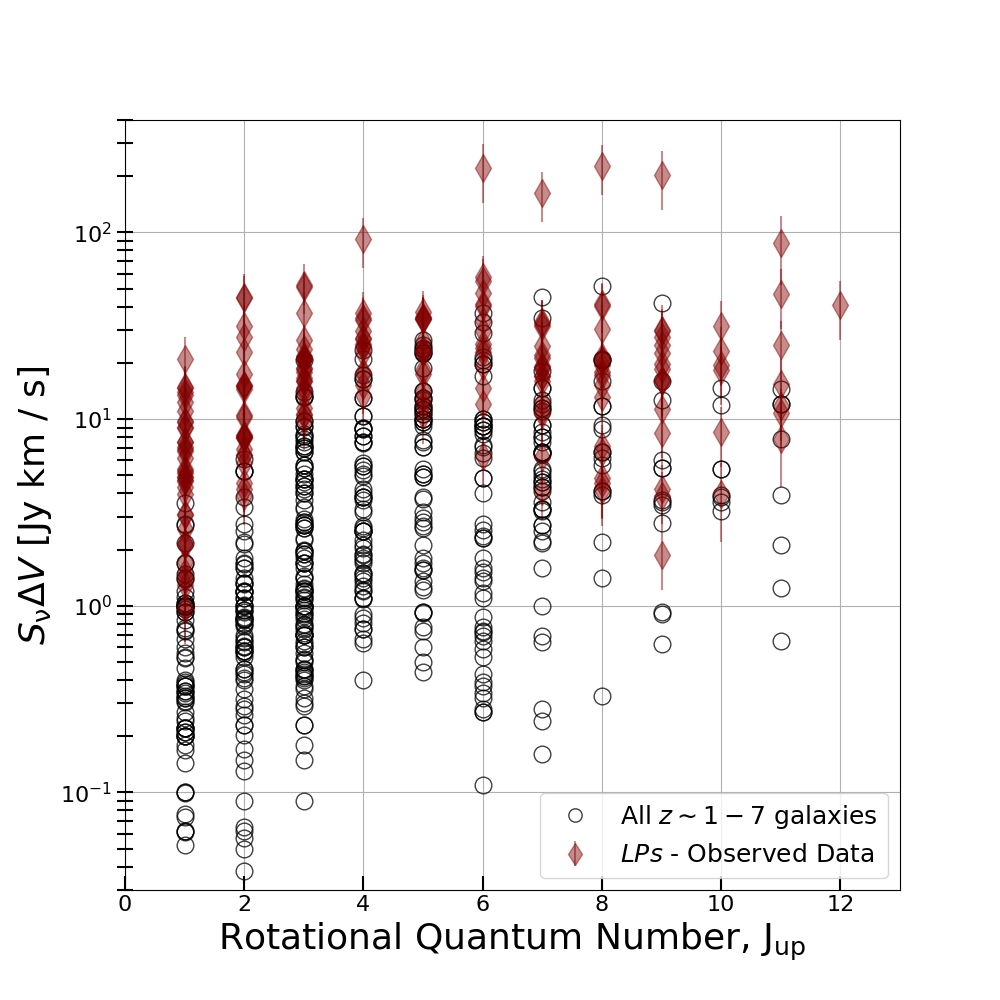}
 }
 \caption{Velocity-integrated flux density (log-scale) measurements plotted versus the CO rotational quantum number, J$_{\rm up}$, for the \LPs\ (red diamond). The black circles show the \LPs\ \& K19 $z \sim 1 - 7$ sample, described in \S \ref{thelines}. }
 \label{fig:snudvJ}
\end{figure}

Fig. \ref{fig:snudvJ} shows the measured velocity-integrated line fluxes, as reported in Table \ref{tab:linemeasurements}, compared to our literature compilation \citep[including ][]{carilli2013, yang2017, canameras2015, kirkpatrick2019}. Many CO line measurements have now probed more than 2 - 3 orders of magnitude in the observed velocity-integrated line flux densities across this large sample of $\sim$ 270 galaxies at $z \sim 1 - 7$. The \LPs\ are among the brightest CO sources on the sky, due to the magnification effects of strong lensing. \\

\begin{figure}
\resizebox{0.499\textwidth}{!}{%
 \includegraphics[width=0.98\textwidth]{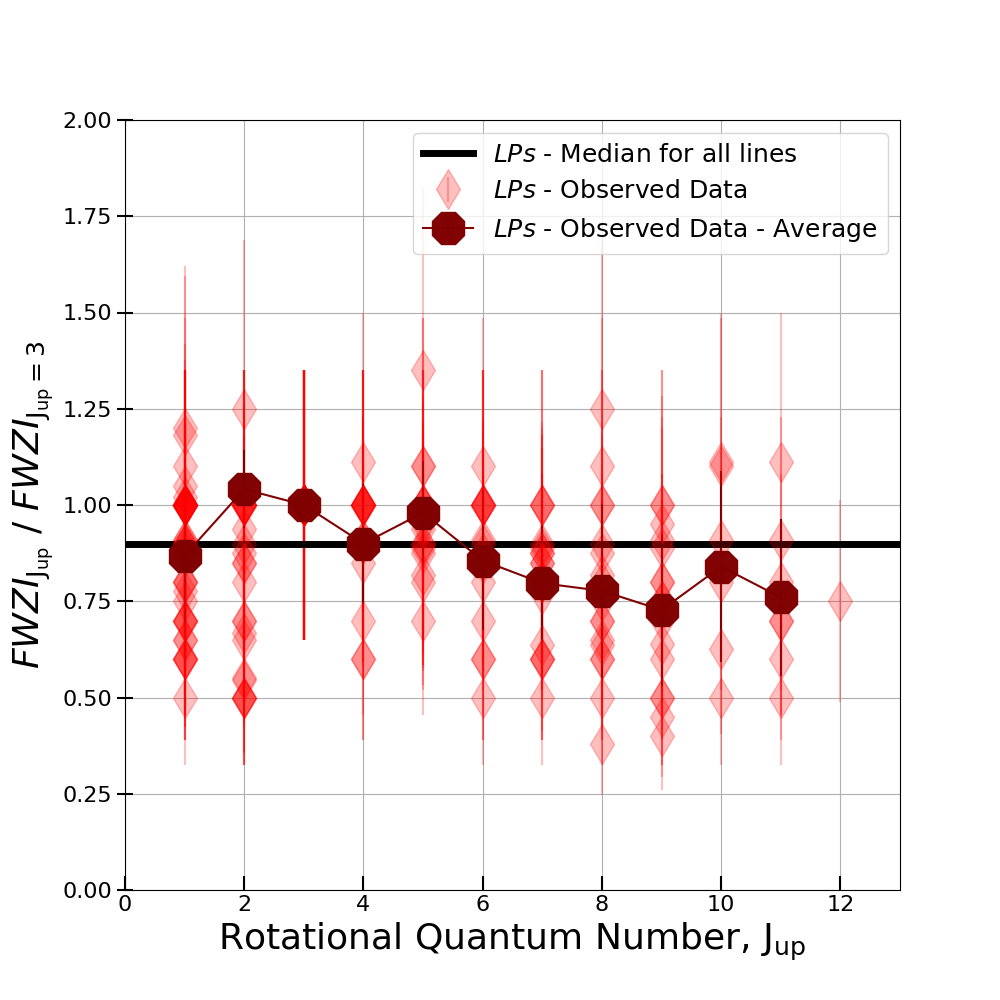}
 }
 \caption{Apparent velocity-integrated flux density measurements plotted versus the CO rotational quantum number, J$_{\rm up}$, for the \LPs\ (red diamond). }
 \label{fig:FWZIvsJ}
\end{figure}

To characterize the velocity integrated flux density we calculate the full line-width at zero intensity (FWZI) for all of the measured line fluxes Table \ref{tab:linemeasurements}. We note that neither all spectra have similar line shapes between the sample of \LPs, nor are all of the lines resolved at relatively high velocity resolution (e.g. 20-50 \kms), so we therefore avoid fitting simple 1-D Gaussian models to analyse the line-profiles. The mean and standard deviation of the FWZI for all of the CO lines for the entire sample of \LPs\ is 851 $\pm$ 183 \kms. In Fig. \ref{fig:FWZIvsJ}, the FWZI is normalized to the FWZI of the CO(3-2) line to examine whether the line profiles may change as a function of J$_{\rm up}$. The average value of this normalized FWZI decreases with increasing J$_{\rm up}$, although the large total uncertainties for the lines do not reveal a statistically significant trend. The \LPs\ with detections of higher-J CO lines show a similar line profile as seen in the emission lines of the lower-J rotational transitions. In some cases the FWZI of the high-J emission lines is narrower than the lower-J emission line profiles at comparable velocity resolution. This is apparent in LPs-J1336, as the CO(1-0; 5-4; 6-5; 8-7; 9-8; 10-9) emission lines have a comparable FWZI, whereas the strong detection of the CO(11-10) line reveals a FWZI that is roughly half. Narrow-line emission in the highest-J lines, compared to the lower-J lines, is observed in LPs-J0116, LPs-J0226, LPs-J1202, LPs-J1544. This is not seen in other systems with such high-J measurements. In 3/24 \LPs\ with large lensing arcs, pointed observations may only partially cover the entire emitting region. Therefore the observed emission is assumed to be representative of the galaxy-scale ISM of the lensed galaxy. In these cases, additional uncertainties have been added to the observed integrated line fluxes (\S \ref{sec:adopteduncertainties}).\\

The lens magnification factor, $\mu_{\rm L}$, may have a different value for the low- and high-J CO lines, and thus may yield a potential differential lensing effect (see Appendix \ref{difflens}). Differential lensing of the diffuse and dense molecular gas traced by [CI] and CO may be negligible in most, if not all, of the \LPs\ as the line profiles would have shown strong variations across the spectrally resolved line profiles. Since the average, normalized FWZI drops slightly (10 - 25\%) from the median value for the higher-J lines, such a change is not statistically significant given error uncertainties. Therefore, we are confident that differential lensing effects do not impact the general trends we present. The strong asymmetric lines may indicate\footnote{Asymmetric lines could also indicate viewing a galactic system at a specific edge on orientation which covers an asymmetric portion of a rotating spheroidal disk or a significantly turbulent environment \citep[e.g. ][]{puschnig2020}.} different magnification factors across the line profile \citep{leung2017}, while the dust and CO may be slightly offset in the source plane \citep{rivera2019}. These values will likely be the same for both the [CI] and CO lines, as the overall line shapes are similar. We discuss differential lensing in greater detail in Appendix \ref{difflens}, and briefly note that \citet{canameras2018b} report a 5-10\% difference in the flux-weighted mean magnification factor derived for the low-J CO and mm dust continuum, respectively, for 5/24 \LPs\ presented in this work. For 2/24 \LPs\ in this work, \citet{canameras2018b} reported less than 30\% differential lensing of the dust and low-J CO. We are unable to de-magnify the sources for the analyses presented in this work, since there a magnification factor for different lines does not exist. Some lensed SPT-selected galaxies have a noted range from negligible differences in the magnification for different CO line transitions, up to a factor two \citep{Apostolovski2019, Dong2019}. \\

\begin{figure*}
 \includegraphics[scale = 0.275]{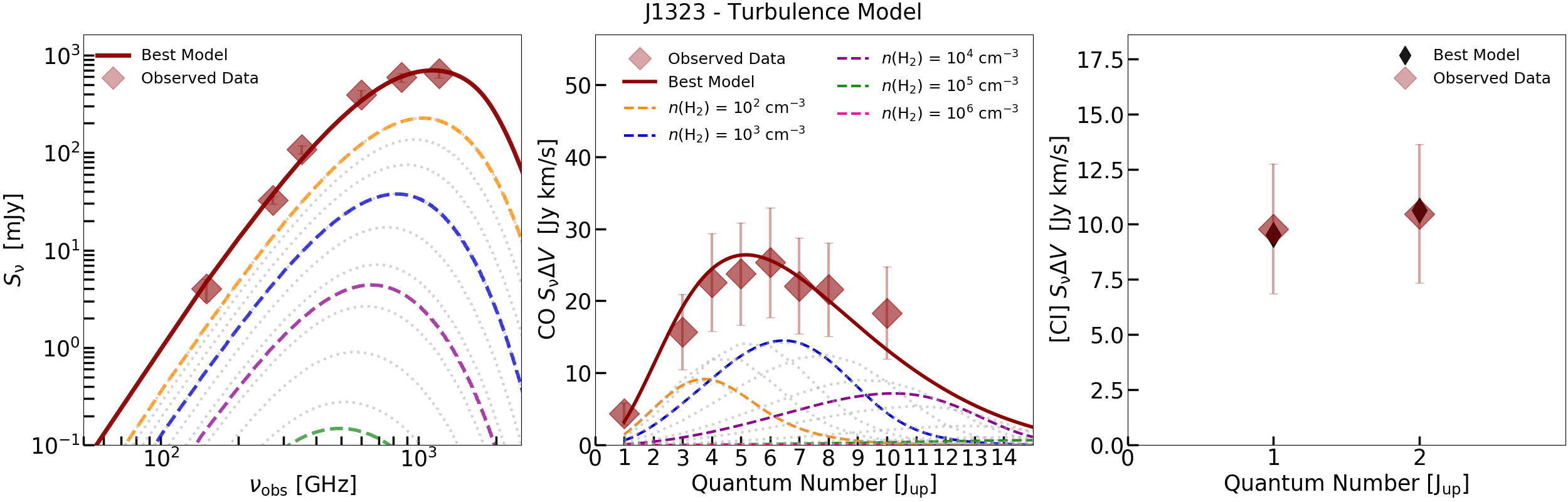}
 \includegraphics[scale = 0.275]{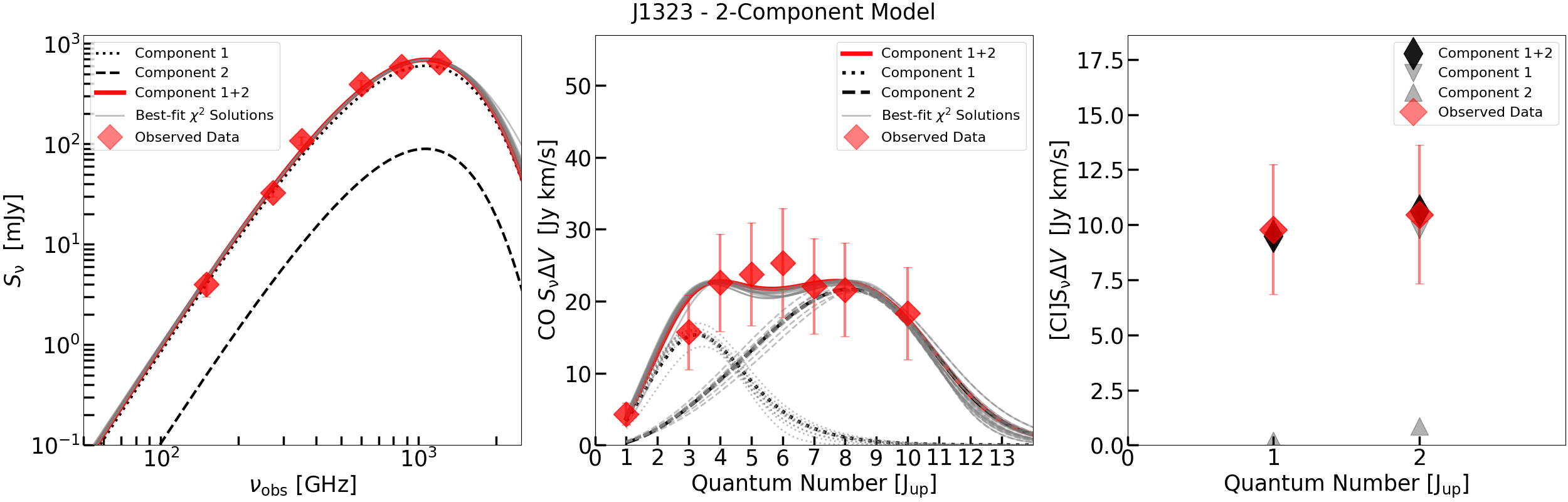}
 \caption{Best-fit, minimum-$\chi^{2}$ model solution for LPs-J1323. \textbf{Top:} \textit{Turbulence} model for the dust SED, CO and [CI] velocity-integrated line fluxes as determined by the best $\chi^{2}$. For clarity, different dashed-colored curves denote the representative contributions to the density PDF for the molecular gas densities of \logden\ = 2 (yellow), 3 (blue), 4 (purple), 5 (green) and 6 (pink) \percc. The gray-dashed lines represent the remaining LVG calculations (from the 50 total samples) which sample the gas density PDF (see \S\ref{modelling}). For the \textit{Turbulence} model, these individual density contributions have a y-axis scaled by a factor 5 for both the dust and line SED to facilitate interpretation of the dominant gas density. All observed data are shown as red diamonds. The best-fit [CI] line fluxes are plotted over the observed data. All solid red lines indicate the total best-fit, minimum-$\chi^{2}$ model. \textbf{Bottom:}  \textit{2-component} model curves for the lower excitation component (black dotted) and higher-excitation component (black dashed). The best-fit [CI] line flux from the lower-excitation component and higher-excitation component are denoted by a downward-facing and upward-facing gray triangle, respectively. }
 \label{fig:bfJ1323}
\end{figure*}


\section{Simultaneous Modeling of Line and Continuum Emission}
\label{modelling}

We utilize two state-of-the-art radiative transfer codes to simultaneously model both the observed line fluxes and measurements of the thermal dust continuum emission. This enables us to study the gas excitation conditions for the \LPs\ using (i.) a widely used approach to model two molecular gas components, and (ii.) a more realistic molecular ISM with a turbulence driven, lognormal density distribution for the gas density. The models we use are primarily derived from the equations presented in \citet{weiss2007}. For the analyses in this work, the primary modification to those models is that the gas and dust are now modelled simultaneously. Below we summarize the main properties of these modelling tools. A future work will enclose all of the details for the models. We will first describe the model which considers two gas components. Afterwards we will summarize the second model, which includes many of the same input parameters as the first model, despite being a more physically motivated, modified version.\\

\subsection{The \textit{2-component} model parameters}

\begin{deluxetable}{cccccccccccc}

\rotate



\tablecaption{Parameter Space}


\tablehead{\colhead{log$_{\rm 10}$ ($n_{\rm H_{\rm 2}}$)} & \colhead{$T_{\rm kin}$} & \colhead{$T_{\rm kin}$ / $T_{\rm dust}$} & \colhead{$\mu$R$_{\rm eff}$} & \colhead{$\Delta$V} & \colhead{$\kappa_{\rm vir}$} & \colhead{GDMR} & \colhead{CO/H$_{\rm 2}$} & \colhead{[CI]/H$_{\rm 2}$} & \colhead{$\beta_{\rm T_{\rm kin}}$} & \colhead{$\beta_{\rm [CI]/H_{\rm 2}}$} & \colhead{$\beta_{\rm T_{d}}$} \\ 
\colhead{(cm$^{-3}$)} & \colhead{(K)} & \colhead{(---)} & \colhead{(pc)} & \colhead{(km s$^{-1}$)} & \colhead{(km s$^{-1}$ pc$^{-1}$ cm$^{3/2}$)} & \colhead{(---)} & \colhead{(---)} & \colhead{(---)} & \colhead{(---)} & \colhead{(---)} & \colhead{(---)} } 

\startdata
1-7  & 15-600  & 0.5-6.0 & 0.1-19999 & 10-200  & 1.0-3.0 & 120-150 & 1 $\times 10^{-4}$-2 $\times 10^{-4}$ & 1 $\times 10^{-5}$-2 $\times 10^{-4}$ & --- & --- & 1.8-2.0 \\
1-7  & 15-600 & 0.5-6.0 & 0.1-19999 & 10-200 & 1.0-3.0 & 120-150 & 1 $\times 10^{-4}$-2 $\times 10^{-4}$ & 1 $\times 10^{-5}$-2 $\times 10^{-4}$ & -0.5-0.05 & -5.0-0.0 & 1.8-2.0 \\
\enddata


\tablecomments{The ranges in parameter space we explore in both models. \textbf{The top row corresponds to the \textit{2-component} model, while the bottom row corresponds to the \textit{Turbulence} model.}}

\label{tab:paramspace}
\end{deluxetable}

Our first model is a simple non-LTE radiative transfer model, referred to as the ``\textit{2-component}'' model. This model can, in principle, take into account an arbitrary number of molecular gas components. Nonetheless, due to the large number of coupled parameters and model degeneracies, we consider only two gas components, each with a unique, constant density. In Table \ref{tab:paramspace} we summarize the range in parameter space we explore for both models. We consider 14 free parameters for the \textit{2-component} model. Each of the two gas components in the \textit{2-component} model has seven free parameters: \logden\ (base 10), $T_{\rm kin}$, $\Delta V_{\rm turb}$, $\kappa_{\rm vir}$, $\sqrt{\mu_{\rm L}}R_{\rm eff}$,  [CI]/\Htwo,  and T$_{\rm kin}$/T$_{\rm d}$. We have specifically restricted the values of the other modelled parameters, i.e. $\beta_{T_{\rm d}}$, CO/\Htwo, $GDMR$. The latter two parameters limit our investigation to solar metallicity environments, as described below. \\

\textit{Effective radius: }\\
We parameterize the size of the emitting region by an effective radius which defines the apparent source solid angle $\Omega_{\rm app} = \mu_{\rm L} \, \frac{\pi R_{\rm eff}^{2}}{D_{\rm ang}^{2}}$  \citep[e.g. ][]{weiss2007}, using the angular diameter distance, $D_{\rm ang}$, $\mu_{\rm L}$ the magnification factor of the observed intensity, and $\sqrt{\mu_{\rm L} }R_{\rm eff}$ the apparent effective radius of the emission region. This apparent, effective disk radius would be equivalent to the intrinsic emitting region if the emission came from a filled aperture for an unlensed, face-on, circular disk \citep{weiss2007}. It is therefore a minimum radius as the emission may come from an un-filled aperture, which may be more widely distributed.  \\

\textit{Molecular gas density and gas/dust temperatures: }\\
For each gas component, we consider a range of molecular gas densities between \logden\ = 1 - 7 \percc. We probe gas kinetic temperatures, $T_{\rm kin}$, ranging between the redshifted CMB radiation temperature and 600 K (corresponding to the highest energy CO transition we model (J$_{\rm up}$ = 15)). We account for the possible decoupling of the gas and dust by setting a limit on the gas kinetic temperature, as $T_{\rm kin} \ge 0.5 T_{\rm d}$. The range we explore for this ratio of the gas kinetic temperature to the dust temperature (T$_{\rm kin}$/T$_{\rm d}$ = 0.5 - 6) is in agreement with theoretical work \citep[see the review by ][]{krumholz2014b}. In well-shielded regions that have \logden\ $ > 4.5$ \percc\ the molecular gas and dust may have a stronger coupling than in lower density environments. According to \citet{krumholz2014b}, within gas densities of \logden\ $>$ 6 \percc\ the gas and dust temperatures are expected to be nearly equivalent. We do not implement an explicit relationship between T$_{\rm kin}$/T$_{\rm d}$ and the gas density. We note that we simply solve for the gas kinetic temperature and dust temperature without modeling any specific heating mechanism. We note that theoretical and observational studies suggest the overall molecular ISM of a starburst/AGN galaxy could also be influenced by cosmic rays or X-rays \citep[e.g.]{meijerink2007}, and these heating mechanisms may be influential in determining the value we derive for the gas kinetic temperature. \\

\textit{Velocity gradient and turbulent velocity dispersion: }\\
The velocity gradient, $\delta v /\delta r $, and the molecular gas density together define dynamically bound or unbound systems, parameterized by the virial parameter, $\kappa_{\rm vir}$. The $\kappa_{\rm vir}$ parameter defines the relationship between turbulent and gravitational energy, and relates the velocity gradient and \Htwo\ gas volume density  \citep[see Eq. 2 of][]{goldsmith2001}. We explore a range corresponding to a physically bound molecular medium, up to a marginally unbound system, with $\kappa_{\rm vir} = 1 - 3$ \citep{greve2009, papadopoulos2012}. The $\Delta V_{\rm turb}$ parameter is the turbulent velocity dispersion for each component in the \textit{2-component} model calculations and is also, by definition, a mathematical term required for dimensional homogeneity. \\

\textit{Gas-phase abundances, metallicity and gas-to-dust-mass ratio: }\\
We leave the carbon abundance as a free parameter, probing ranges consistent with diffuse to dense giant molecular clouds (GMC) abundances of [CI]/\Htwo\ = 1$\times 10^{-5}$ - 2$\times 10^{-4}$. The wide range in [CI]/\Htwo\ reflects the chemistry in the dense gas. There is an expected decrease in the value of the carbon abundance for increasing molecular gas density, as chemical network calculations show that atomic carbon quickly disappears from the gas phase, and is transformed into other molecules \citep[see e.g. ][]{hollenbach1999, glover2012, goldbaum2016}. We assume an average Galactic disk value for the CO gas-phase abundance in the range of CO/\Htwo\ = 1 - 2$\times 10^{-4}$ to be consistent with the typical molecular abundance of giant molecular clouds in the Milky Way \citep{scoville1983, wilson1986, blake1987}. Instead of allowing the gas-to-dust-mass ratio parameter to range freely, we restrict this to a value between $GDMR = 120 - 150$ \citep{draine2011}; i.e., consistent with the observed value in the Milky Way \citep{draine2011}. Recent studies \citep[][and references therein]{casey2014} suggest that massive star-forming galaxies represent high density cosmic regions at $z = 1 - 3$. The fiducial value of solar metallicity, Milky-Way type values, are supported by our selection criteria to study extremely dusty star-forming galaxies with sufficient metal enrichment at high-\z. We may therefore expect the \LPs\ to have already accumulated at least a near-solar metallicity in a relatively short amount of time \citep{cen1999, bothwell2016}. Some derived quantities, such as the $GDMR$ and the $\alpha$ conversion factor, will depend on metallicity \citep{narayanan2012}. For galaxies with $2 - 3\, \times$ solar metallicity, the total gas mass comparisons would be impacted by a relative linear decrease in both the $GDMR$ and overall gas mass estimates from [CI] and CO. \\

\subsection{Computing the line and continuum fluxes}
\label{sec:linecontfluxes}
We model the line fluxes of the CO(1-0) to CO(15-14) transitions, corresponding to upper state energy levels $E_{\rm u} =$ 5.5 - 663.4 K. We use the collisional rate coefficients from \citet{flower2001} to solve for the balance of excitation and de-excitation from and to a given energy state. To compute the CO (and CI if detected) level populations, it is important to take the continuum radiation fields into account. We include: i.) the CMB radiation at the respective redshift, with $T_{\rm CMB} = 2.73 \times (1 + z_{\rm source})$ K, and ii.) the IR radiation field. \\

The apparent line flux densities, $\mu_{\rm L} S_{\rm CO/[CI]}$, are directly proportional to the physical apparent source solid angle and line brightness temperatures, $T_{\rm b}$. In the non-LTE, LVG framework we calculate the full radiative description of the $T_{\rm b}$, although it is classically defined by its equivalent representation on the Rayleigh-Jeans side of the emitting spectrum, i.e. $h\nu_{\rm obs} << kT_{\rm b} $. The values of $T_{\rm b}$ we compute depend on the gas volume density, the kinetic temperature, and the gas-phase abundance per velocity gradient in the LVG description. Altogether, the values of $T_{\rm b}$ are used to model the observed line fluxes:

\begin{equation}
 \mu_{\rm L} S_{\rm CO/[CI]} = \frac{T_{\rm b} \,2 k \nu_{\rm obs}^{2} \Omega_{\rm app} }{c^{2}\,(1 + z) },
\label{eq:eqTb}
\end{equation}
with $c$ the speed of light, $z$ redshift, $\nu_{\rm obs}$ the observed frequency of the CO or [CI] line and $k$ the Boltzmann constant.\\

We parameterize the observed dust continuum radiation field by the dust temperature, $T_{\rm d}$, dust emissivity index, $\beta_{\rm T_{d}}$, apparent dust mass and source solid angle. The latter three parameters also characterize the wavelength at which the dust opacity becomes unity, $\lambda_{\rm 0}$\footnote{$\lambda_{\rm 0}$ is directly proportional to the dust column density.}. We note that $\lambda_{\rm 0}$ is not a free parameter in our model, but can be computed from the apparent dust mass and source solid angle via Eq.\ref{eq:eqTaudust}. For simplicity, we restrict the $\beta_{\rm T_{d}}$ to a value between $\beta_{T_{\rm d}} = 1.8 - 2.0$ in both models to be consistent with previous studies of the \LPs\ \citep{canameras2015, harrington2016}. These values are also in agreement with the Milky Way average \citep{planckcollaboration2011b} and other studies of local and high redshift star-forming galaxies \citep{casey2014}\footnote{$\beta_{\rm T_{d}}$ is subject to large uncertainties in dust grain size distributions \citep{draine2011}.}.  \\

We then compute the full radiative transfer analysis for the two components in the \textit{2-component} model to derive both the dust opacity and line opacities in each calculation. The larger component can overlap with the more compact component, and we therefore take into account this difference when computing the overall dust SED \citep[see e.g.][]{Downes2007}. We keep the same frequency dependent dust emissivity index, $\beta_{T_{\rm d}}$, for each component. The dust optical depth is calculated using equations 2 and 3 of \citet{weiss2007}, assuming a frequency dependent dust mass absorption coefficient, $ \kappa_{\rm d}$ [cm$^{2}$ g$^{-1}$] \citep{krugel1994}; yielding 

\begin{equation}
\centering
\tau_{\rm \nu} = \frac{\kappa_{\rm d} \mu_{\rm L} M_{\rm d}}{\Omega_{\rm app} D_{\rm ang}^{2}} = \frac{0.4 (\nu_{\rm r}/250 {\rm GHz})^{\beta_{T_{\rm d}}} \,  \mu_{\rm L}M_{\rm d}}{\Omega_{\rm app} D_{\rm ang}^{2}}. 
\label{eq:eqTaudust}
\end{equation}

We connect the modelled line and continuum fluxes by using the derived dust opacity and inferred CO (or [CI]) gas column density to calculate the \Htwo\ gas column density, using equation 7 of \citet{weiss2007}. The $GDMR$ parameter is ultimately used to link the overall line fluxes and dust continuum in a self-consistent manner. We recall that the T$_{\rm kin}$/T$_{\rm d}$ parameter also links the line and continuum emission properties. We applied a prior for some of the \LPs (see best-fit model plots in the supplemental figures), with dust photometry limited mostly to \Planck\ measurements, so that the dust SED turns over beyond rest-frame flux-densities of $\sim S_{\rm > 6000 GHz,rest} $ (i.e. rest-frame $\sim $50$\mu$m for the $z \sim 2 - 3$ \LPs). This is in agreement with the physical conditions with which our model is sensitive to, i.e. the rest-frame FIR to mm wavelengths -- rather than near- and mid-IR wavelengths. This restriction also prevents a largely unconstrained (and also unphysical) solution space, with extremely high apparent FIR luminosity ($\mu_{\rm L}L_{\rm FIR: 40-120\mu m} > 10^{16}$ L$_{\rm \odot}$)\footnote{This is comparable to setting an upper limit for the dust temperature.}. Note, dusty high-\z\ star-forming galaxies, with full coverage of their thermal dust SED, fully support this prior \citep{strandet2016}. \\

\subsection{The \textit{Turbulence} model parameters}
The second model, hereafter the ``\textit{Turbulence}'' model, is more sophisticated in describing the molecular ISM. It has nine free parameters, and is represented as a single gas component described by a gas density PDF. We also model the line and continuum emission simultaneously in this model, including all of the same input parameters. For the \textit{Turbulence} model, the effective radius connects the source solid angle to the gas density PDF, which makes this model distinct from the \textit{2-component} model. In contrast, the \textit{2-component} model simply treats the gas density and the source solid angle as completely independent parameters, and also does not draw an explicit connection between the gas density and gas kinetic temperature.\\

We explore a broad range of values for the \Htwo\ gas volume density in the \textit{Turbulence} model, \logden\ = 1 - 10 \percc, with a restricted range for the mean density of the density PDF to values of \logden\ = 1 - 7 \percc. The mean molecular gas density thereby determines the other model parameter to describe the global mean ISM properties of the \LPs. We sample the best-fit, minimum-$\chi^{2}$ \textit{Turbulence} model density PDF by 50 bins. Each of the 50 gas densities are proportional to a solid angle that is occupied by that specific density bin. Therefore, each density corresponds to a radius -- such that the sum of all areas is normalized to the input source solid angle.  This implies that the model fit values for the dust and line SEDs are the sum of 50 individual LVG calculations which have used the value for each of those densities to calculate the relative emission properties. These altogether sum to the total line and continuum emission that has been measured.\\

There are two unique parameters for the \textit{Turbulence} model,  $\beta_{\rm [CI]}$ and $\beta_{\rm T_{\rm kin}}$. The power-law index, $\beta_{\rm [CI]}$, constrains the value of the carbon abundance relative to [CI]/\Htwo\ \citep{weiss2003}. We express $\beta_{\rm [CI]}$ as a power-law of the density. We further explore this parameter in \S \ref{sec:carbonprops}. The $\beta_{\rm T_{\rm kin}}$ parameter couples the gas kinetic temperature to the gas volume density by a power-law index, $\beta_{\rm T_{\rm kin}}$, as $T_{\rm kin} \propto $ \logden\ $^{\beta_{\rm T_{\rm kin}}}$, such that the more diffuse gas tends to have higher gas kinetic temperatures. This functional behavior has been well-studied in magneto-hydrodynamical simulations \citep{krumholz2014b}. The modelled galaxy-wide turbulent velocity dispersion, $\Delta V_{\rm turb}$, is a free parameter in the \textit{Turbulence} model. Although similar to the \textit{2-component} model, here it determines the width of the log-normal gas density PDF that is centered on the mean molecular gas density. 


\subsection{Fitting}
\label{sec:fit}
We use the parameter ranges in Table \ref{tab:paramspace} to model the observed data using a Bee Algorithm optimisation procedure \citep{Pham2009}, as used in \citet{strandet2017}. In this optimisation procedure, each model iteration attempts to solve for the observed data by exploring a number of model calculations ( hereafter 'bees') based on the free parameters. These 'bees' have a random initialisation within the defined parameter space, and record the model parameters with the best reduced $\chi^{2}$ value, as determined by the dust, CO and [CI] data. The parameter space is further explored by 'bees' which provide a fine sampling around the best $\chi^{2}$ regions, while other 'bees' continue to evaluate the parameter space randomly to avoid being trapped in a local minimum during each iteration. We evaluate $\sim 10^{5}$ models in each modelling procedure (for either the \textit{2-component} or \textit{Turbulence} model). To avoid repeatedly obtaining the same best-fit, minimum-$\chi^{2}$ values, and also to avoid remaining fixed in a narrow solution-space within the posterior probability distribution of each parameter, we re-generate this entire procedure multiple times, resulting in $\sim$2 million model evaluations per galaxy for each of the \textit{2-component} or \textit{Turbulence} models. To describe the mean, global gas excitation conditions of the \LPs, we refer primarily to the mean quantities and the standard deviation for the sample (Table \ref{tab:means}). The quantities presented in Table \ref{tab:means} are based on the total $\chi^{2}$-weighted mean parameter values for each of the individual \LPs, as evaluated for each parameter value from the $\sim$2 million models. The general trends and conclusions are not affected by the choice of the total $\chi^{2}$-weighted mean and standard deviation of the global properties we derive, as opposed to, e.g. the median (50th percentile) values. Since we present the modelling of spatially unresolved, galaxy integrated measurements, with large absolute calibration errors (\S \ref{sec:adopteduncertainties}), we adopt this mean quantity to reflect the average galaxy-wide properties based on the limitations of our data.\\ 


\section{Model Results}
\label{modelresults}
The best-fit, minimum-$\chi^{2}$ model plots for all of the \LPs\ can be accessed online, and we show an example below, for LPs-J1323 (Fig.~\ref{fig:bfJ1323}). We plot this best-fit model values for the dust SED, CO spectral line energy distribution (line SED) and both ground-state [CI] velocity-integrated line fluxes. For the dust SED and CO line SEDs in the  \textit{Turbulence} model we also plot the relative contribution from each of the density PDFs. To facilitate comparison to the total observed best-fit model, we arbitrarily increase the y-axis value for each density component to scale the individual LVG calculations. These calculations, representative of different densities sampling the gas density PDF, are shown in different colors to visualize which mean density dominates the observed intensities. The accompanying figure for the best-fit \textit{2-component} model shows the true y-axis values for the relative contributions from component one and component two -- which add to the total observed data points. \\

\subsection{CO Line SEDs}
\label{bestfit}

\begin{figure*}
 \includegraphics[scale = 0.349]{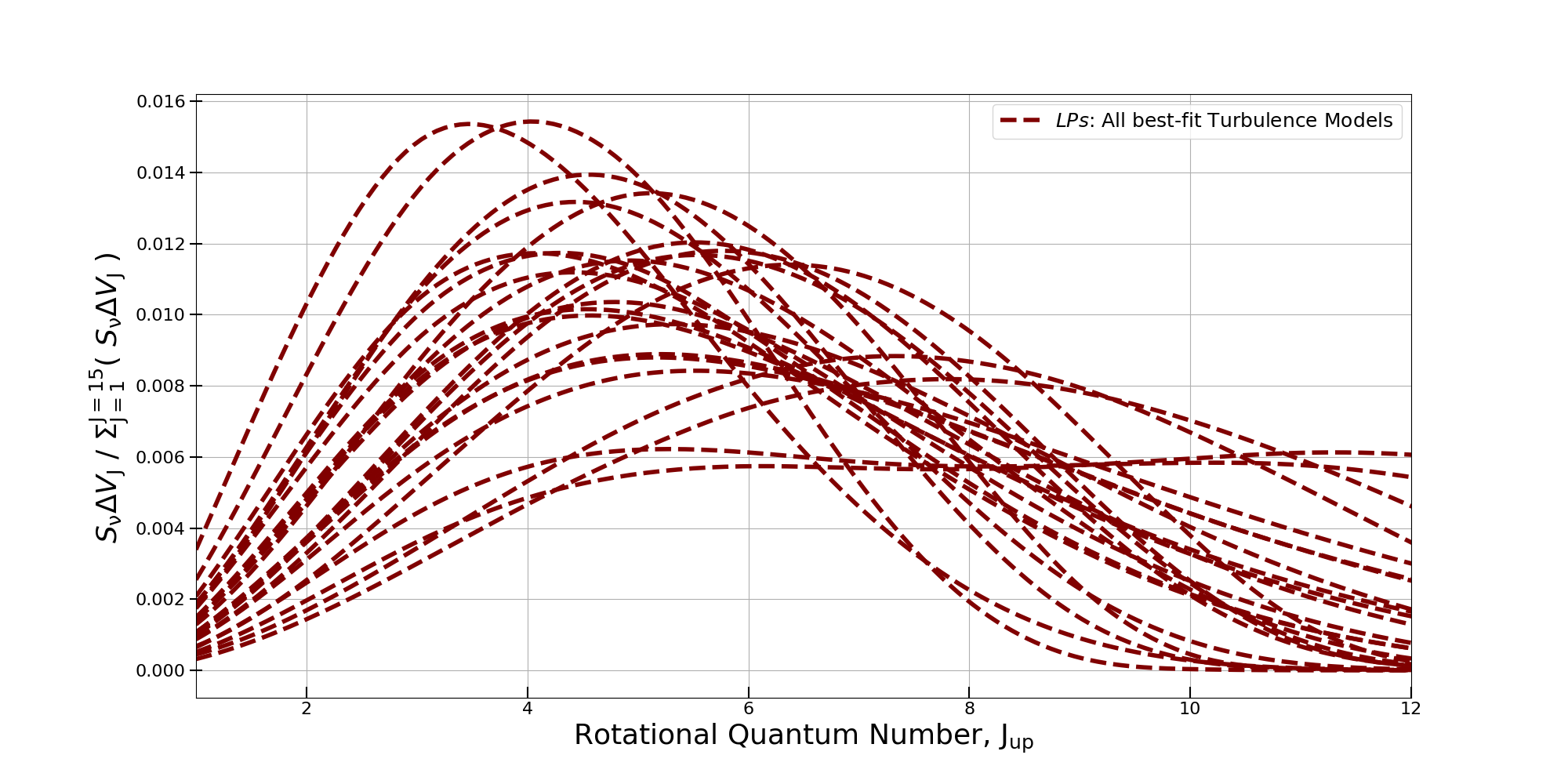}
 \caption{Best-fit, minimum-$\chi^{2}$ \textit{Turbulence} models for the CO velocity-integrated line fluxes, normalized by the sum of all J$_{\rm up} = 1 - 15$ velocity-integrated line fluxes, for all of the \LPs. }
 \label{fig:bfAll}
\end{figure*}

The majority of the \LPs\ show a broad peak in the CO line SED at J$_{\rm up} = 4-6$. The observed dust emission arises from molecular gas with \logden\ = 2-3 \percc, while the observed CO excitation ladders are dominated by \logden\ = 3-4 \percc. There is sustained CO excitation out to J$_{\rm up} = 9 - 11$ in most of the \LPs. For this emission, molecular gas with \logden\ = 4-5 \percc\ is dominant (see e.g. LPs-J0209, LPs-J1329, LPs-J1138). \\


To examine the dispersion in gas excitation conditions, Fig. \ref{fig:bfAll} shows all of the best-fit, minimum-$\chi^{2}$ \textit{Turbulence} model-derived CO velocity-integrated line fluxes, normalized by the sum of all J$_{\rm up} = 1 - 15$ velocity-integrated line fluxes. This normalisation indicates the relative strength between various line transitions among the sample, and we will present a more quantitative classification of the broad range in gas excitation conditions in \S \ref{sec:classes}. Using the velocity integrated line fluxes, we can calculate the line luminosity of CO or [CI] in terms of the area-integrated line surface brightness, $L'_{\rm CO/[CI]}$. We calculate this value using the standard equations presented in \citet{solomon1997}. The ratio of this value for any two CO transitions will provide an average estimate of the intrinsic brightness temperature, $T_{b}$, ratio within the CO emitting gas. For the two lowest rotational transitions, this ratio is often close to unity for active star-forming systems \citep{carilli2013}, assuming that the two lines have the same spatial extent on average with the same $T_{b}$, such that the two lines are thermalized. Table \ref{tab:Lprimeratio} shows the best-fit, minimum-$\chi^{2}$ model derived line ratios from our physically motivated \textit{Turbulence} model, yielding systematically derived values for the brightness temperature ratios corresponding to the ratio of $L'_{ \rm  CO(    J_{\rm up}-(J_{\rm up}-1)    )   }$/$L'_{\rm CO(1-0)}$, denoted as R$_{\rm J_{\rm up},1} = 1$. Table~\ref{tab:Lprimeratio} summarizes the mean and standard deviation of the \LPs\ sample from these best-fit, minimum-$\chi^{2}$ \textit{Turbulence} models, and we will discuss Table~\ref{tab:Lprimeratio} in more detail in \S \ref{sec:highzgas}.\\

\begin{deluxetable*}{cccccc}
\label{tab:Lprimeratio}



\tablecaption{Line brightness temperature ratios}


\tablehead{\colhead{J$_{\rm up}$} & \colhead{\LPs\ } & \colhead{\LPs\ \& K19} & \colhead{K19 All Sources''} & \colhead{CW13 SMGs} & \colhead{CW13 QSOs} \\ 
\colhead{(---)} & \colhead{(---)} & \colhead{(---)} & \colhead{(---)} & \colhead{(---)} & \colhead{(---)} } 

\startdata
1 & 1 & 1 & 1 & 1 & 1 \\
2 & 0.88 $\pm$ 0.07 & 0.73 $\pm$ 0.10 & 0.78 $\pm$ 0.13 & 0.85 & 0.99 \\
3 & 0.69 $\pm$ 0.12 & 0.75 $\pm$ 0.11 & 0.78 $\pm$ 0.14 & 0.66 & 0.97 \\
4 & 0.52 $\pm$ 0.14 & 0.46 $\pm$ 0.07 & 0.49 $\pm$ 0.10 & 0.46 & 0.87 \\
5 & 0.37 $\pm$ 0.15 & 0.36 $\pm$ 0.06 & 0.34 $\pm$ 0.07 & 0.39 & 0.69 \\
6 & 0.25 $\pm$ 0.14 & 0.28 $\pm$ 0.04 & 0.31 $\pm$ 0.06 & --- & --- \\
7 & 0.17 $\pm$ 0.12 & 0.18 $\pm$ 0.03 & 0.21 $\pm$ 0.04 & --- & --- \\
8 & 0.11 $\pm$ 0.09 & 0.08 $\pm$ 0.02 & 0.11 $\pm$ 0.03 & --- & --- \\
9 & 0.07 $\pm$ 0.07 & 0.07 $\pm$ 0.02 & 0.14 $\pm$ 0.04 & --- & --- \\
10 & 0.04 $\pm$ 0.05 & 0.07 $\pm$ 0.02 & 0.08 $\pm$ 0.04 & --- & --- \\
11 & 0.02 $\pm$ 0.02 & 0.05 $\pm$ 0.02 & 0.12 $\pm$ 0.03 & --- & --- \\
12 & 0.02 $\pm$ 0.03 & 0.02 $\pm$ 0.01 & --- & --- & --- \\
\enddata


\tablecomments{Mean and 1-$\sigma$ standard deviation of the line brightness temperature ratios among the sample of 24 \LPs\ based on the best-fit, minimum-$\chi^{2}$ \textit{Turbulence} models. The \LPs\ \& K19 sample catalog of all observed CO lines, in $\sim$270 $z \sim 1 - 7$ galaxies, is used to derive the median line ratios, as described in \S \label{sec:highzgas}. The All sources'' ($z$ = 1 - 7) sample of heterogenously selected galaxies with CO line detections \citet{kirkpatrick2019}. The quoted values in the last two columns are from \citet{carilli2013} for the average values for both high-\z\ (sub)mm bright CO emitters (SMG) and quasars (QSO). }


\end{deluxetable*}

\subsection{Physical gas properties of the \LPs\ }

\subsubsection{CO and [CI] Line Opacities }
\label{opacity}

We now focus on the best-fit values for the CO and [CI] line opacities, as derived in the \textit{2-component} model. In the LVG approximation, we consider an emitting region of gas that is excited due to both the collisional interactions and the external radiation field. The observed line fluxes are computed using the line opacities, $\tau$ and the standard LVG assumption of the escape probability method formalism, which defines the probability of a photon escaping or entering the medium. As noted in other studies \citep{scoville1974}, this probability is proportional to $(1 + \tau)^{-1}$. 

The observed line and continuum fluxes are determined by the relevant gas-phase abundance(s), volume density and the molecular gas kinetic temperature (i.e. the Maxwellian velocity distribution). Overall, these effects shape the value of the line opacity, specifically as

\begin{equation}
\tau_{\rm CO/CI} \propto \frac{\rm N(mol) [cm^{-2}] }{\rm \Delta V_{\rm turb} {\rm [ km s^{-1}]} } \propto \frac{\rm n({\rm H_{\rm 2} }) {\rm [cm^{-3}]} \times [mol]/[H_{\rm 2}] }{\delta v/\delta r {\rm [km\, s^{-1}/pc]} },
\label{eq:eqTauline}
\end{equation}

where `N(mol)' is, here, the CO or [CI] gas column density, $\Delta V_{\rm turb}$ is the galaxy-wide turbulent velocity, $\delta v/\delta r$ is the large-scale, systemic velocity gradient, of the molecular/atomic gas, and `$n{\rm (H_{\rm 2})}$' is the \Htwo\ gas volume density. \\

\begin{figure*}
 \includegraphics[width=\columnwidth]{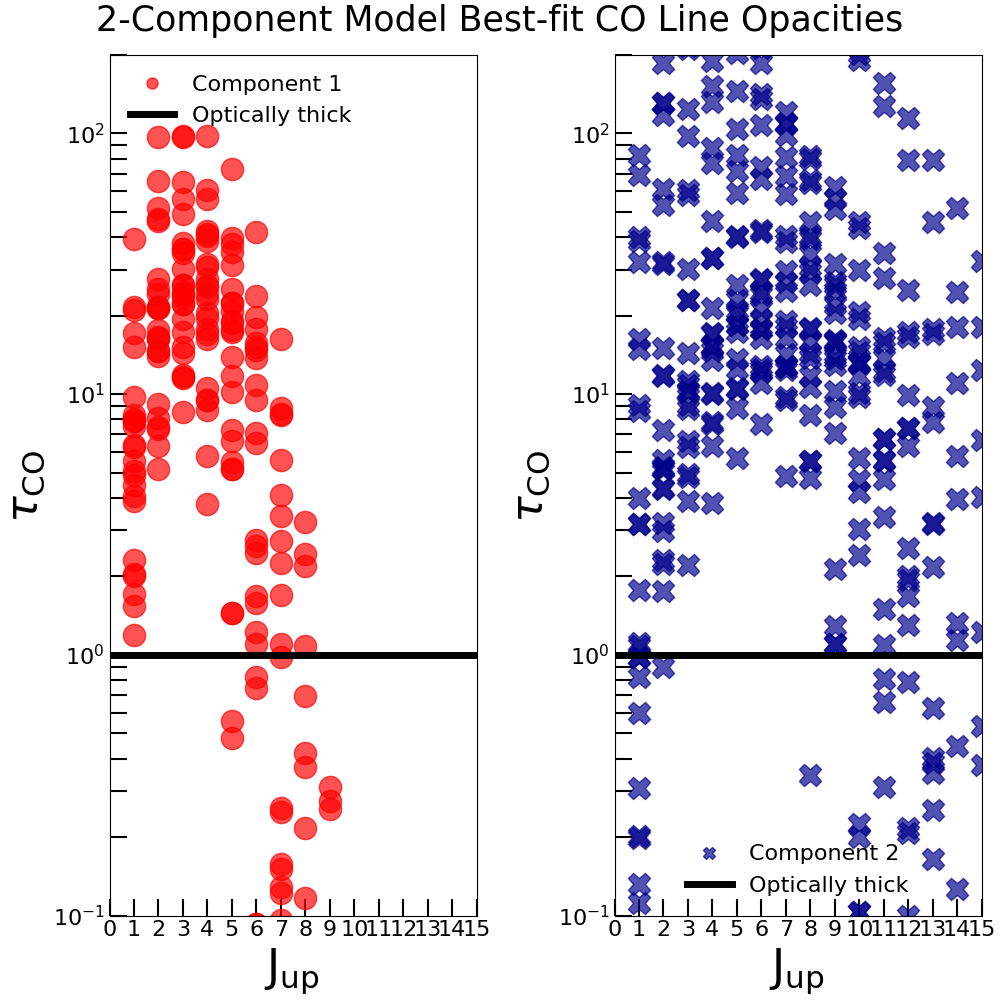}
 \includegraphics[width=\columnwidth]{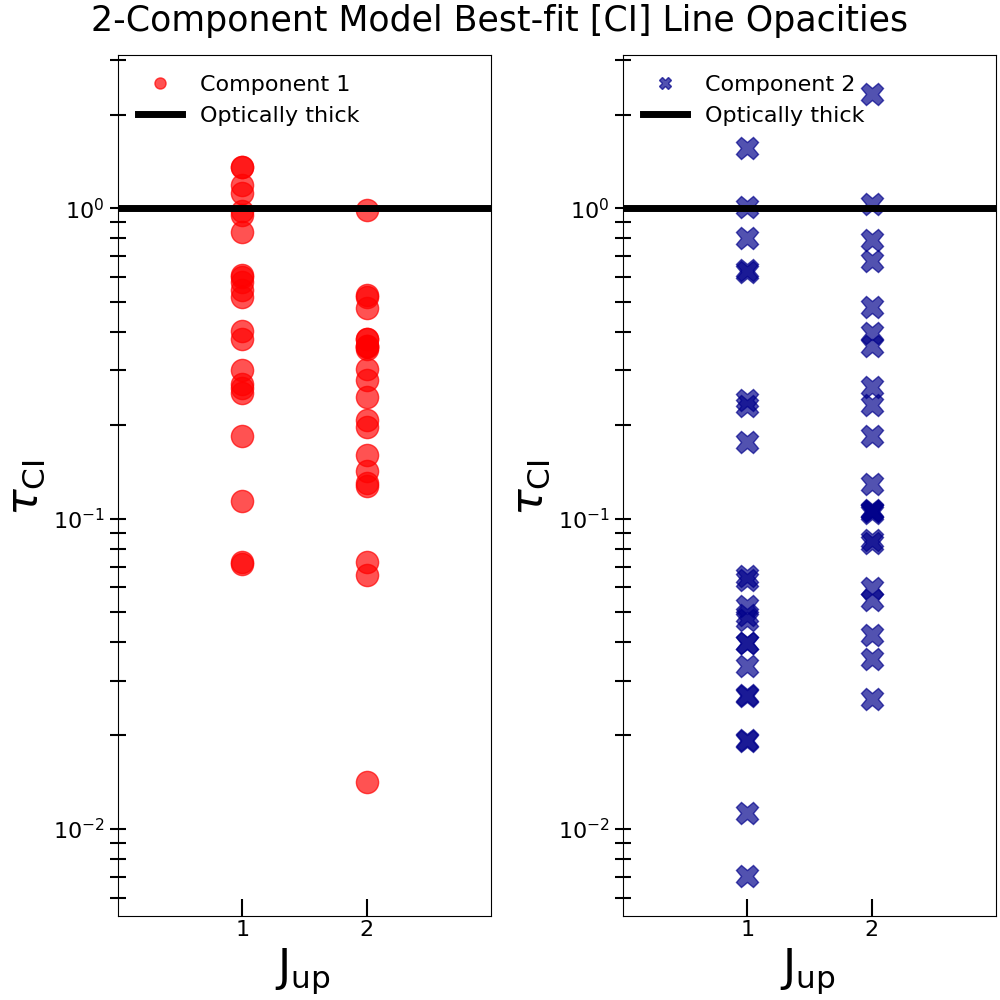}
 \caption{Best-fit results from the \textit{2-component} model for the calculated line opacities. \textbf{Left:} CO line opacity versus rotational quantum number, J$_{\rm up}$. \textbf{Right:} [CI] line opacity versus quantum level number. The solid black line indicates an optical depth of unity. The average best-fit, minimum-$\chi^{2}$ model uncertainty is smaller than the marker size.}
 \label{fig:lineopacity}
\end{figure*}

Figure \ref{fig:lineopacity} shows, for all \LPs, the line opacities we derive for the upper state levels for each line transition from the best-fit, minimum-$\chi^{2}$ \textit{2-component} model results. We confirm the common assumption that the CO lines are optically thick and the atomic carbon fine-structure lines are optically thin. The CO line opacity depends on both the level population in the upper energy level state (the effective CO column density) and the galaxy-wide turbulent velocity dispersion. For a fixed column density, the higher the turbulent velocity, the lower the line opacity \citep[see e.g. ][]{narayanan2014}. As shown for both components in each of the \LPs, the CO line opacity first increases with J$_{\rm up}$ before it decreases progressively as the individual level populations are less frequently excited out to higher-J. \\

Fig.~\ref{fig:lineopacity} also shows that the CO lines often do not freely radiate their emission, i.e. they are still optically thick, until J$_{\rm up} = 6 - 8$ and J$_{\rm up} = 8 - 15$ in components one and two, respectively. The more highly excited second component remains optically thick beyond J$_{\rm up} = 15$ in some cases. Also, as the second component is warmer and denser, the J$_{\rm up}$ = 0 and J$_{\rm up}$  = 1 levels are less populated, thereby more systems may exhibit optically thin CO(1-0) line emission in this more highly excited component. In general, beyond CO(8-7), the contribution to the total CO partition drops significantly and molecules will populate those higher states less frequently on average.\\

Importantly, the gas does not need to be diffuse in order to be optically thin. In fact, the second component, which we discuss later to be the denser component (\S \ref{meanprops}), has more instances of the CO(1-0) line being optically thin. The lower density gas has a higher opacity, and a CO partition function that is weighed heavily by the lower-J lines. Our results show that as the density increases, the lines become more distributed across the CO partition function, which results in optically thin CO(1-0) line emission in the denser gas. This is consistent with theoretical work of \citet{narayanan2014}, which had utilized both hydrodynamic simulations and radiative transfer analyses in order to calculate the CO line excitation for various idealized disk and merger galaxies at $z > 1$.    \\

\subsubsection{Characterizing the molecular ISM properties}
\label{meanprops}


\begin{deluxetable}{cccccccccc}

\rotate



\tablecaption{Mean parameter values for the sample \LPs }


\tablehead{\colhead{log$_{10}$($ n_{   H_{ \rm 2}  } $)} & \colhead{$T_{\rm k}$} & \colhead{$T_{\rm d}$} & \colhead{$T_{\rm k}$ / $T_{\rm d}$} & \colhead{$\kappa_{\rm vir}$} & \colhead{$\Delta V_{\rm turb}$} & \colhead{$\sqrt{\mu_{\rm L}}R_{\rm eff}$} & \colhead{$GDMR$} & \colhead{$M_{\rm ISM}$} & \colhead{[CI]/H$_{\rm 2}$} \\ 
\colhead{(cm$^{-3}$)} & \colhead{(K)} & \colhead{(K)} & \colhead{(---)} & \colhead{(km s$^{-1}$ pc$^{-1}$ cm$^{3/2}$)} & \colhead{(km s$^{-1}$)} & \colhead{(pc)} & \colhead{(---)} & \colhead{(M$_{\rm \odot}$)} & \colhead{(---)} } 

\startdata
4.31 $\pm$ 0.88 & 119 $\pm$ 77 & 44.7 $\pm$ 9.75 & 2.56 $\pm$ 1.30 & 1.45 $\pm$ 0.36 & 125 $\pm$ 40 & 13534 $\pm$ 3147 & 130 $\pm$ 4.2 & 2.68E+12 $\pm$ 1.28E+12 & 6.82E-05 $\pm$ 3.04E-05 \\
\enddata


\tablecomments{The mean and standard deviation for all $\chi^{2}$-weighted mean parameter values across the sample of \LPs, as evaluated for each parameter value from $\sim$2 million model evaluations.}

\label{tab:means}
\end{deluxetable}

Table ~\ref{tab:means} shows the mean and standard deviation value across the sample of \LPs\ for the main free parameters. Note, the median values for the sample of \LPs\ does not differ within the uncertainties. The \LPs\ have a mean \Htwo\ gas density and galaxy-wide mean turbulent velocity of $<$\logden$> = 4.3 \pm 0.9 $ \percc, and $< \Delta V_{\rm turb} > = 125 \pm 40 $ \kms, respectively. The mean value of the gas kinetic temperatures for both components in the \textit{2-component} model are roughly equivalent to the mean kinetic temperature using the \textit{Turbulence} model -- despite the inherent differences in the physical assumptions of each model. 

\begin{figure}
\centering
\resizebox{0.487\textwidth}{!}{%
 \includegraphics[scale = 0.3]{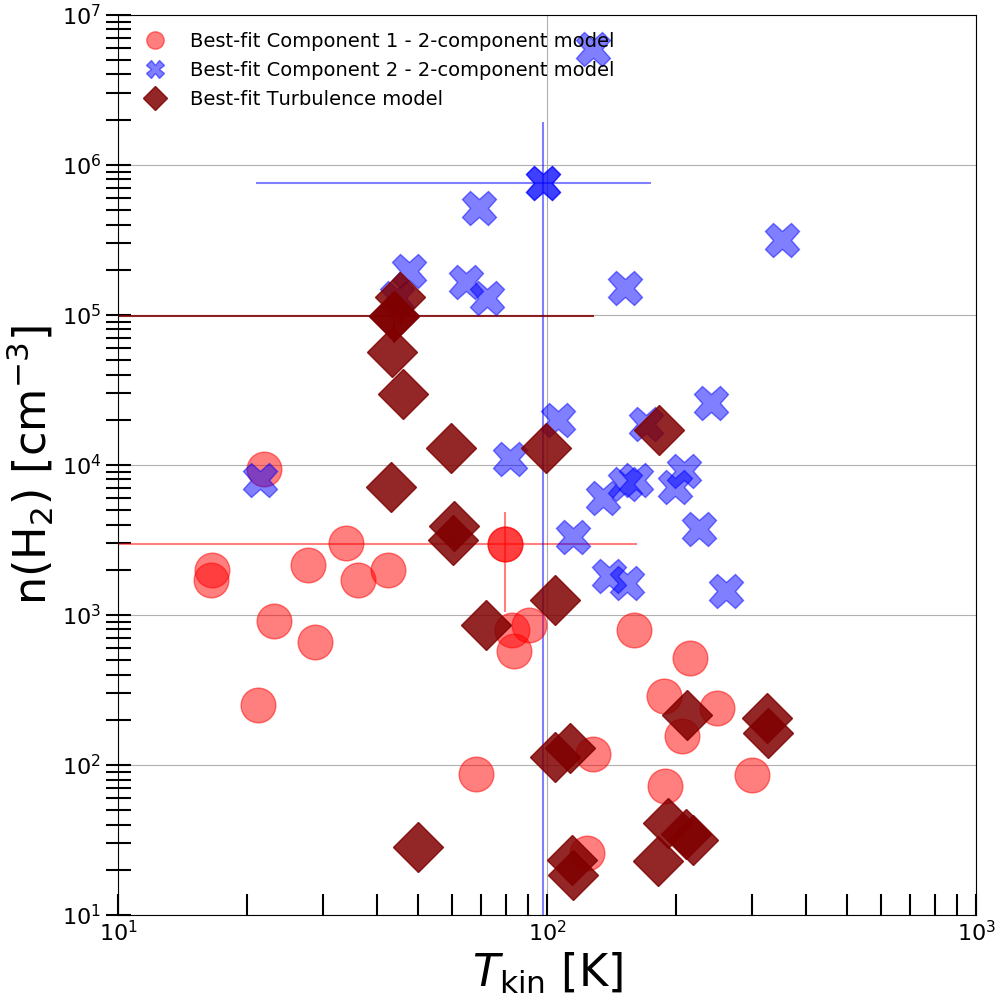}
}\\
 \caption{Best-fit, minimum-$\chi^{2}$ solutions for the \Htwo\ gas volume density and gas kinetic temperature derived for the \textit{2-component} and \textit{Turbulence} models (log-log-scale). We also show representative errors for both models. }
 \label{fig:TkinDen}
\end{figure}

Using the derived dust temperature, we find the mean ratio of $ < T_{\rm kin}/ T_{\rm d} > = 2.6 \pm 1.3$. For mean densities above $\sim$10$^{4.5}$ \percc , the value of $ T_{\rm kin}/ T_{\rm d} $ converges closer to unity as the molecular gas and dust become coupled \citep{goldsmith2001}. This is indeed derived for the denser component, i.e. component two, which has a higher mean density, but lower $ T_{\rm kin}/ T_{\rm d} $, for the \textit{2-component} model. This is also consistent in the \textit{Turbulence} model results, as the \LPs\ with higher mean density have lower values of $ T_{\rm kin}/ T_{\rm d} $. The mean apparent radius of the \LPs\ is found to be $\sqrt{\mu_{\rm L}}R_{\rm eff} \sim 10 - 15 $ kpc for the \textit{Turbulence} model. For the \textit{2-component} model, we derive a mean effective radius of $\sqrt{\mu_{\rm L}}R_{\rm eff} \sim 8.5$ kpc and $\sim 3.3 $kpc, for the more diffuse and denser components, respectively.\\

We explore the relationship between the value of $ T_{\rm kin}$ and \Htwo\ gas density in Fig. \ref{fig:TkinDen}. We plot the best-fit, minimum-$\chi^{2}$ solutions obtained for both components in the \textit{2-component} model. We also compare to the best-fit, minimum-$\chi^{2}$ model results from the more physically motivated \textit{Turbulence} model. In general, for the \textit{2-component} model, the first component tends to have a lower \Htwo\ gas density than the second component. There is a large dispersion in the value of $ T_{\rm kin}$ for both components, although the first component tends to have more values at lower $ T_{\rm kin}$. Since we find a similar range in $ T_{\rm kin}$  for both components, it is clear that indeed the higher-J CO lines are driven mostly by the fact that the densities are higher, and that the $ T_{\rm kin}$  plays a secondary role. This relation between the second component of the \textit{2-component} model, and the observed high-J CO transitions is shown for all of the \LPs\ in the online supplemental version to this manuscript. As shown, the second component may be largely unconstrained and have best-fit, minimum-$\chi^{2}$ solutions for the density which are unlikely based upon examination of the more realistic results for the \textit{Turbulence} model. Fig. \ref{fig:TkinDen} also shows that for the best-fit, minimum-$\chi^{2}$ values of the \Htwo\ gas density for the \textit{2-component} model, the second component is always denser than the first. We find that the dominant emitting component associated with the excitation of the lower-J CO lines, has a mean volume density \logden\ $ = 2.2 - 3.7 $ \percc, while the second component has a mean volume density between \logden\ $= 3.2 - 6.4 $ \percc, consistent with the observed trends in the line opacities seen in Fig. \ref{fig:lineopacity}. Altogether, the \LPs\ have a pervasive, dense, and highly active ISM with an average gas kinetic temperature $< T_{\rm kin} > = 120 \pm 77$ K. The median values for both components in the \textit{2-component} model are $\sim 81$ K and $\sim 137$ K, for component one and two respectively. The Kendall's tau coefficient $\tau = 0.05$ indicates the gas kinetic temperature of component one and two are uncorrelated. This suggests that the diffuse (component one) and dense (component two) gas, although both relatively warm, share a distinct range of temperatures. The 2-sample Kolmogorov-Smirnov test also confirms that the gas kinetic temperatures likely share a different distribution for component one and two, respectively, with a $p$-value of 0.06\footnote{A $p$-value of 0.05 or less allows one to reject the null hypothesis that the two samples of kinetic temperatures come from the same distribution.}.\\
\\

To evaluate the apparent FIR luminosities we compute the integrated rest-frame FIR (40-120$\mu m$) dust SED. As noted above in \S\ref{modelling}, in the \textit{Turbulence} model we add together the 50 LVG calculations which sample the mean gas density PDF (constant $GDMR$ for each calculation), to further derive the total dust SED. We calculate a wide range in apparent FIR luminosities among the \LPs\ of $\mu_{\rm L} L_{\rm FIR} = 8 - 470 \times 10^{12} {\rm L}_{\rm \odot}$, with the corresponding dust temperatures of $\sim$40 - 50 K. The contribution, per component, of the FIR luminosity is approximately divided among the \LPs\ for the \textit{2-component} model. The more highly excited component contributes $\sim$50\% of the total $\mu_{\rm L} L_{\rm FIR}$, on average, with a large dispersion. Following the traditional method in \citet{kennicutt2012} we integrate the total IR luminosity between 8-1000$\mu m$ to derive a mean apparent $\mu_{\rm L} SFR = 35.6 \pm 4.4 \times 10^{3} $ M$_{\rm \odot } {\rm yr}^{-1}$. With an average magnification factor of 20, this would correspond to an intrinsic mean SFR for the \LPs\ of order 1500 M$_{\rm \odot } {\rm yr}^{-1}$.\\

The dust opacity for the \LPs\ becomes unity for each component at wavelengths comparable to what is expected, i.e. $\geq$ rest-frame 100 $\mu m$. This is consistent with other studies of optically thick dust within the ISM of local and high-\z\ star-forming systems \citep{blain2003, huang2014, lutz2016, greve2012, spilker2016, hodge2016, simpson2017}. Overall, there is a range of values for $\lambda_{\rm 0}$, from a few tens of $\mu m$ to $\sim$100-300 $\mu m$ for both component one and component two, although the latter \footnote{We note that the value of $\lambda_{\rm 0}$ for each component are in agreement for the \LPs\ which we have applied a dust SED prior (see \S\ref{sec:linecontfluxes}).}. We find a range of values between $N_{\rm H_{\rm 2}} \sim $ 1 - 10$\times 10^{23}$cm$^{-2}$ and $N_{\rm H_{\rm 2}} \sim $ 0.5 - 50$\times 10^{24}$cm$^{-2}$ for component one and component two, respectively. We estimate the effective optical extinction, $A_{\rm V}$ (in magnitudes), using the result for the Milky Way from \citet{guver2009}, i.e.

\begin{equation}
{\rm N_{\rm H} ({\rm cm^{-2} })} = 2.2\times10^{21} A_{\rm V}.
\end{equation}
We find a value of $A_{\rm V} > 450$, for a fiducial value of $N_{\rm H_{\rm 2}} \sim $ 1 $\times 10^{24}$cm$^{-2}$ for the \LPs. The \Htwo\ gas column densities in the second component of the most extreme \LPs\ resembles regions similar to local starbursts and even comparable to the rare, highly dust enshrouded local starbursts exceeding $N_{\rm H_{\rm 2}} = $ 10$^{24-25}$cm$^{-2}$ \citep{sanders1996, scoville2017arp220}). The \textit{2-component} model does not recover small regions, and therefore any emission from such compact, high gas column density, galactic nucleus regions would not dominate the total emission. In fact, the \textit{Turbulence} model indicates the smallest regions, corresponding to the highest gas density, contribute a negligible amount to the total line and dust SEDs (Fig. \ref{fig:bfJ1323}). It is often assumed that for dusty star-forming galaxies, with SFRs $> 100 - 1000$ M$_{\rm \odot }$ yr$^{-1}$, that the thermal dust emission transitions from optically thick to optically thin beyond wavelengths of $\sim$ 100$\mu m$ or more \citep{casey2018a, casey2018b}. This assumption is also verified in local star-forming systems \citep[e.g ][]{Downes1993,scoville2014, scoville2016}, but is difficult to constrain based on limited observations of individual high-\z\ systems. \\

\subsubsection{Total molecular ISM mass estimates}
\label{totalmass}
We define the total, apparent, molecular gas mass\footnote{Corrected by the Helium abundance, a factor 1.36 \citep{allen1973}. }, $\mu_{\rm L} M_{\rm ISM}$, based on our non-LTE radiative transfer LVG calculations of the \Htwo\ gas column density and the effective radius. The \Htwo\ gas column density is directly proportional to the volume density, $n(H_{\rm 2})$, multiplied by the equivalent path-length of the molecules, i.e. $\Delta V_{\rm turb} (\delta v/ \delta r)^{-1}$, and therefore yields, together with the effective radius,

\begin{equation}
\mu_{\rm L} M_{\rm ISM} \propto \frac{\mu_{\rm L} \, R_{\rm eff}^{2}  \, [{\rm pc}^{2}]  \times n (H_{\rm 2}) \, [{\rm cm}^{-3}] \times \Delta V_{\rm turb} \, [{\rm km s^{-1}} ]}{\delta v/ \delta r \, [{\rm km \, s^{-1}/pc}] }.
\label{eq:eqMass}
\end{equation}

The velocity gradient, $\delta v \,/ \delta r $, is averaged across the modelled molecular gas component, which is assumed to fill the source solid angle. This corresponds to the mass of each component, i.e. the total mass for the \textit{2-component} model is the sum of both components. The \textit{Turbulence} model density PDF is sampled by 50 density bins, each of which has an associated solid angle and allows for 50 individual mass calculations. The total $\mu_{\rm L} M_{\rm ISM}$ is the sum of all of the masses corresponding to the full density PDF. We use the value of $\mu_{\rm L}M_{\rm ISM}$ derived in the \textit{Turbulence} model to estimate a range for the mean total molecular ISM mass of $\mu_{\rm L} M_{\rm ISM} = $ 3.6$\times 10^{11}$ -- 1.6$\times 10^{13}$ \Msun. The \textit{2-component} model-derived $\mu_{\rm L}M_{\rm ISM}$  is broadly consistent with the \textit{Turbulence} model, although the latter tends to be larger up to a factor $\sim$1.5. The inherent power-law dependence between density and gas kinetic temperature of the \textit{Turbulence} model prevents over-dense solutions for the mean density in the log-normal PDF, and thus drives a realistic turnover in the CO line intensity at higher-J. Therefore, in the \textit{Turbulence} model, there is more diffuse gas, on average, which contributes a larger fraction to the total molecular ISM mass. In contrast, the \textit{2-component} model tends to fit the higher-J lines with a stronger contribution from the second, denser component, which contributes less to the total mass. \\

The mass of the more highly excited component in the \textit{2-component} model (i.e. component two) can be thought of as a tracer of dense molecular gas \citep{gowardhan2017}. The best-fit, minimum-$\chi^{2}$ model solutions in our sample of \LPs\ indicate a median and mean value of $M_{\rm ISM, c2}/M_{\rm ISM, total} = $ 25 and 30\%, respectively, for this proxy for the dense molecular gas fraction believed to be more closely associated with current SF. Therefore the more diffuse/less-excited component carries most of the mass. The larger mass fraction for component one is due to the larger size of that component, which scales non-linearly with the mass (Eq. \ref{eq:eqMass}).\\

The \LPs\ have an average apparent total molecular gas mass $\mu_{\rm L}M_{\rm ISM} =  2.5 \times$10$^{12}$ \Msun. If this would be transformed into stars, this theoretical gas depletion time, $\tau_{\rm dep}$, results in a timescale on the order of  $\tau_{\rm dep} = < \mu_{\rm L}M_{\rm ISM} >/< \mu_{\rm L}SFR > = ( 2.5 \times 10^{12} )/( 35.4 \times 10^{3} ) \sim  $ 70 Myr. This rapid depletion timescale is an order of magnitude lower than the 1 Gyr gas depletion time observed in local star-forming galaxies \citep{leroy2013, saintonge2013,saintonge2016}, in agreement with the strong redshift dependence summarized by \citet{tacconi2018, tacconi2020}. This further supports the notion that these systems lie above the main-sequence for star-forming galaxies at $z \sim 2- 3$ \citep{whitaker2012a}. We use the derived values for $\sqrt{\mu_{\rm L}}R_{\rm eff}$ to also calculate the surface gas mass density $\Sigma_{M_{\rm ISM}} = \mu_{\rm L}M_{\rm ISM}/\pi (\sqrt{\mu_{\rm L}}R_{\rm eff})^{2} \sim $ 800 - 22000 \Msun \, pc$^{-2}$. Note the magnification factors cancel to first-order. The total mean values of some of the \LPs\ may be largely unconstrained, due to the lack of either ancillary [CI] data, strong dust photometric support (other than \Planck data alone), or an insufficient amount of CO line detections. These \LPs\ sample the higher-end of the observed molecular gas mass surface densities when compared to local star-forming galaxies \citep[][also see \S \ref{sec:alphas}]{schmidt1959, Downes1998, kennicutt2012, bolatto2013}. The active star-forming regions of the \LPs\ are, however, extended by 25 - 100$\times$ larger in area, with an intrinsic emitting size radius of order a few kpc ($\sim$3 kpc; see \S \ref{sec:sizes}). \\

\subsection{Atomic Carbon Gas Excitation}
\label{sec:carbonprops}

\begin{figure*}
 \includegraphics[scale = 0.35]{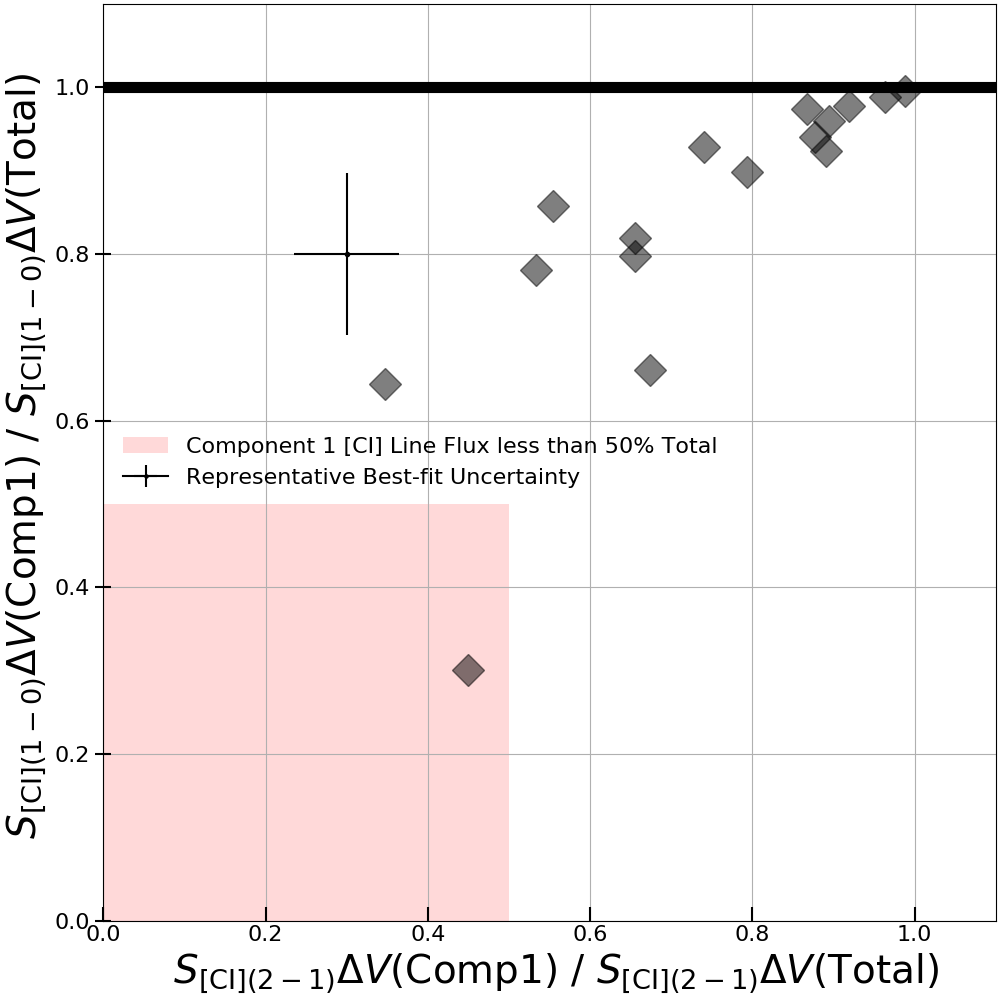}
  \includegraphics[scale = 0.35]{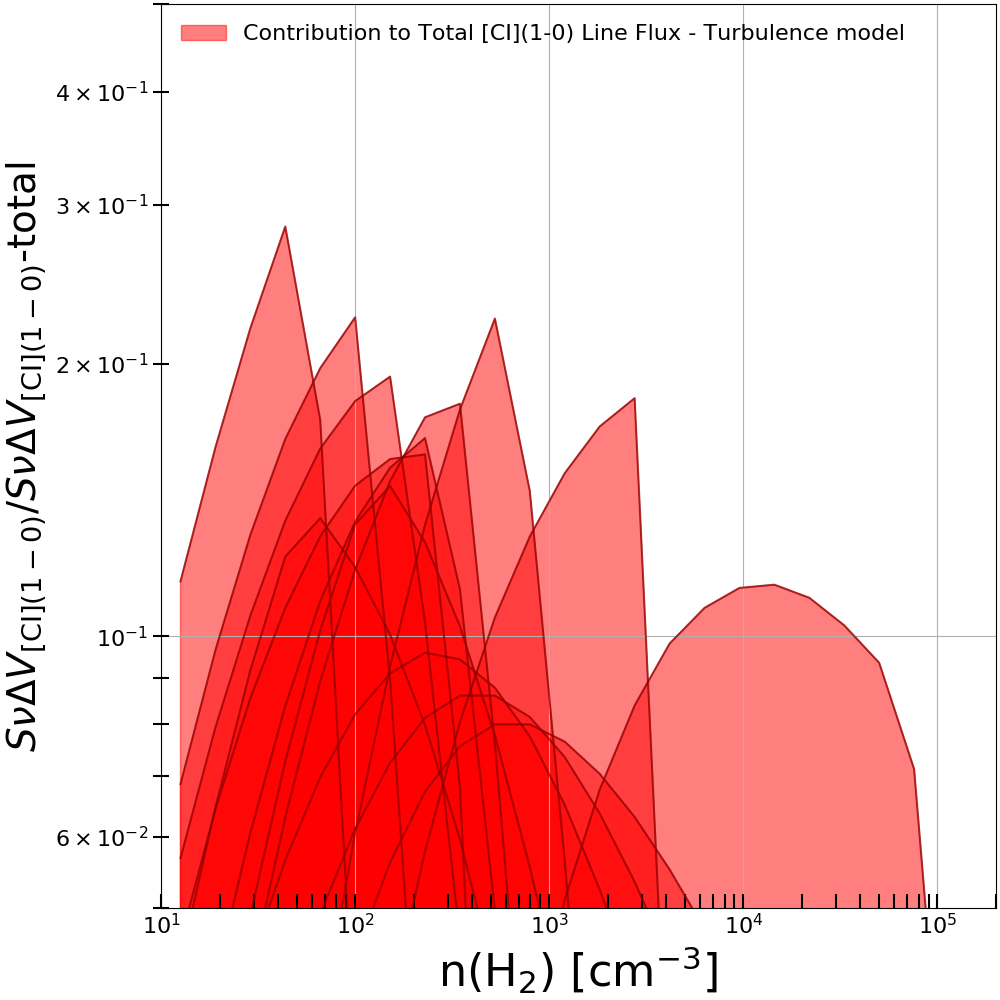}
 \caption{\textbf{Left:}Relative [CI](1-0) and [CI](2-1) line flux densities for component 1 versus the combined line flux densities for both components in the \textit{2-component} model. The red-square indicates that both [CI] lines predominantly arize from the denser component, i.e. component 2. \textbf{Right: } For each of the \LPs\ we plot (log-log-scale) the relative contribution to the total [CI](1-0) line flux derived in the \textit{Turbulence} model, from the 50 individual [CI](1-0) line fluxes corresponding to the 50 \Htwo\ densities which sample the density PDF.}
 \label{fig:CIflux}
\end{figure*}


In total, 21/24 \LPs\ have one or both of the [CI] emission lines detected. Only 5/24 \LPs\ have a single carbon line detection, while the remaining 16/24 \LPs\ have measurements of both fine-structure lines. The [CI] measurements of the \LPs\ represent the brightest apparent [CI] line fluxes reported at high-\z\ \citep{brown1992, barvainis1997, weiss2003, weiss2005carbon, walter2011,Alaghband-Zadeh2013, bothwell2017, yang2017, andreani2018}. Strong turbulent mixing of the cool neutral media in the ISM may likely occur in these turbulent \LPs\ \citep{Xie1995}, allowing for enriched [CI]/\Htwo\ abundances in the interiors of molecular clouds, and thus strong [CI] emission. Both theoretical and observational studies have demonstrated the reliability of using [CI] to trace the overall kinematics of the cold gas, as well as to determine the total molecular gas mass \citep{papadopoulos2004a, papadopoulos2004b, weiss2005carbon, tomassetti2014, glover2014, glover2015, israel2015, israel2020}. The latter requires knowledge of the atomic carbon excitation temperature, $T_{\rm exc}$, and gas-phase abundance, [CI]/\Htwo, to accurately convert the [CI] line emission to the atomic carbon mass, $M_{\rm C}$, and further to the total molecular gas mass $M_{\rm ISM}$ \citep{weiss2003, weiss2005carbon}. \\

We first examine which density phase the [CI] line emission predominantly arises from. The left-hand side of Fig. \ref{fig:CIflux} plots the relative integrated flux values for the [CI](1-0) and [CI](2-1) lines from component one, with respect to the total integrated flux value for both components combined, for the \textit{2-component} model. We remove from the figure the 5 \LPs\ with only a single [CI] line to avoid mis-interpreting our results. Almost all of the [CI](1-0) and [CI](2-1) line emission in the \LPs\ comes from the first component. In general, the first component is best-traced by the [CI](1-0) line. In general, this indicates that the carbon lines can be reliable tracers of the bulk gas mass, since we have shown in \S \ref{totalmass} that the first component in the \textit{2-component} model carries most of the total mass. One of the \LPs, LPs-J1139, seems to have a significant contribution from the denser component as indicated by the low contribution to the overall [CI] line emission from component one. This is due to its unusually high [CI](2-1) to [CI](1-0) line ratio \citep{nesvadba2019}, although the CO(7-6) line reported by \citet{canameras2018} appears to consistently under-predict the model-derived line flux density. The right-hand side of Fig. \ref{fig:CIflux} further reveals the relationship between the [CI](1-0) velocity integrated line fluxes and \Htwo\ density based on the \textit{Turbulence} model. As suggested by the more simplistic \textit{2-component} model, the \textit{Turbulence} model shows that the diffuse gas, with \logden\ = 2 - 3 \percc, is primarily responsible for the [CI](1-0) line emission. For such active star-forming systems, this implies that the carbon lines are well-suited to predominantly trace the diffuse molecular gas. \\

\begin{figure}
\centering
\resizebox{0.485\textwidth}{!}{%
 \includegraphics[scale = 0.5]{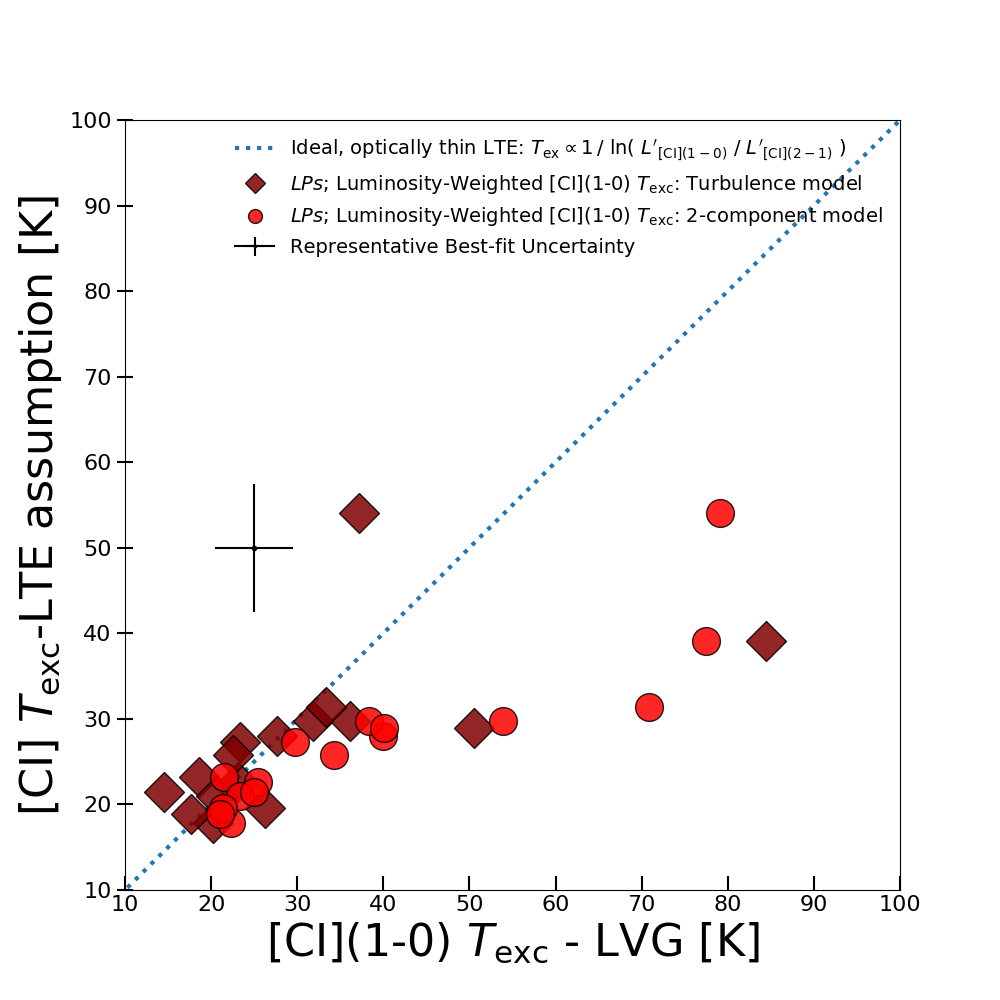}}\\
 \caption{Luminosity-weighted excitation temperature of the atomic carbon [CI](1-0) line based on both the \textit{2-component} model (red circle) and \textit{Turbulence} model (maroon diamond) are shown on the x-axis. These values are compared to the predicted value based on the assumption of optically thin, LTE gas conditions \citep{schneider2003n}.}
 \label{fig:ciTex}
\end{figure}

Next we investigate the atomic carbon $T_{\rm exc}$ for this diffuse molecular gas phase traced by the [CI] lines. In particular, our non-LTE analyses enables us to test the validity of the optically thin, LTE assumption framework that is commonly applied to detections of [CI] lines in star-forming galaxies. The measurement of both ground-state [CI] lines can, in principle, provide an independent estimate to constrain the carbon $T_{\rm exc}$ \citep{stutzki1997, schneider2003n}. The use of the carbon line ratio, R$_{\rm [CI]}$ = $L'_{\rm [CI](2-1)}/L'_{\rm [CI](1-0)}$, to determine $T_{\rm exc}$ requires two major assumptions: i.) that the line emission is optically thin and ii.) the lines are excited under LTE conditions, as defined in the Boltzmann Equation: $T_{\rm exc, LTE} = 38.8/{\rm ln}(2.11/R( {\rm [CI])} )$ K  \citep{stutzki1997, schneider2003n}. The value of $T_{\rm exc}$ can differ for each of the line transitions, i.e. the temperature needed to recover the relative populations of the upper/lower levels from a Boltzmann distribution.\\

We briefly return to Fig.~\ref{fig:lineopacity}, based on our best-fit, minimum-$\chi^{2}$ \textit{2-component} models, to recall that the majority of the \LPs\ have optically thin [CI](1-0) and [CI](2-1) lines. This also supports the capability of using the optically thin [CI] lines as a strong tracer of the bulk atomic carbon column density. We further calculate the value of the flux-weighted [CI](1-0) excitation temperature, $T_{\rm exc}$([CI](1-0)) using the relative integrated flux values from both components of the \textit{2-component} model. In addition, we also calculate the equivalent flux-weighted [CI](1-0) line  $T_{\rm exc}$ using the individual gas properties corresponding to the 50 density bins used to sample the density PDF derived in the \textit{Turbulence} model.\\

We find that the average luminosity-weighted excitation temperature for $T_{\rm exc}$([CI](1-0)) $\sim $ 40 K for the \textit{2-component} model and $T_{\rm exc}$([CI](1-0)) $\sim $ 32 K for the \textit{Turbulence} model. Note, the mean value for $T_{\rm exc}$([CI](2-1)) $\sim $ 29 K for the \textit{Turbulence} model. In the LTE assumption, $T_{\rm exc} := $ $T_{\rm exc}$[CI](1-0) = $T_{\rm exc}$[CI](2-1). The \LPs\ have a systematically lower value of $T_{\rm exc}$[CI](2-1) than $T_{\rm exc}$[CI](1-0), up to 25-30\% in some cases, reflecting those sub-thermal gas excitation conditions. Fig.~\ref{fig:ciTex} compares our derived, flux-weighted value of $T_{\rm exc}$[CI](1-0) using both models to the optically thin, LTE assumed value for the $T_{\rm exc}$, as presented in \citet{stutzki1997,schneider2003n}. In general, our model-derived values agree with the ideal framework of assuming the atomic carbon excitation occurs within optically thin, LTE gas conditions, yet we find that some of the \LPs\ would have had systematically under-predicted values of the carbon $T_{\rm exc}$ under these ideal assumptions. This emphasizes the importance in non-LTE modelling to better understand the sub-thermal excitation of the cold atomic and molecular ISM in star-forming galaxies. Our results for $T_{\rm exc}$ can also be compared with the recent large compilation of all local and high-\z\ star-forming systems with [CI] detections \citep{valentino2020}. They assume the same LTE assumptions as in \citet{schneider2003n}, yielding $T_{\rm exc}$ of $\sim$ 25 K with a moderate dispersion. This includes the high-\z\ starburst/quasar sample of \citet{walter2011}, which had an average excitation of $\sim$ 30 K when using the same optically thin, LTE assumptions.  \\

There are 5/16 \LPs\ with both [CI] detections for which the \textit{Turbulence} model still predicts low excitation temperatures of the [CI](1-0) line of $\leq $20 K. These are therefore over-predicted according to the simplified LTE assumption. We note that the low values of $T_{\rm exc}$ in these \LPs\ are accompanied by their relatively low line ratios between the [CI](2-1) and [CI](1-0). At values of $T_{\rm exc} \leq $ 20 K, \citet{weiss2005carbon} demonstrated that the atomic carbon mass estimate will increase exponentially. Many of the \LPs\ show substantially sub-thermal gas excitation, as shown from the $T_{\rm exc}$[CI](1-0) values. This may be due to the enhanced molecular gas kinetic temperatures with respect to the carbon excitation temperature, $T_{\rm kin}/T_{\rm exc}$[CI](1-0) $\sim $ 4. Therefore, blind LTE assumptions would have strongly impacted the inferred total carbon mass and relative abundance in those galaxies by about an order of magnitude. If the [CI] lines are completely dominated by sub-thermal gas excitation, the model-derived value of $T_{\rm exc}$ will be higher by up to a factor of 2-3 in the \textit{2-component} model. There is considerably less scatter in the \textit{Turbulence} model, which is likely due to the differences in these models when deriving the carbon gas-phase abundance.\\

We recall that we have restricted the CO/\Htwo\ abundance to a Milky Way value of $\sim$10$^{-4}$, close to that of local star-forming systems with CO/\Htwo\ = 0.5 - 1 $\times10^{-4}$. We do, however, allow the value of [CI]/\Htwo\ to vary as a free parameter in both models. We find, for the \textit{Turbulence} model, the sample mean for the \LPs\ of $< {\rm [CI]}/{\rm H_{\rm 2}} > = 6.82 \pm 3.04 \times10^{-5}$. To better understand the \textit{2-component} model, we note that if the value of [CI]/\Htwo\ were increased in the second component to match that of the first component, then the [CI] emission would always be close to LTE (since the \Htwo\ density of component two is always higher than the first component (Fig. \ref{fig:TkinDen}). The \textit{2-component} model solves this by reducing the value of [CI]/\Htwo\ in the second component, so that the emission is dominated by the sub-thermally excited emission from the first component. Thus, the value of [CI]/\Htwo\ in the second component must be lower, otherwise the models would result in higher than observed line ratios. This is one of the main criteria for the \textit{Turbulence} modelling procedure, which realistically forces the gas-phase carbon abundance to decrease with increasing \Htwo\ densities according to a power-law relation. \\

At high-\z, knowledge of the excitation conditions and abundance of carbon is often the main source of uncertainty. The mean value of [CI]/\Htwo\ we find for the \LPs\ is comparable to previous estimates by local/high-\z\ studies, although typically this has been achieved via the inferred \Htwo\ mass from single (low-J) CO transitions \citep{weiss2005carbon, walter2011, valentino2020}. We caution that there is an order of magnitude dispersion in [CI]/\Htwo\ among the \LPs, which has strong implications for the inferred conversion from the [CI] line luminosity to M$_{\rm ISM}$ in star-forming galaxies (as discussed in \S \ref{sec:alphas}). Some high-\z\ carbon gas-phase abundance estimates are a few $\times$10$^{-5}$ \citep{walter2011}, which are broadly consistent with that of low-\z\ galaxies. This suggests that the starbursts and QSO have at least solar gas-phase metallicities or higher \citep{gerin2000, weiss2001, israel2002, israel2003}. Overall, we find values often lower than the abundance derived in the solar neighborhood of [CI]/\Htwo\  $\sim 3.5\times10^{-4} $ \citep{Anders1989}. In some cases, the \LPs\ show similar [CI]/\Htwo\ abundances to the cold Milky Way CO-faint clouds, with $\sim$ $1-2\times10^{-5} $ \citep{Frerking1989, Keene1997}.\\

\section{Discussion}
\subsection{Molecular gas excitation at high-\z\ }

\subsubsection{Classifying the gas physical conditions in the \LPs}
\label{sec:classes}

\begin{figure}-
\centering
\resizebox{0.487\textwidth}{!}{%
 \includegraphics[scale = 0.3]{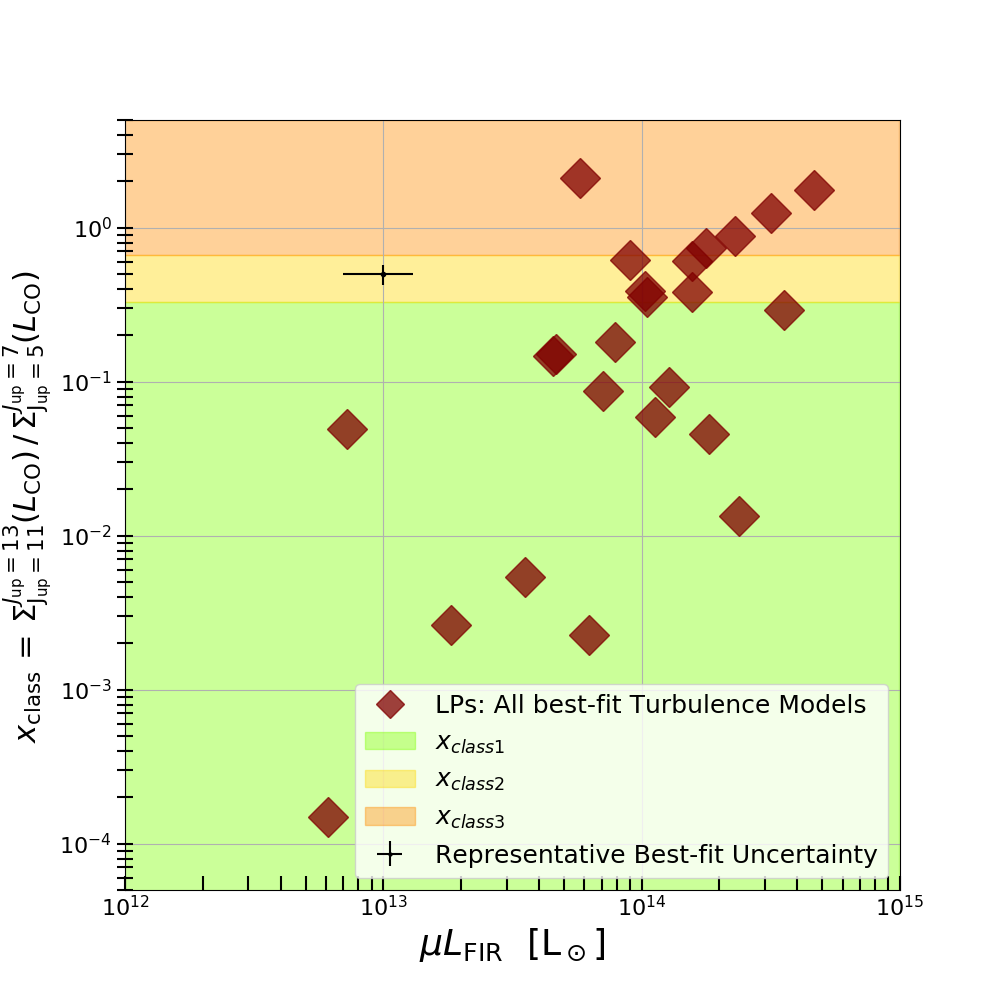}
}\\
 \caption{Plot (log-log-scale) between $x_{\rm class} $ and FIR luminosity, based on the best-fit, minimum-$\chi^{2}$ \textit{Turbulence} models. For details on the classification scheme, see \S \ref{sec:highzgas} and Eq. \ref{eq:xclass}. }
 \label{fig:xclasses}
\end{figure}

Fig.~\ref{fig:bfAll} shows that the \LPs\ offer a rich perspective into the wide range of gas excitation properties of CO for high-\z\ star-forming galaxies. Overall, there seems to be a continuous distribution in gas excitation conditions for this sample of \LPs. The high magnification therefore allows us to probe an intrinsically heterogeneous mix of dusty star-forming galaxies. Following the classification scheme defined by \citet{rosenberg2015} for the diverse sample of local IR-luminous star-forming galaxies, we apply the parameter, hereafter ``$x_{\rm class}$'', to quantify the range of excitation conditions in these 24 \LPs. This parameter specifically characterizes the drop-off slope, after the expected peak of the CO line SED of J$_{\rm up}$ = 5 - 7. For each individual galaxy, we compare the relative line luminosity strength (in \Lsun) of the higher-J CO(J$_{\rm up}$ = 11 - 13) lines versus the mid-J CO(J$_{\rm up}$ = 5 - 7) lines:

\begin{equation}
x_{\rm class} = \frac{L_{\rm CO(11-10)} + L_{\rm CO(12-11)} + L_{\rm CO(13-12)}}{L_{\rm CO(5-4)} + L_{\rm CO(6-5)} + L_{\rm CO(7-6)}},
\label{eq:xclass}
\end{equation}

with three excitation classes defined as $x_{\rm class1} = [ < 0.33]$, $x_{\rm class2} = [0.33, 0.66]$, and $x_{\rm class3} = [ > 0.66]$. The sample of \LPs\ indeed shows a broad range of excitation conditions based on these three different CO line SED classifications, with a continuum of $x_{\rm class}$ values. In total, there are 14 \LPs\ within $x_{\rm class1} $, five \LPs\ within $x_{\rm class2} $ and five \LPs\ within $x_{\rm class3} $. We note that the three $z \sim 1$ \LPs\ are all in $x_{\rm class1} $, however the distribution of $z$ and classification is well-mixed for the $z = 2 - 3.5$ sub-sample. Fig. \ref{fig:xclasses} shows the classification versus the FIR luminosity, indicating a clear relation between these two quantities, despite the strong incompleteness in our flux-limited sample of \LPs, based on their selection criteria. The correlation we observe, between $x_{\rm class}$, or more broadly the CO line ratios, and the FIR luminosity is consistent with previous studies of star-forming galaxies \citep{greve2014, rosenberg2015}. \\


\subsubsection{Mean CO brightness temperature ratios at $z > 1$}
\label{sec:highzgas}

\begin{figure*}
\centering
\resizebox{0.997\textwidth}{!}{%
 \includegraphics[scale = 0.394]{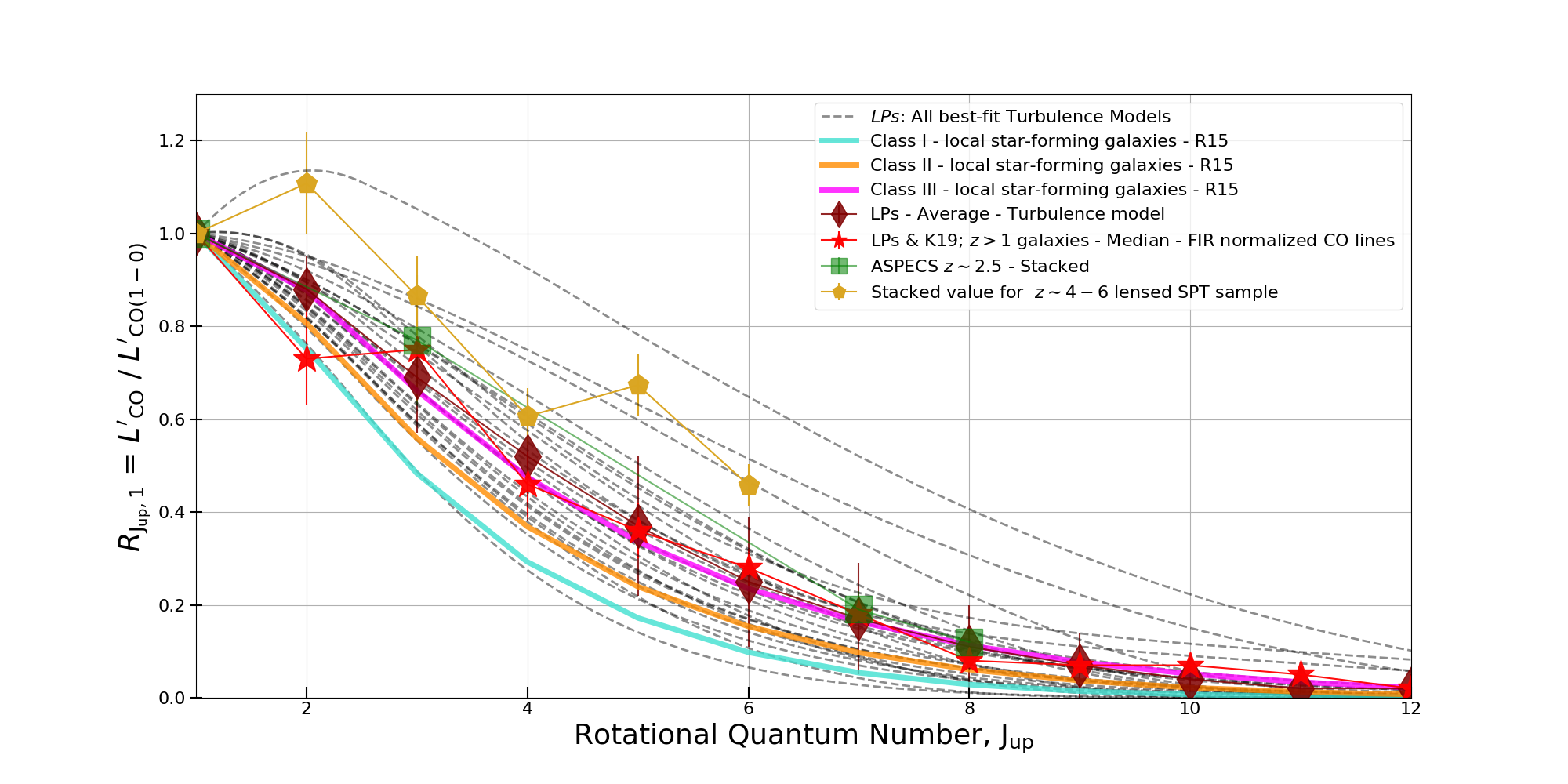}
 }\\
 \caption{ \textit{Turbulence} model results for the best-fit, minimum-$\chi^{2}$ model solutions for the CO line luminosities, $L'_{\rm CO}$, normalized to the CO(1-0) line, for the \LPs. The average line ratios derived using the \textit{Turbulence} model (maroon diamonds) are compared to the values from the \LPs\ \& K19 sample (red star), which are purely based on observations, with median line ratio values determined by normalizing each line by a common FIR luminosity. In addition, we plot representative local IR-bright star-forming galaxies from \citet{rosenberg2015}, with increasing classifications determined by their excitation conditions, from lowest to highest: Class I (light blue solid line), Class II (orange solid line) and Class III (magenta solid line). Yellow pentagons represent stacked values for the lensed SPT galaxies \citep{spilker2016}, compared to the stacked value of $z \sim 2.5$ main-sequence star-forming galaxies (green squares) from the ASPECS sample \citep[][]{walter2016,riechers2020, boogaard2020}.        }
 \label{fig:coladder}
\end{figure*}

The CO line luminosities, $L'_{\rm CO}$, normalized to the CO(1-0) line, are often used to study the global gas excitation conditions of a galaxy and the line SEDs \citep[e.g.][]{bothwell2013}. Here we use our dataset of the \LPs, based on $\sim$ 4 - 6 CO lines for each of the \LPs, to better understand the CO line SEDs of such IR-bright, $z > 1$ galaxies. These $L'_{\rm CO}$ ratios are also used to scale the higher-J CO lines to the ground-state transition to infer the total molecular gas mass \citep{carilli2013}. The sample mean and standard deviation for the \LPs\ are reported in Table \ref{tab:Lprimeratio}. The \LPs\ have line brightness temperature ratios that are often well-below unity (i.e. sub-thermal line excitation). \\

In Fig.~\ref{fig:coladder} we use the \textit{Turbulence} model to show all of the best-fit CO line luminosities, $L'_{\rm CO}$, normalized to the CO(1-0) line. We also show the stacked brightness temperature ratios of the lensed SPT sample \citep{spilker2016}, and the stacked values of the unlensed, $z\sim 2.5$ main-sequence star-forming galaxies \citep[][]{riechers2020, boogaard2020}. The former tends to represent the higher gas excitation seen in a subset of the \LPs. The latter, which considers only 7 sources in the stacked value, tends to agree with both the model-derived average values for R$_{3,1}$, R$_{7,1}$, and R$_{8,1}$. The \LPs\ also show significantly higher line ratios than the $z \sim 1 - 2$, main-sequence star-forming galaxies in COSMOS field, for the available values of R$_{4,1} = 0.27$, R$_{5,1} = 0.21$, and R$_{7,1} = 0.06$ \citep{valentino2020co}\footnote{Note, we have used a fiducial value of R$_{2,1} = 0.75$ to make this comparison, since only CO(2-1) line measurements are available.}. Galaxies on the brightest end of the QSO luminosity function, at $z \sim 2 - 4$, tend to have R$_{5,1} \sim 1$ \citep[e.g.][]{weiss2007, Bischetti2020}, which suggests that there may be a differing dominant gas excitation mechanism for the global ISM of the \LPs\, which have a mean value of R$_{5,1} \sim 0.37$.  \\
The local IR-bright star-forming galaxies from \citet{rosenberg2015} are also shown in Fig.~\ref{fig:coladder}, according to their increasing classifications determined by their gas excitation conditions, from lowest to highest: Class I, Class II and Class. For reference the Milky Way has a global average value of R$_{3,1}$ of 0.28 $\pm$ 0.17 \citep{fixsen1999}.  The \LPs\ show a broad range of gas excitation, although most have higher line ratios than both Class II and Class III galaxies, which are representative of most local (U)LIRGs \citep[see also][]{papadopoulos2012a, lu2014, kamenetzky2016}. The \LPs\ show systematically higher excitation than most local IR-bright, spiral galaxies, as well as the more highly excited local radio/X-ray AGN host galaxies \citep{vanderWerf2010, papadopoulos2012a, spinoglio2012, meijerink2013, rosenberg2015, liu2015, kamenetzky2016}. \citet{papadopoulos2011} report values of R$_{3,1}$  = 0.67 and R$_{6,1} = 0.2 - 1.6$ for local (U)LIRGs \citep{papadopoulos2012}. This is, overall, consistent with earlier studies and the value we obtain for the \LPs. Local star-forming systems with $9 < {\rm log(} L_{\rm IR} {\rm )} < 12 $ show a median value of R$_{3,1}$ to be close to 0.5 \citep{mauersberger1999}, with some 60 local barred galaxies and starbursts having an average value of R$_{3,1}$ close to $0.9 \pm 0.1$ \citep{yao2003}. Overall, the line brightness temperature ratios for the \LPs\ are usually not as high as one of the most IR luminous local starburst galaxies, M82 \citep{weiss2005m82}, which has a global average R$_{2,1}$, R$_{3,1}$, R$_{4,1}$ and R$_{5,1}$ of 0.98, 0.93, 0.85 and 0.75, respectively.   \\

\citet{bothwell2013} previously interpreted the mean CO line SED of 32 $z > 1$ star-forming galaxies, by normalizing their CO line luminosities by a common FIR luminosity, while others normalized to a common measurement of the dust continuum at 1.4mm \citep{spilker2016}. Based on our simultaneous modelling of the lines and continuum, we can test whether or not this is a valid method for building a mean line SED shape. It has been well-known that the FIR luminosity and CO line luminosity tend to increase with one another proportionally \citep{greve2012, liu2015, valentino2020co}, and Fig. \ref{fig:xclasses} also shows that the CO excitation is correlated with the FIR luminosity. Hence, there is still a bias when using the FIR luminosity to compute a mean CO line SED. Although the line ratios are sensitive to the FIR luminosity, different subsets of high-\z\ sources do not have every line detected. There are also different selection biases, therefore a common FIR luminosity may be used to normalize each line detection in an attempt to remove these biases and to avoid physically misleading line ratios when comparing large samples. \\

To do this, we use the \LPs\ \& K19 $z \sim 1 - 7$ sample (hereafter ``\LPs\ \& K19 sample''; N$_{gal}$ = 269). To construct this compilation, we used our dataset for the \LPs\, including the database compiled in \citet{kirkpatrick2019}, which includes a vast majority of the heterogeneously selected samples of high-\z\ galaxies with CO line detections \citep[including ][]{carilli2013, pope2013, aravena2014, sharon2016, yang2017, frayer2018, perna2018, kirkpatrick2019}. Following previous studies, each CO line is normalized by a common FIR luminosity ($L_{\rm FIR} = 1\times10^{12.5}$ \Lsun). The values listed in Table \ref{tab:Lprimeratio} are derived using the \LPs\ \& K19 $z \sim 1 - 7$ CO line compilation, which consists of 90 CO(1-0) lines, 86 CO(2-1) lines, 128 CO(3-2) lines, 80 CO(4-3) lines, 68 CO(5-4), 73 CO(6-5), 63 CO(7-6), 39 CO(8-7), 33 CO(9-8), 14 CO(10-9), 15 CO(11-10) and 2 CO(12-11) high-\z\ line measurements. \\

As shown in both Table \ref{tab:Lprimeratio} and Fig. \ref{fig:coladder}, the average results from our best-fit, minimum-$\chi^{2}$ \textit{Turbulence} models are strikingly similar to the median brightness temperature ratios derived from the FIR luminosity-normalized lines we calculate from the CO line observations alone using the \LPs\ \& K19 sample, or the ``All sources'' sample of \citet{kirkpatrick2019}, which considers all $z > 1$ galaxies with CO line detections. In Table ~\ref{tab:Lprimeratio} we also reference the average line ratios, up to the R$_{5,1}$ ratio, from \citet{carilli2013}, which is based on the available data for (sub)mm bright, star-forming galaxies (SMGs) and QSOs at the time. We also quote in Table ~\ref{tab:Lprimeratio} the comparable, and more recent values reported by \citet{kirkpatrick2019}. We note that the values from the latter are consistent with the low-mid-J CO line ratios reported by other recent studies of lensed, SMGs at $z > 1$ \citep{yang2017, canameras2018}. \\

At first glance, our results would seem to support this method of using the FIR luminosity to normalize the CO lines to derive a mean CO line SED, however the sparse amount of well-sampled CO line SEDs per galaxy in the literature suggests that the three or more orders of magnitude dispersion in the observed line intensities (Fig. \ref{fig:snudvJ}) all average out. The \LPs\ \& K19 $z \sim 1 - 7$ sample includes some of the brightest (sub)mm selected galaxies with multiple CO line detections, which further exaggerates the effects of averaging large samples of $z > 1$ galaxies \citet{bothwell2013, spilker2016, yang2017, canameras2018}. As noted by \citet{narayanan2014}, there can be a factor 5-10 difference in the CO line SEDs, at mid- to high-J transitions, for similarly selected, (sub)mm bright galaxies with the same integrated FIR luminosity. The broad range of excitation conditions and average line ratios we present for the \LPs\ further highlights the notion that it is unlikely for their to be a template CO line SED for any $z > 1$ galaxy population. As noted throughout this work, our results agree with the theoretical models of \citet{narayanan2014} for the CO line SEDs of $z > 1$ star-forming galaxies, as parameterized by the SFR surface density. This suggests that in the absence of multiple CO line measurements, a SFR surface density estimate may be combined with limited line data, and further use the theoretical models of \citet{narayanan2014} to estimate the CO line SED.  \\



\subsection{Molecular gas mass estimates }

\subsubsection{Intrinsic Emitting Size Regions}
\label{sec:sizes}
One way to constrain whether or not the \LPs\ are not only some of the most massive, gas-rich star-forming galaxies, but also perhaps the largest in size, is to cross-examine the model-derived radius with the intrinsic source size -- as expected from studies of star-forming galaxies at $z > 1$. As presented in \S \ref{meanprops}, the mean value we derive from the \textit{Turbulence} model for the radius of the modelled source emitting region is $ \sqrt{\mu_{\rm L}}R_{\rm eff} \sim 10-15 $ kpc. The intrinsic source size may differ. When comparing to other observed sources, we assume that the observed emission corresponds to a filled, face-on circular disk \citep[see e.g.][]{weiss2007} with an effective radius. This sets a lower limit to the true source size, as there is no information of how the gas and dust emission is distributed within the source solid angle. We recall again the lens magnification factor estimates for the \LPs\ in Table \ref{tab:summaryLPs}, as they will be used to estimate the intrinsic emitting size. For a reference, we derive an average value of $\mu_{\rm L}$ for the \LPs\ using the upper limit value of $\mu_{\rm L} $ in Table \ref{tab:summaryLPs}. For galaxies without a published value of $\mu_{\rm L}$, we use the value based on the ``Tully-Fischer'' argument presented in \citet{harris2012} and our CO(1-0) line measurements (see Appendix \ref{difflens}). The average lens magnification factor has a value of $\mu_{\rm L}  \sim 20$. The expected, \textit{intrinsic}, total line and continuum emitting size radius for all of the \LPs\ $R_{\rm eff}/\sqrt{\mu_{\rm L}} = 13.5/\sqrt{20.4} \, \sim 3 $ kpc. This size is consistent overall with the size of the dust continuum emission from massive star-forming galaxies at $z = 1 - 3$ \citep[1 - 5 kpc,][]{simpson2015a, hodge2016, oteo2016a, oteo2017b, barro2016, rujopakarn2016, fujimoto2017, jimenez-andrade2019, hodge2020}. \\

We can further test the reliability of our \textit{Turbulence} model if we consider two of the \LPs\ with estimates of the intrinsic, lens model reconstruction of the source size, based on high-angular resolution data: LPs-J105353 \citep{canameras2017b} and LPs-J0209 \citep{geach2018, rivera2019}. As summarized in \citet{harrington2019} for LPs-J0209\footnote{Also referred to as the Red Radio Ring.}, the lens model reconstruction from the low-J CO line image presented in both \citet{geach2018, rivera2019} suggests a molecular gas reservoir with an emitting radius of $\sim$ 1 - 2 kpc. The flux-weighted mean magnification factor derived for this source is $\sim 15$, corresponding to an expected \textit{intrinsic} emitting radius for LPs-J0209 of $R_{\rm eff}/\sqrt{\mu_{\rm L}} = 16.1/\sqrt{14.7} \, \sim 4 $ kpc. Note that we are modelling the full extent of the CO(1-0) to CO(15-14) line emission. Therefore, our modelling is consistent with the independently derived source-plane radius derived from both the CO(4-3) and CO(3-2) emission lines within the overall uncertainties. Based on CO(1-0) line observations of high-\z\ galaxies, the factor two difference can be accounted for, as the easily excited CO(1-0) line emission is expected to be more extended on average \citep{emonts2014, casey2018ngvla, spingola2020}. LPs-J105353\footnote{Also referred to as PLCK G244.8+54.9.} is proposed to consist of two independent regions (roughly 1.5 kpc in length along the major axis) in the reconstructed source plane CO(4-3) line image \citep{canameras2017b}, each separated by $\sim$ 500 pc, corresponding to an intrinsic emitting size radius of $\sim 2$ kpc. Our modelling would suggest an expected \textit{intrinsic} emitting radius for LPs-J105353 of $R_{\rm eff}/\sqrt{\mu_{\rm L}} = 14.5/\sqrt{26.8} \, \sim 3 $ kpc, which is consistent within the uncertainties considering both the lens model and our radiative transfer model. Thus, the intrinsic size may play a role in understanding the high apparent fluxes. A magnification factor of 50 would be required to reduce the mean value we derive from the \textit{Turbulence} model, i.e. $\sqrt{\mu_{\rm L}}R_{\rm eff} \sim 10 - 15 $ kpc, to match the more common size expected for $z > 1$ star-forming galaxies of $\sim 2$ kpc. Since the average magnification factor for the \LPs\ is $\sim$ 20, the size of these systems may be one of the primary physical parameters responsible for explaining their extreme IR luminosity. \\


\subsubsection{Converting $L'_{\rm CO}$ and $L'_{\rm [CI]}$ to $M_{\rm ISM}$}
\label{sec:alphas}

\begin{figure}
\centering
\resizebox{0.487\textwidth}{!}{%
 \includegraphics[scale = 0.394]{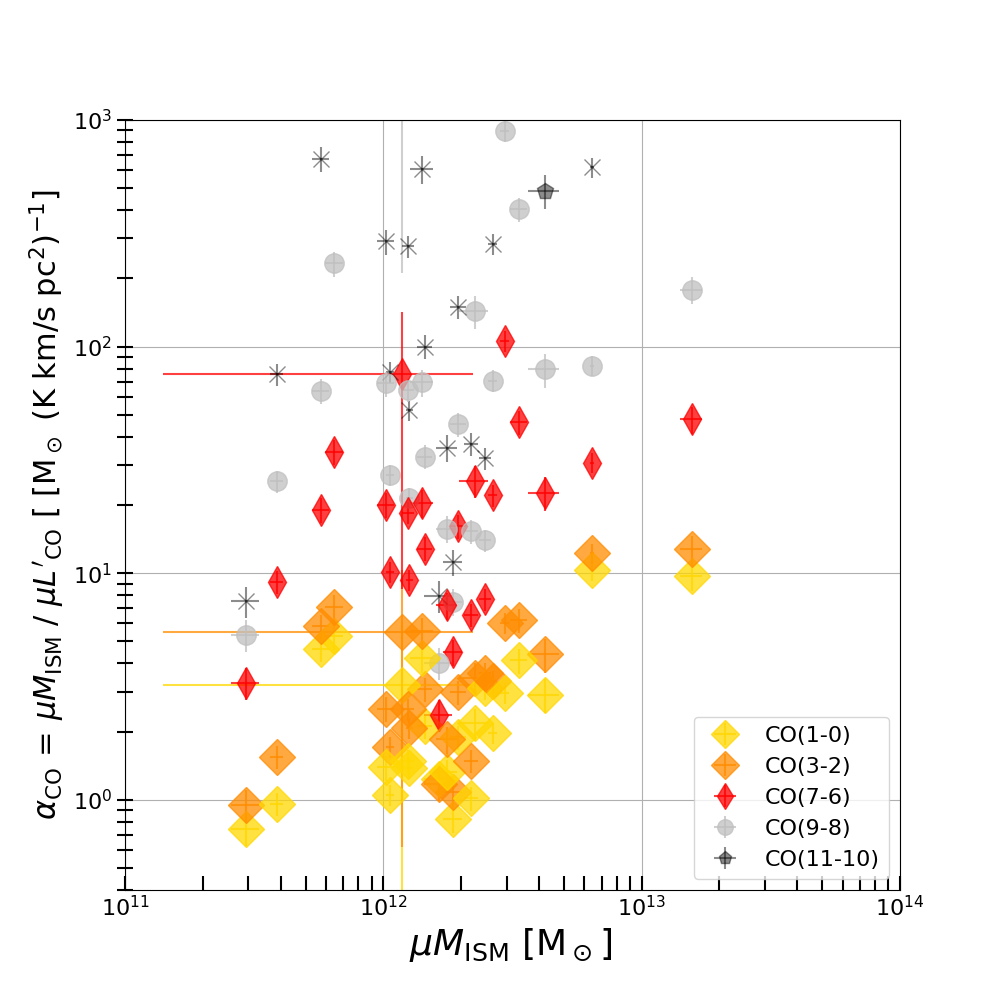}
 }\\
 \caption{CO line to molecular gas mass conversion factor, $\alpha_{\rm CO }$, value versus $\mu M_{\rm ISM}$ for the CO(1-0) (gold diamond), CO(4-3) (orange diamond), CO(7-6) (thin red diamond), CO(9-8) (gray circle) and CO(11-10) (black cross) lines, as derived from the best-fit, minimum-$\chi^{2}$ \textit{Turbulence} models (log-log-scale). }
 \label{fig:alphaperJ}
\end{figure}

We use the results of the more realistic \textit{Turbulence} model (\S \ref{modelresults}) to calculate the derived values of both $\alpha_{\rm CO}$ and $\alpha_{\rm [CI]}$ [hereafter without units attached to ease readability; \alphaunits ] conversion factors between the line luminosity to total molecular gas mass, $M_{\rm ISM}$. We refer to these factors with respect to the ground-state transitions, CO(1-0) and [CI](1-0), unless otherwise noted. Fig.~\ref{fig:alphaperJ} shows the results for the value of $\alpha_{\rm CO}$ based on a representative set of low-to-high-J CO transitions, versus the minimum-$\chi^{2}$ model solution for the total molecular gas mass, for all of the \LPs. The CO(1-0) and CO(3-2) derived $\alpha_{\rm CO}$ values are less scattered than those derived from CO(9-8) and CO(11-10) transitions. This demonstrates that the lower-J lines are more reliable tracers of $M_{\rm ISM}$, as expected.\\

To consider the use of [CI] line emission as a tracer of $M_{\rm ISM}$, we recall the results in \S \ref{sec:carbonprops}, in particular: i.) the [CI] lines are optically thin, and ii.) most of the carbon emission arises from the more diffuse line emitting region (i.e. component one from the \textit{2-component} model) with \logden\ $\sim 2 - 3$ \percc\ -- which is the cold gas component responsible for the bulk of the total $M_{\rm ISM}$ (also see Fig. \ref{fig:CIflux}). The value of $\alpha_{\rm [CI](1-0)}$ can be used to convert the optically thin $L'_{\rm [CI](1-0)}$ measurement to $M_{\rm ISM}$, which we find an average value of $< \alpha_{\rm CI} > = 16.2 \pm 7.9$ for the \LPs\ with both [CI] lines detected. \citet{crocker2019} recently studied a sample of 22 local spiral galaxies, representing a wide range of SFR and stellar masses, using spatially resolved \Herschel\ SPIRE observations of CO and [CI]. They found a lower value of $\alpha_{\rm [CI](1-0)} = 7.9 $, with a factor 1.5-2 uncertainty. There are differences in these values of $\alpha_{\rm [CI](1-0)}$ for the \LPs, as compared to these less extreme local star-forming galaxies, yet this may be due to differences in the calibration. Our current work performs a full radiative transfer analysis, while \citet{crocker2019} had used an intensity-intensity correlation between the low-J CO and [CI] line emission and an assumed value of $\alpha_{\rm CO}$. The latter was determined previously by \citep{sandstrom2013} based on an assumed range in the $GDMR$ and assumed line ratios. Another recent study used an alternative method to first measure [CI] and \Htwo\ absorption lines from gamma-ray burst and QSO objects \citep{heintz2020}. Thereafter they determine the value of $\alpha_{\rm [CI](1-0)} \sim 21$ for an assumed solar metallicity, which is consistent with our derived values.  \\

Based on the \textit{Turbulence} model, we find an average value for the \LPs \, of $< \alpha_{\rm CO(1-0)} > = 3.4 \pm 2.1$, with a factor 10 dispersion and a mean value of $ < \alpha_{\rm CO(1-0)} > = 4.2$ for the galaxies with the best dust photometry and CO/[CI] line coverage.  Note, the scatter in the conversion factors come from the scatter in the line luminosities and total molecular gas masses used to derive the mass-to-light conversion for each of the \LPs. Therefore, a single conversion factor value would not reflect this intrinsic dispersion among the sample. \\

Almost all of the \LPs\ have a value of $\alpha_{\rm CO}$ which is higher than the local IR (ultra)luminous star-forming galaxy (a.k.a ``ULIRG'') conversion factor of $\alpha_{\rm CO} = 0.8$ \citep{Downes1998}. The ULIRG value was derived using similar LVG approximations using only the available low-J CO lines, as well as dynamical mass measurements of the concentrated nuclear starburst regions. The warm, diffuse and dense molecular gas that pervades the molecular medium of local IR luminous star-forming galaxies was not fully traced by the CO(1-0) and CO(2-1) lines \citet{Downes1998}. The lower-J lines only trace the diffuse and warm gas under these active star-forming conditions, and therefore they correspond to lower values of $\alpha_{\rm CO}$. We have also shown this in our best-fit, minimum-$\chi^{2}$ model CO line SEDs, where the CO(1-0) line emission arises mostly from the diffuse molecular gas with density \logden\ $\sim$ 2 \percc, while the gas at these lower densities has higher kinetic temperatures. A lower value of $\alpha_{\rm CO}$ would be inferred for the \LPs\ if limited to only the low-J CO lines tracing this warm and diffuse phase, and thereby neglecting the higher-density gas which is required to excite the higher-J CO lines. Therefore, our physical results are still consistent with the general conclusions in modelling the local ULIRGs \citep{Downes1998}, yet we have modelled the substantial contributions to the overall CO line SED from warm and dense gas of the order of $T_{\rm kin} \sim $ 120 K and \logden\ = 4 - 5 \percc. This explains why our overall result reflects higher values of $\alpha_{\rm CO} \sim 3.4$ for the \LPs.\\

\begin{figure}
\resizebox{0.496\textwidth}{!}{%
 \includegraphics[scale=0.34]{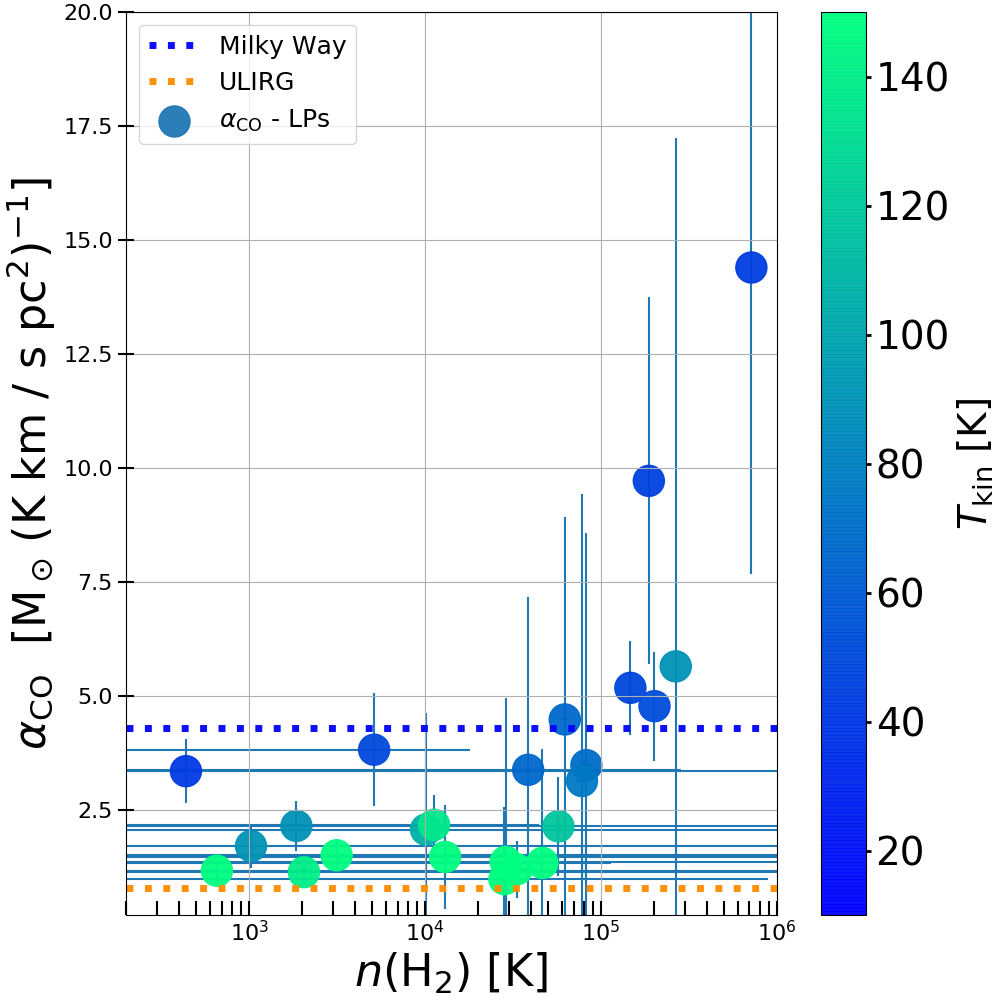} 
 }\\
 \caption{\textit{Turbulence} model derived $\alpha_{\rm CO}$ factor versus \Htwo\ gas density (log-x scale). The colorbar denotes the gas kinetic temperature. These are the total $\chi^{2}$-weighted parameter mean and standard deviation values derived from the $\sim$2 million model calculations. The canonical ULIRG and Milky Way values for $\alpha_{\rm CO}$ are $\alpha_{\rm CO} = 0.8$ and $4.3$ \alphaunits, respectively.}
 \label{fig:AlphaDenTk}
\end{figure}

Fig. \ref{fig:AlphaDenTk} shows our derived conversion factors as a function of both the mean \Htwo\ gas density and gas kinetic temperature. Almost half of the 24 \LPs\ have low values of $\alpha_{\rm CO} = 1-2$, and are associated with higher gas kinetic temperatures ($T_{\rm kin} > 120 $K). The left-hand side of Fig. \ref{fig:AlphaDenTk} shows a non-linear decrease in $\alpha_{\rm CO}$ as a function of increasing $T_{\rm kin}$, whereas the right-hand side of Fig. \ref{fig:AlphaDenTk} suggests a strong rise in $\alpha_{\rm CO}$ for increasing \Htwo\ gas density. We note that the system with one of the highest values is the known radio-AGN/starburst, LPs-J0209, and the highest redshift source in our sample, LPs-J105322 ($z \sim 3.5$) both have well sampled dust photometry and CO lines out to J$_{\rm up} = 11$ and both [CI] lines, yet we still derive the highest values of $\alpha_{\rm CO(1-0)} \geq 10$, albeit with larger uncertainty. In future work we will investigate the statistical relationship between $\alpha_{\rm CO}$ and these parameters. Since the LPs-J0209 source is known to have a compact radio AGN \citep{geach2015}, future studies may explore the reasons whether or not star-forming galaxies with coeval AGN activity \citep{harrington2019} may tend to show higher values of $\alpha_{\rm CO}$. In fact, the AGN APM0827, has a relatively high value of $\alpha_{\rm CO} \sim 5$ and \logden\ $\sim$ 5 \percc\ \citep{weiss2007}.\\

The large range observed for the \LPs\ suggests a strong diversity in gas excitation properties among this relatively small sample. It seems incorrect to assume a common value for high-\z\ star-forming galaxies, as discussed previously in the context of a continuity of gas excitation conditions and the corresponding variation in the conversion factor \citep[e.g.][]{casey2014}. The two different values commonly applied have been based on a simplified bi-modal population of star-forming galaxies \citep[e.g.][]{daddi2010a}, yet more complex two-population models are successfully reproducing several observed properties in observed high-\z\ galaxies \citep{sargent2014}. The latter may be tested with assumptions of a continuity in gas excitation conditions for different galaxy populations. Variations of the overall CO gas-phase abundance and variations across a turbulent star-forming disk will alter the conversion factor \citep{wolfire2010, glover2011,shetty2011a, shetty2011b, narayanan2012, bolatto2013a, narayanan2014}, therefore it is possible for a wide variation in $\alpha_{\rm CO}$ to exist for a given gas column density. \\

\begin{figure}
\resizebox{0.497\textwidth}{!}{%
 \includegraphics[scale=0.34]{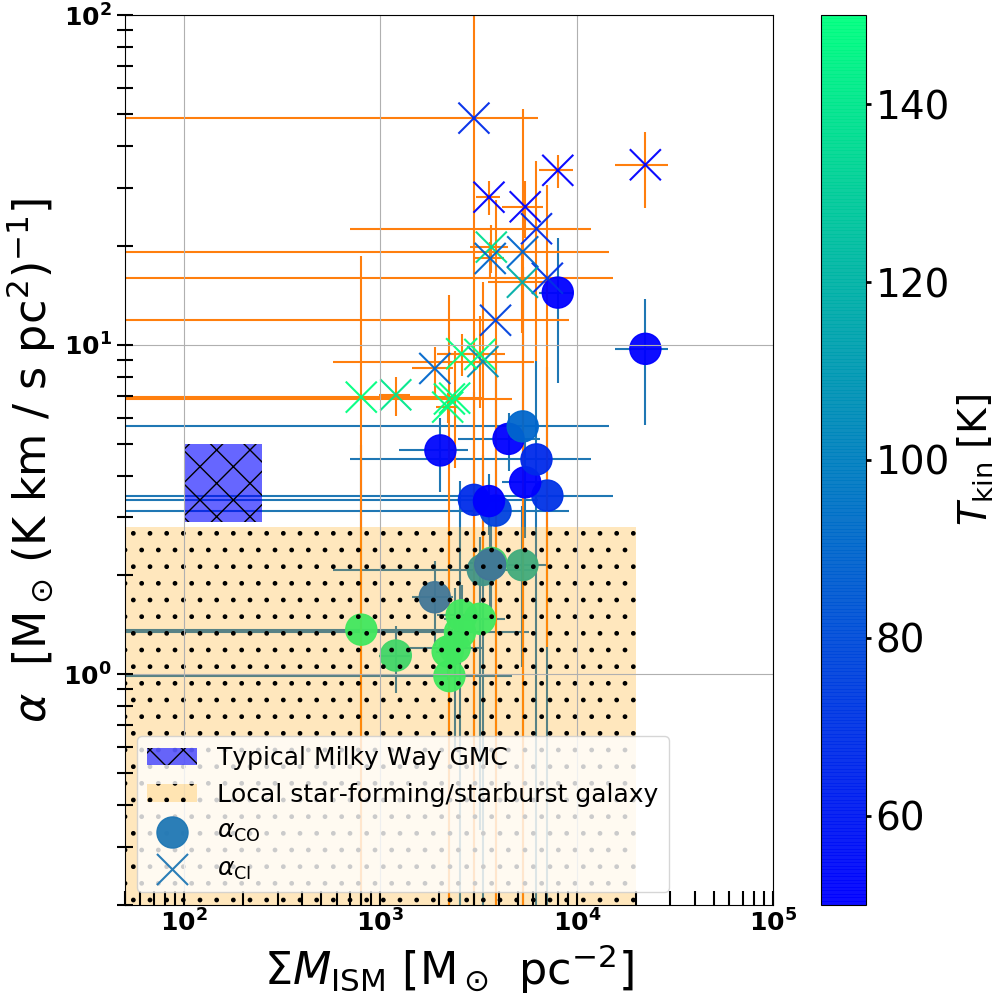}
 }\\
 \caption{\textit{Turbulence} model derived CO and [CI] line to molecular ISM mass conversion factors, plotted against the molecular ISM gas mass surface density, with $\alpha_{\rm CO}$ and $\alpha_{\rm [CI]}$ denoted as circles and crosses, respectively (log-log-scale). The colorbar denotes the gas kinetic temperature, $T_{\rm kin}$. Also shown are the values for $\alpha_{\rm CO}$ derived in the Milky Way and the local IR luminous star-forming systems \citep[see ][]{bolatto2013}. Both plots report the total $\chi^{2}$-weighted parameter mean and standard deviation values derived from the $\sim$2 million model calculations. }
 \label{fig:AlphaSigMass}
\end{figure}

In Fig.~\ref{fig:AlphaSigMass}, we examine the relation between the derived $\alpha_{\rm CO}$ and $\alpha_{\rm [CI]}$ conversion factors with the gas mass surface density in the sample of \LPs. The rather extreme galaxy-integrated surface gas mass densities for the \LPs\ is accompanied by a wide range in the conversion factor we derive. Previous works \citep{tacconi2008} have suggested that as the gas mass surface density goes beyond 100 \Msun pc$^{-2}$, i.e. comparable to GMCs in the Milky Way and nearby galaxies, the values of $\alpha_{\rm CO}$ decreases. Fig.~\ref{fig:AlphaSigMass} indicates that our analyses suggest the possibility of a factor 10 dispersion of $\alpha_{\rm CO}$ between $\Sigma_{M_{\rm ISM}} = 10^{3 - 4}$ \Msun pc$^{-2}$, yet a positive trend. Since the CI emission is optically thin the conversion factor must increase. This drives the value we derive for $\alpha_{\rm [CI]}$ to higher values, which are comparable to the value of $\alpha_{\rm CO}$ for those low-metallicity nearby galaxies. The average value of $\alpha_{\rm CO}$ is slightly lower than the Galactic value \citep[][]{bolatto2013}. Overall, the sample mean value for the \LPs\ is similar to the value of $\alpha_{\rm CO} = 3.6 $ derived using clumpy disk dynamical mass model calculations for near-IR selected galaxies at $z \sim 1.5$ \citep{daddi2010a}.  In general, lower $T_{\rm kin}$, and both higher $\Sigma_{M_{\rm ISM}}$ and \logden, seem to increase the value of $\alpha_{\rm CO}$ based on our full radiative transfer modelling of the \LPs. Comparable spatially resolved datasets would allow for such robust tests of functional forms between these parameters to enable a thorough comparison among this sample of \LPs. For example, \citet{maloney1988} considered $\alpha_{\rm CO} \propto \sqrt{n(H_{\rm 2}) }/T_{\rm kin}$ \citep{scoville2012}, while others have proposed a weaker dependence on the gas kinetic temperature, i.e. $\alpha_{\rm CO} \propto \sqrt{n(H_{\rm 2})/T_{\rm kin} } $ \citet{dickman1985, shetty2011b}.\\

\subsubsection{Comparisons between $M_{\rm ISM}$ estimates derived from optically thin, dust continuum methods}
\label{sec:Mass1mm}
The simultaneous modelling of the CO, [CI] and dust continuum emission enables a robust comparison to the inferred value of total molecular ISM mass,  $M_{\rm ISM}$,  based only on the properties of the dust SED. Recently, there has been a growing set of methods using dust continuum measurements to infer the $M_{\rm ISM}$ in star-forming galaxies, although observations of dust and its effects have been used to derive gas column densities over many years. Previous studies have integrated information about the visual extinction, gas-to-dust-mass ratio, the CO line luminosity to $M_{\rm ISM}$ conversion factor, $\alpha_{\rm CO}$, and observations of the thermal dust continuum along the assumed optically thin, Rayleigh-Jeans, side of the emission spectrum \citep{lilley1955, heiles1967, savage1970, aannestad1973, emerson1973, hildebrand1977, hildebrand1983, young1986, lonsdale1987, solomon1988, scoville1991, young1991, blain1993, kruegel1994, young1996, solomon1997,calzetti2000, genzel2010, leroy2011, magnelli2012, magdis2012, scoville2014,scoville2016, scoville2017, tacconi2018, schreiber2018, coogan2019, Kaasinen2019}. \\

\begin{figure}
\resizebox{0.497\textwidth}{!}{%
 \includegraphics[scale = 0.2]{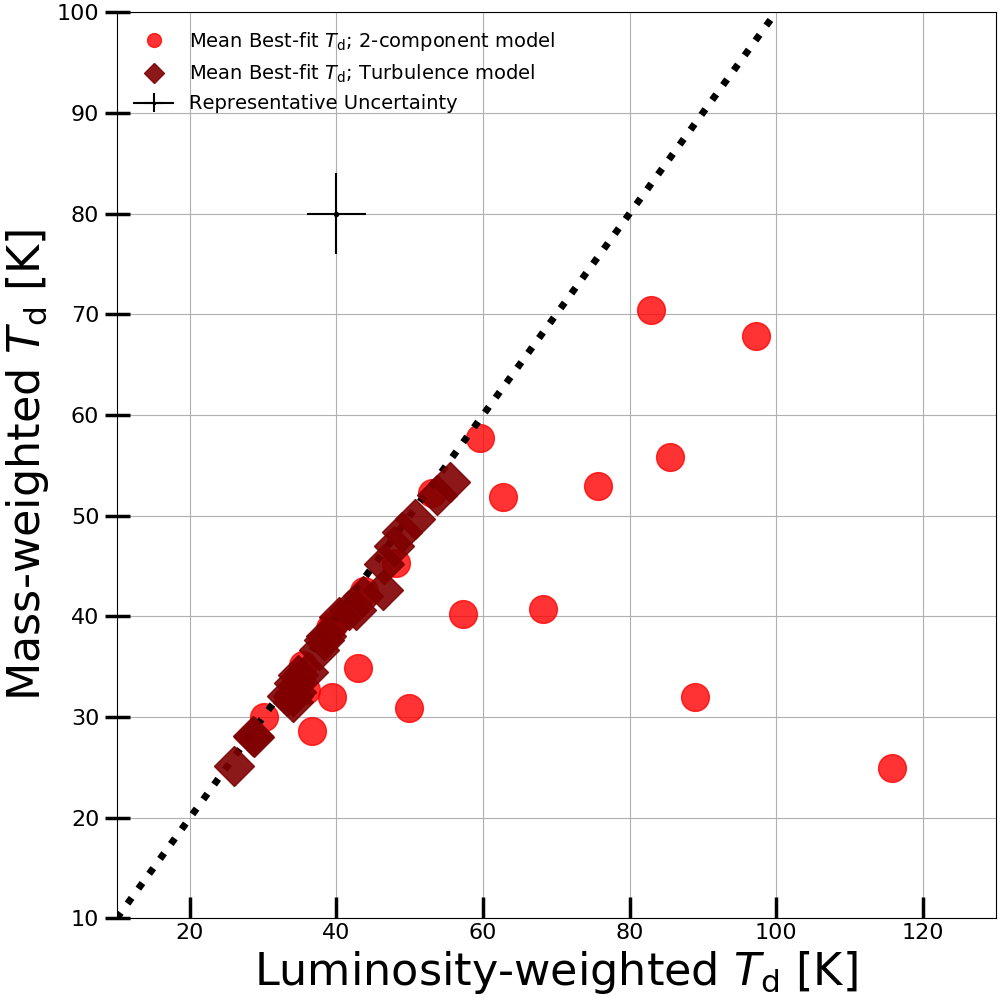}
 }\\
 \caption{$L_{\rm 850 \mu m}$,weighted dust temperature, $T_{\rm d}$, versus the mass-weighted value of $T_{\rm d}$ for both the \textit{2-component} model and the \textit{Turbulence} model. The calculations are described in \S \ref{sec:Mass1mm}}.
 \label{fig:lummassweightTd}
\end{figure}

The use of a single-band dust continuum measurement to estimate $M_{\rm ISM}$ is based on the assumption that the rest-frame dust continuum emission is optically thin beyond $\lambda_{\rm rest} \geq 250 \mu$m \citep[recently highlighted by ][]{scoville2014, scoville2016, scoville2017}. The method of \citet{scoville2014, scoville2016, scoville2017} (hereafter the ``\textit{1-mm method}'') uses the inferred rest-frame 850$\mu$m continuum emission, and derives an empirical calibration of the dust opacity per unit ISM mass. The estimate is increasing for the total $M_{\rm ISM}$ based on larger samples with CO(1-0) line measurements, enabling further calibrations of this method \citep[e.g.][]{Kaasinen2019}. In the \textit{1-mm method}, a Milky Way value of $\alpha_{\rm CO} = 6.5$ \citep{scoville1987, scoville2017} is applied for the absolute scaling from CO(1-0) line luminosity to the total $M_{\rm ISM}$. The \textit{1-mm method} uses a cold, and so-called ``mass-weighted'' dust temperature, $T_{\rm d}  = 25$ K \citep{scoville2014, liang2018, liang2019}, rather than the so-called `luminosity-weighted' value determined from template or modified blackbody SED fit results. Although these values are not inherently weighted by the luminosity, \citet{scoville2014, scoville2016, scoville2017} advocate that such template/graybody fitting will be driven by the warmest and brightest dust components. In addition, \citet{scoville2014, scoville2016, scoville2017} assume that most of the dust mass is in the colder phases, and will therefore be unaccounted for if the total molecular gas mass is derived using the results for the dust temperature based on a template/graybody SED fit. So far, this assumption has not been demonstrated explicitly using combined non-LTE radiative transfer modelling of the line and continuum SEDs. Here we explore whether or not the bulk dust mass likely arises from dust with colder temperatures. In fact, we suspect that both the warm and diffuse dust will contribute both to the total dust mass and SED. Both of the best-fit models (see e.g. Fig \ref{fig:bfJ1323}) suggest that the dust SED is dominated by warm and relatively diffuse gas with densities between \logden\ $\sim 2 - 4$ \percc, which correspond to the densities contributing most of the molecular ISM mass (see \S \ref{totalmass}). In the following, we test the hypothesis of the \textit{1-mm method}, which assumes that the global values of the mass-weighted and luminosity-weighted dust temperatures are different, using our model results. Since our models provide the dust temperature, mass and the luminosity for each gas component, we can compute a mass- and luminosity-weighted dust temperatures for each galaxy as:

\begin{equation}
\label{eq:mwTd}
    {\rm T_{\rm d,weighted}} =  \frac{ \Sigma_{\rm i = 1}^{i = N} \, W_{\rm i} T_{d, i}}{\Sigma_{\rm i = 1}^{i = N} W_{i} },
\end{equation}

where $T_{\rm d,weighted}$ is the mass-weighted or luminosity-weighted value, $N = 2$ for the \textit{2-component} model and $N = 50$ for the \textit{Turbulence} model, since we have 50 individual calculations of the dust temperature and mass. A similar approach in deriving a mass-weighted value of the dust temperature is reported by \citet{schreiber2018}, albeit they implemented a dust mass weight based on the value from an assembly of templates. The results of our calculations are shown in Fig.\ref{fig:lummassweightTd}, using the rest-frame $L_{\rm 850 \mu m}$ value to calculate our luminosity-weighted dust temperature. Here we use the rest-frame $L_{\rm 850 \mu m}$ (see also e.g. \citet{spilker2016}), however our following conclusions remain valid if we use rest-frame $L_{\rm 500 \mu m}$. We note that our definition of luminosity-weighted dust temperature is not the same as that referred to in \citet{scoville2014, scoville2016, scoville2017}, as the mean value from template/graybody fitting is considered the luminosity-weighted average. The dust temperatures from our models, before any weighting, are consistent, within 1-$\sigma$ uncertainties, with previously determined values of the dust temperatures from modified black-body fits \citep{canameras2015, harrington2016}. Overall, the values we derive for the \LPs\ span a range common to high-\z\ dusty star-forming galaxies. Therefore the high apparent IR luminosity for the \LPs\ does not bias our results to sources with higher dust temperatures. \\

\begin{figure}
\centering
\resizebox{0.497\textwidth}{!}{%
 \includegraphics[scale = 0.394]{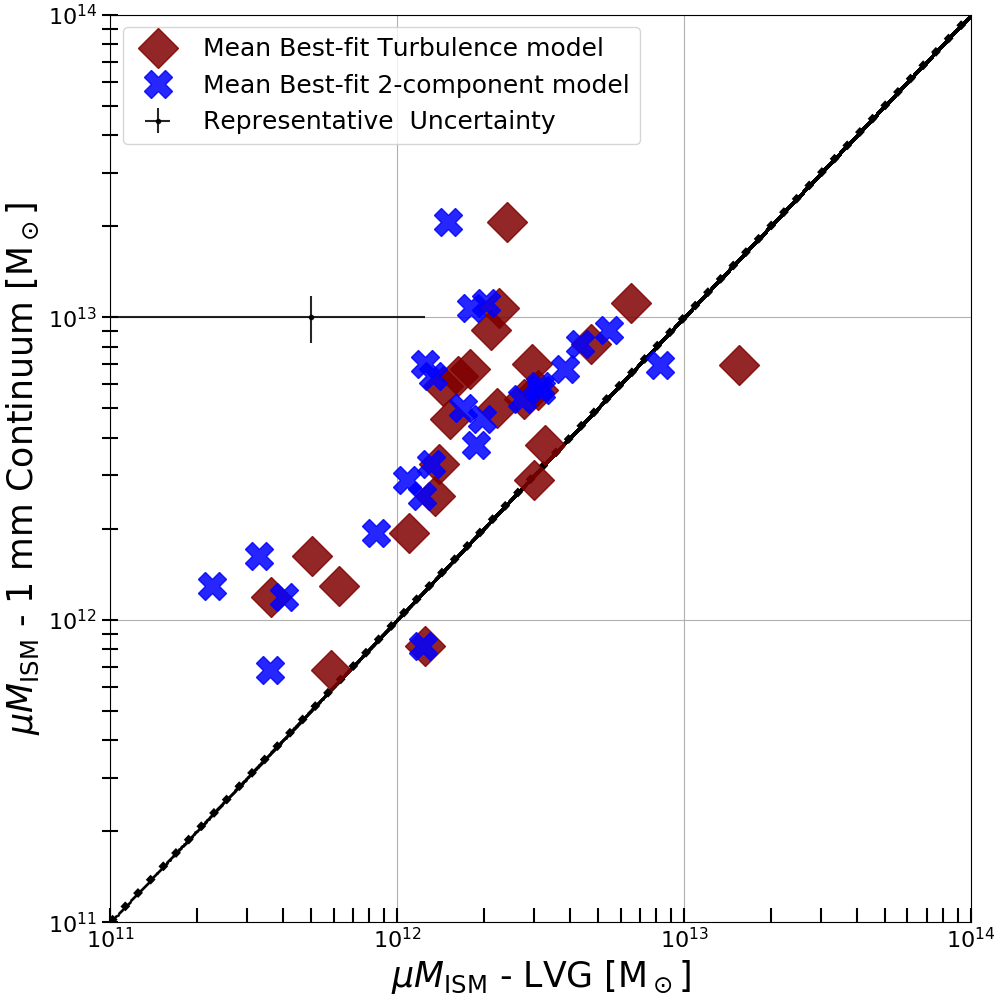}
 }\\
 \caption{Mass-mass plot (log-log-scale) of apparent total molecular gas mass, $\mu_{\rm L}M_{\rm ISM}$, derived using three methods. The y-axis shows the result using a single continuum measurement of the thermal dust emission and an assumed mass-weighted $T_{\rm d}  = 25$ K based on the scaling methods described in \citet{scoville2017}. The x-axis indicates the values of $\mu_{\rm L}M_{\rm ISM}$, as derived for both the \textit{2-component} model and the \textit{Turbulence} model.}
 \label{fig:mass-mass}
\end{figure}

Two immediate outcomes from this exercise are: 1.) the mass-weighted $T_{\rm d}$ value and luminosity-weighted $T_{\rm d}$ value for both models are remarkably consistent with one another, with a one-to-one correlation observed for all of the \LPs\ when using the more realistic \textit{Turbulence} model, and 2.) the mass-weighted $T_{\rm d}$ value for both models is consistently higher than the advocated value of $T_{\rm d} = 25$ K in the \textit{1-mm method} for star-forming galaxies. The mass-weighted value was justified by \citet{scoville2014} to reflect the fact that the dust grains exposed to strong radiative heating would represent a so-called luminosity-weighted dust temperature. We demonstrate, \textit{using two separate models}, that it is instead justified to use the effective luminosity-weighted dust temperature, e.g. as derived in a dust SED fit, when using the \textit{1-mm method} to calculate $M_{\rm ISM}$. We find a one-to-one correlation among the mass-weighted and luminosity-weighted $T_{\rm d}$ values, indicating the dust mass in the \LPs\ primarily consists of relatively warm diffuse gas, on contrast to the assumed cold diffuse dust content in the Milky Way. We stress, however, that the often-used optically thin MBB fit should not be used since it underestimates $T_{\rm d}$ compared to models with a realistic transition wavelength between the optically thin and optically thick emission regimes of the dust \citep{jin2019, cortzen2020}. \\

We further demonstrate the effects this may have in Fig. \ref{fig:mass-mass}, which shows the value of $\mu_{\rm L} M_{\rm ISM}$ derived using three separate methods. The first two methods are our full radiative transfer calculations of $\mu_{\rm L} M_{\rm ISM}$, as computed using the \textit{2-component} model and the \textit{Turbulence} model. We compute the final estimate of $\mu_{\rm L} M_{\rm ISM}$  for the \LPs, using the \textit{1-mm method} and the respective AzTEC or ALMA $\sim$1-mm dust continuum measurements \citep[][ Berman et al. in prep.]{harrington2016}, a mass-weighted $T_{\rm d}  = 25$ K and a value of $\beta_{T_{\rm d}} = 1.8$ (which is consistent with our modelling procedure, \S \ref{modelling}). Overall, the \textit{1-mm method} systematically over-predicts $\mu_{\rm L} M_{\rm ISM}$, consistent with other work \citep{schreiber2018,liu2019b}. This confirms the assumed mass-weighted $T_{\rm d}$ in the \textit{1-mm method} are too low to explain the \LPs, as the value we use for the $GDMR$ in our modelling is comparable to that used in the empirical calibration presented most recently in \citet{scoville2017}. In some cases, the discrepancy may be larger than a factor two, and this is likely due to the large dispersion we find for the \LPs\ for the average value of $\alpha_{\rm CO}$. In a recent study by \citet{Kaasinen2019}, the rest-frame 850$\mu$m luminosity was also used to cross-calibrate the mass estimate. They found a factor of two discrepancy in the total molecular gas mass estimates using both spatially resolved CO(1-0) line emission, with an assumed $\alpha_{\rm CO} = 6.5$ \alphaunits\, and a spatially resolved dust continuum measurement used to infer the rest-frame 850$\mu$m emission and the \textit{1-mm method} described above (assuming the same value of $\beta_{T_{\rm d}} = 1.8$ as we have used).\\

For many of the \LPs, the value of $\alpha_{\rm CO}$ is less than the standard value used in the \textit{1-mm method}. For global comparisons between galaxy populations, a single value was deemed appropriate, however, the estimate of $ M_{\rm ISM}$ for high-\z\ star-forming galaxies may be over- or under-estimated on average if the value of $\alpha_{\rm CO}$ is undetermined. The molecular gas and dust are, presumably, well-mixed in such turbulent star-forming systems \citep{krumholz2018}. Since they are believed to trace one another, the single-band $\sim$ 1-mm, dust continuum method to derive the $ M_{\rm ISM}$ has a clear advantage because it is more feasible to obtain a (sub)mm continuum detection of a high-\z\ galaxy than it is to detect multiple CO lines and perform a non-LTE radiative transfer analysis to explicitly derive an estimate of $ M_{\rm ISM}$. The strong dependencies on the assumed dust SED and gas excitation conditions are important to consider when estimating $ M_{\rm ISM}$, and therefore motivate further benchmarking between the various methods to derive $M_{\rm ISM}$ in high-\z.\\


\subsection{Heating, cooling and turbulence-regulated SF}
\label{sec:heatandcool}

SF occurs deep within molecular clouds, and this process requires cooling to aid gravitational collapse. At relatively high gas column densities, the kinetic energy transferred to CO molecules is converted to line photons which then radiate that energy away. Therefore we expect that CO line cooling is the dominant cooling process of the molecular gas-rich star-forming regions within the \LPs. At lower column densities the far-IR fine-structure lines of singly ionized carbon, [CII], and neutral and doubly ionized oxygen, [OI] and [OIII], are often considered as the dominant coolants of the star-forming ISM \citep{hollenbach1985, rosenberg2015, diaz-santos2016}. The highly ionized and/or high temperature (and density) regions traced by these FIR fine-structure emission lines are not expected to contribute to the cooling of the gas and dust rich molecular gas traced by the CO line measurements of the \LPs. The contribution of collisionally excited [CII] line cooling arises from neutral gas within dense PDRs, corresponding to \logden\ $\sim$ 3 \percc\ and T = 100 K \citep{hollenbach1997, goldsmith2012}. Within the denser molecular gas phase we model in this work, we expect that collisional excitations between molecules is believed to play a stronger role as a gas heating term, as opposed to FUV heating from  photodissociation regions (PDR) which lie between the HII regions and the cold molecular gas \citep{Tielens1985}. In addition, \citet{meijerink2011} demonstrate in a pure PDR model that the CO cooling fraction is 3\% – 5\%, while \citet{rosenberg2015} noticed similar cooling power from the CO lines up to tens of percent of the total cooling budget\footnote{Without further information on the FIR fine-structure lines for the sample of \LPs, we are unable to make a full comparison using the CO cooling budget alone.}. They observed a strong CO cooling fraction, which does not show a deficit as observed in the FIR fine-structure lines. \\

To explore the nature, and possible source(s), of the energy for the total CO line cooling in the \LPs\, we first calculate the sum $\Sigma_{\rm J_{\rm up} = 1}^{\rm J_{\rm up} = 15 } (\mu_{\rm L}L_{\rm CO_{\rm J_{\rm up}}})$ for each CO line luminosity using the best-fit, minimum-$\chi^{2}$ values from the \textit{Turbulence} model (see Table~\ref{tab:modelfluxesratios}), and derive a range of values for the apparent CO cooling power for the \LPs\ between $\Sigma$ ($\mu_{\rm L}L_{\rm CO_{\rm J_{\rm up}}}$)  $\sim 7 \times 10^{42}$ to $\sim 5 \times 10^{44}$ ergs s$^{-1}$.  Our analyses of the \textit{Turbulence} model results suggest that the global molecular ISM in the \LPs\ often has an \Htwo\ gas density \logden\ $>$ 4 \percc, and gas kinetic temperatures between 60 - 150 K. This implies that our estimate is specifically connected to the total cooling budget of this dense molecular gas phase. If we consider this cooling power to be continuous over the mean molecular gas depletion time, of the order of 70 Myr, the total energy emitted is of the order of $E_{\rm CO-70 Myr} \sim 10^{59-60}$ ergs. \\

We can also estimate the turbulent kinetic energy of the molecular gas using the mean, intrinsic molecular gas mass we derived in \S~\ref{totalmass}, and the mean turbulent velocity dispersion, resulting in $E_{\rm turb} = 0.5 (M_{\rm ISM}/{<\mu_{\rm L}> = 20}) \times \Delta V_{\rm turb}^{2} \sim 10^{54-55}$ ergs. We recall our results for the \textit{Turbulence} model, with the sample mean galaxy-wide, turbulent velocity dispersions for the \LPs\ of $< \Delta V_{\rm turb} > = 125 \pm 40 $ \kms, consistent with the second-moment velocity dispersion maps of line images for high-\z\ star-forming systems \citep{leung2019b, Yang:Gavazzi:2019, talia2018, venemans2019, tadaki2020, neri2020, jimenez-andrade2020}. This fiducial value of $E_{\rm turb}$ for a given mass and instantaneous turbulent velocity dispersion is 4-5 orders of magnitude lower than $E_{\rm CO-70 Myr} \sim 10^{59-60}$ ergs. Therefore, over the 70 Myr, a considerable amount of energy is responsible for sustaining the continuous CO line emission. \\

As noted in \citet{rosenberg2014}, local IR luminous systems may have various sources of turbulent energy, e.g., merger activity, AGN and powerful outflows. Studies predict the coexistence of starburst and AGN activity, particularly at $z \sim 2$ \citep{hopkins2008}, yet the \LPs\ have been shown to be strongly powered by SF rather than an AGN \citep[e.g.][]{harrington2016}. Nonetheless, we can estimate the relative contribution of mechanical feedback energy from AGN outflow activity. As a reference, one of the most powerful radio-loud QSOs, 3C82, has a jet power of the order of $10^{47}$ ergs s$^{-1}$ \citep{punsly2020}. Theoretical studies which aim to reproduce the formation of local massive elliptical galaxies have indicated an AGN mechanical outflow energy estimate of the order of $10^{42 - 43}$ ergs, with a rate of $\sim$ 3$\times 10^{-5}$ Myr$^{-1}$ \citep{gaspari2012}. If we use this AGN episodic rate from \citet{gaspari2012} and the jet power of 3C82, we estimate a total power of the order of $10^{58}$ ergs in 70 Myr -- i.e. $\le 10$\% of the total energy radiated away by CO line emission. Note, it is unlikely that all of the mechanical jet power is continuous, nor is it directly transferred into the molecular gas of the ISM, since 3C82 has a biconical outflow orientation. The mechanical power from AGN jets may also impart a significant amount of energy in the form of galactic outflows, of the order of $10^{44 - 46}$ erg s$^{-1}$ \citep{veilleux2009, sharma2012, mcnamara2014, russell2016, veilleux2020}. The unconstrained nature of galactic outflows at high-\z\ is still to be determined, as the mean gas densities may exceed the jet densities by four or five orders of magnitude \citep{mcnamara2016}, and therefore is a caveat in this interpretation. \\

AGN may also produce a significant amount of X-ray heating \citep{meijerink2006, meijerink2007}. X-ray absorption effects between the rest-frame 2 - 30 keV energy range are pronounced at higher energies. The gas column densities we derive in \S \ref{meanprops} suggests that we are often in the Compton-thick regime beyond $N_{\rm H} > 10^{24}$ cm$^{-2}$ \citep{hickox2018}. Currently, there is no constraint on the X-ray luminosity in the \LPs, and we are aware of only one example of a radio-AGN, i.e. LPs-J0209 \citep{geach2015, harrington2019}. We also cannot rule out the possibility of a heavily dust obscured AGN, yet our sample of \LPs\ have a strong selection function biased away from identifying strong QSOs \citep{yun2008, harrington2016}. Although there are limited X-ray studies of dusty star-forming galaxies at $z > 1 $, we are able to estimate a possible energy from an assumed apparent X-ray luminosity, as derived from the apparent SFR for the \LPs\ and a local ratio of $L_{\rm X: 0.5 - 8 keV}/SFR$ \citep{mineo2014}. We use an assumed intrinsic spectral shape for a non-AGN X-ray contribution, with an X-ray absorbing gas column density of $N_{\rm H} \sim  10^{22}$ cm$^{-2}$, and infer an apparent X-ray luminosity of the order of $10^{43}$ ergs s$^{-1}$. This energy is consistent with other, $z \sim 2- 3$, FIR detected X-ray AGN galaxies \citep{mullaney2011, mullaney2012}.\\

Since we expect much of the intervening gas column densities to be up to three orders of magnitude higher than this assumed value, this likely reflects an extreme upper limit. This corresponds to an apparent X-ray energy of the order of $10^{59}$ ergs. Although this is equivalent to the CO line cooling, when integrated across the fiducial 70 Myr timescale, it is a strong assumption for the X-ray luminosity to be continuous. Since this may be a strong upper limit, we can also conclude that X-ray luminosity from a non-AGN component is unlikely to be the primary heating mechanism to excite the gas-rich molecular ISM in the \LPs. \\

Cosmic rays, unless inhomegenously distributed, are unlikely to regulate star-forming gas at physical scales larger than 100 pc -- although the distribution and random diffusion of cosmic rays is highly uncertain \citep{thompson2006, zweibel2013}. Cosmic ray heating \citep[see the review by][]{krumholz2014b}, is primarily one of the strongest gas heating mechanisms within cloud interiors when the dust temperatures are $\sim$ 20 K and the gas densities are $\sim$100-1000 \percc. Beyond gas densities of 10$^{4}$ \percc, the effects of dust grain-molecular gas energy exchange (via the IR radiation field and/or grain-molecule collisions) are predicted to become stronger, if not dominant over cosmic ray and far-UV (FUV) heating \citep{goldsmith2001, krumholz2011b, narayanan2014}. It appears that cosmic ray heating may not be pervasive throughout the ISM of the \LPs, since their mean densities are above 10$^{4}$ \percc. Both observational and theoretical work suggests that the more diffuse gas, which can extend out to $\sim$ 10 kpc, may be strongly influenced by cosmic rays in dusty star-forming galaxies at $z \sim 2 - 3$ \citep{papadopoulos2004b, acciari2009, abdo2010, Bisbas2015, falgarone2017, indriolo2018}. The relative role of cosmic rays in driving the heating in high-\z\ star-forming galaxies is currently unconstrained, which remains a caveat in this analysis. There is evidence to suggest that the UV radiation field strength determines the relative cosmic ray ionization rate, and for such galaxies it may be likely that the majority of the cosmic rays are confined to the local star-forming regions within the ISM because the UV radiation decreases faster than the inverse square of the distance from the ionizing source \citep{indriolo2018}.  \\

Our results for the sample mean value of $T_{\rm kin}/T_{\rm d} = 2 - 3 $, on average, suggests galaxy-wide, turbulence-driven, mechanical heating as a signpost for the significantly high SF activity in these high-\z\ galaxies. The $T_{\rm kin}/T_{\rm d}$ ratio parameter, to zeroth order, may be an interesting parameter to better understand the relative level of mechanical (traced by $T_{\rm kin}$) versus photoelectric heating (traced by $T_{\rm d}$). The local starburst galaxy, NGC6240, has a large CO line to continuum ratio driven by galaxy wide shocks \citep[i.e. mechanical energy input; ][]{papadopoulos2014}. This scenario seems to be consistent with the large total line-widths (\S \ref{thelines}) and highly turbulent star-forming medium inferred for the \LPs. Therefore some form of kinetic activity must be responsible to drive this ratio to higher values, which implies some form of kinetic energy density must sustained to distribute the significant molecular gas content within the ISM of the \LPs. The values of $T_{\rm kin}/T_{\rm d} = 2-3$ are also found in the Milky Way regions with strong interstellar radiation field strengths, $G_{\rm 0} = 10^{5}$, such as the Orion PDR regions (a peak density of \logden\ $\sim$ 5 \percc). These regions have a typical visual extinction $A_{\rm V} < 4$ (in mag). These optical extinctions correspond to low column densities, where most of the FUV radiation is absorbed \citep[][; Fig. 16]{hollenbach1999}. We have already shown in \S \ref{meanprops} that the \LPs\ have extinction values of the order of many hundreds of magnitude. Deeper within the PDR structure of Orion, corresponding to $A_{\rm V} > 4 $, the $T_{\rm kin}/T_{\rm d} $ values decrease towards a value of unity or less \citep[][; Fig. 16]{hollenbach1999}. Since we do not have values close to, or less than, $T_{\rm kin}/T_{\rm d}  = 1 $, we can conclude that FUV heating from PDRs is likely not the primary heating mechanism in the ISM of the \LPs.\\

Other forms of heating mechanisms, therefore, seem to be required for the more intense star-forming galaxies like the \LPs. \citet{rosenberg2014} introduce an additional form of mechanical heating\footnote{Note, \citet{rosenberg2014} scale this simple mechanical heating term between a normalized value of 0 and 1, and combine this with their PDR model in order to match the LVG derived values of the $T_{\rm kin}$.} to their PDR models in order to fit the observed line SED for the local starburst galaxy, NGC253. In fact, this mechanical heating term is required to account for more than two-thirds of the observed mid-high-J CO line fluxes. It is also required to also recover the solutions to their alternative LVG models, one of which corresponds to a molecular phase with \logden\ $> $ 3.5 \percc, and $T_{\rm kin} = 60$ K. The PDR models alone could only reproduce a maximum gas kinetic temperature of 18 K when considering the gas density for the LVG model as an PDR input value. They argue the radiation field required to heat the gas photodissociates the CO molecules in the PDR in these models, the result is a factor of three lower value for the gas kinetic temperature between the PDR and LVG model results. \citet{rosenberg2014} therefore argue there is a need for this additional mechanical form of heating. \citet{rosenberg2014} do not fit for the dust temperature and $T_{\rm kin}$ simultaneously, so there is no direct comparison to our modelling procedure. Nevertheless, it is clear that PDRs are physically unlikely to be able to excite the observed line fluxes for the \LPs. This is because the \LPs\ have dust temperatures higher than the PDR-derived, maximum gas kinetic temperature of 18 K (in the case of \logden\ = 3.5 \percc). We therefore infer that the $T_{\rm kin}/T_{\rm d}$ parameter reflects a strong mechanical heating mechanism within the molecular ISM of the \LPs, despite the fact that our model inexplicably accounts for the source of this mechanical energy. \\

\begin{figure*}
\resizebox{0.997\textwidth}{!}{%
 \includegraphics[scale=0.34]{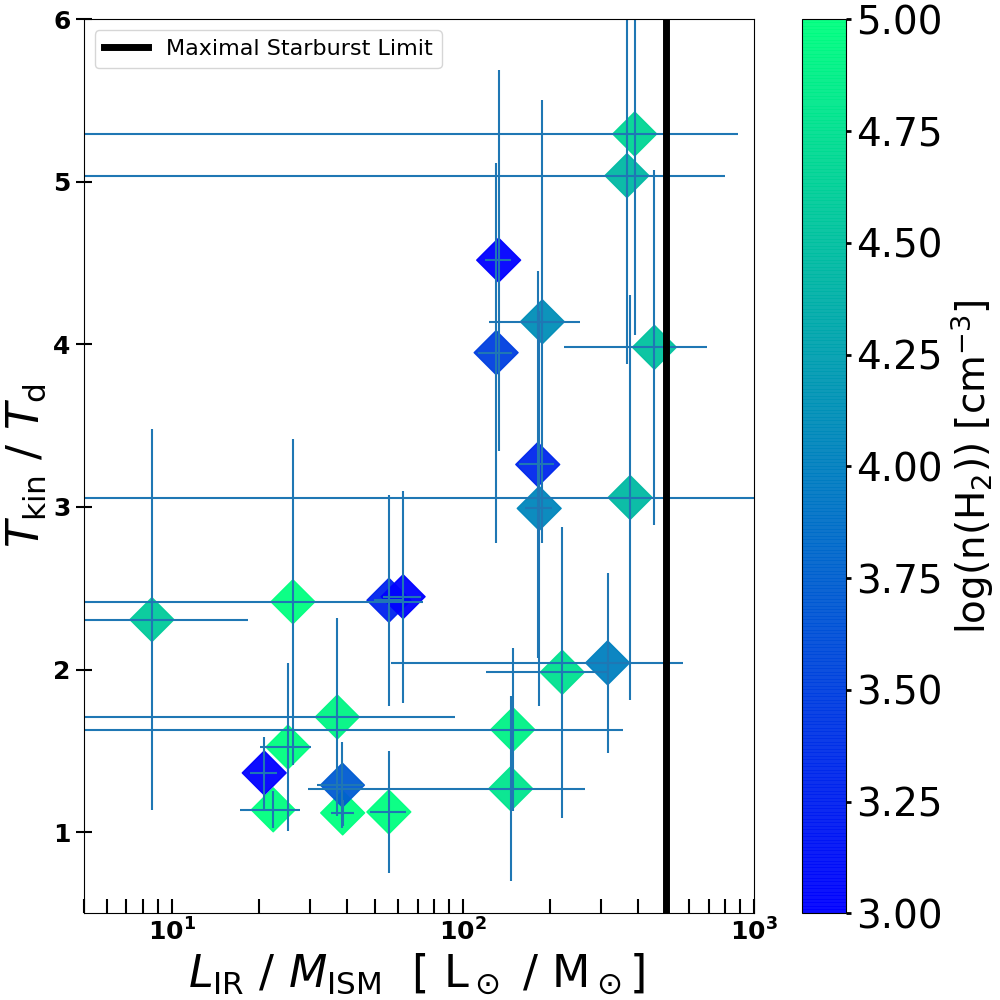}
 \includegraphics[scale=0.34]{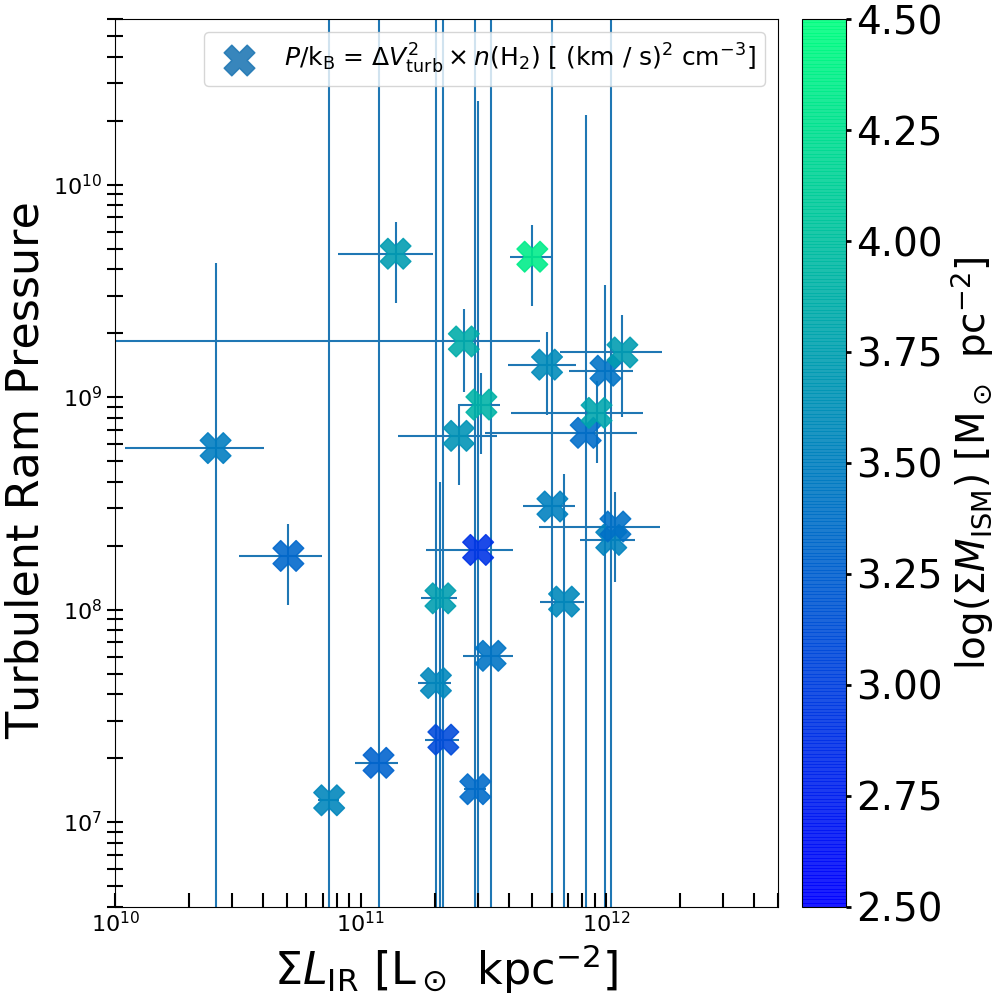} 
 }\\
 \caption{\textbf{Left: }$T_{\rm kin}/T_{\rm d}$ parameter versus the SF efficiency proxy, i.e. the total IR luminosity to molecular ISM mass ratio, $L_{\rm IR}/ M_{\rm ISM}$, derived using the \textit{Turbulence} model, with the derived mean \Htwo\ gas density in the colorbar axis (log-x scale).\textbf{ Right: }Turbulent ram pressure versus the IR luminosity surface density, with the molecular gas mass surface density in the colorbar axis (log-log-scale). Both plots report the total $\chi^{2}$-weighted parameter mean and standard deviation values derived from the $\sim$2 million model calculations. }
 \label{fig:SFE}
\end{figure*}

Fig.~\ref{fig:SFE} shows the $T_{\rm kin}/T_{\rm d}$ parameter to the ratio of the total IR luminosity to molecular gas mass  (i.e. a proxy of SFR per unit gas mass, hereafter SF efficiency ``SFE''). The SFE is believed to strongly increase with increasing redshift, accompanied by a similarly strong evolution in the mass accretion rate for field galaxies at $z > 1$ \citep{scoville2017, tacconi2020}, as well as increased turbulent velocity dispersions and SN rates \citep{Joung2009, krumholz2018}. The combined feedback from various stellar evolution processes are likely captured in the SFE parameter, therefore we expect that the value of $T_{\rm kin}/T_{\rm d}$ will increase with higher values of SFE. The left-hand side of Fig.~\ref{fig:SFE} shows some of the \LPs\ approaching the limits of a maximal starburst \footnote{A maximal starburst is a system within which radiation pressure overcomes the SF episode by disrupting ongoing SF activity, with a theoretical limit of $\mu_{\rm L} L_{\rm IR}/ \mu_{\rm L} M_{\rm H_{\rm 2}} = 500 $ L$_{\rm \odot}$/M$_{\rm \odot}$ \citep{thompson2005, andrews2011}. }. Overall there is a range of values for the SFE of the \LPs, corresponding to galaxies which would be considered main-sequence $z \sim 2 - 3$ star-forming galaxies, as well as the extreme outlier starburst galaxies \citep[e.g. ][]{genzel2010, genzel2015, tacconi2018, tacconi2020}. Higher values of the IR luminosity are proportional to the increase in SFR, and may therefore be an indicator for the mechanical energy input required to increase the values of $T_{\rm kin}/T_{\rm d}$. Indeed, Fig.~\ref{fig:SFE} shows that as the SFE-proxy increases, there are more \LPs\ with higher values of $T_{\rm kin}/T_{\rm d}$. The increased SFR in the \LPs\ is expected to contribute a significant amount of energy from stellar feedback, both in the form of massive proto-stellar outflows and/or supernovae (SN) explosion shocks \citep{mckee1989, norman1980, draine1993, krumholz2006a, nakamura2007, krumholz2009,hopkins2011, leung2019c, keller2020, Haid2019, Haid2018, Seifried2017} -- all of which dissipate energy through a turbulent energy cascade towards smaller physical scales \citep{dobbs2013,vanLoo2013}. The \LPs\ have a significant quantity of molecular gas, therefore we expect that this will dominate over the radiation pressure within the ISM\footnote{Due to, e.g., Rayleigh-Taylor instabilities.}. Therefore we expect that mechanical stellar feedback (rather than radiative feedback) will likely have a strong contribution to the pervasive turbulent gas conditions within the ISM \citep{jacquet2011, krumholz2012}. \\

The total wind energy of an O-type star on the main-sequence of stellar evolution, combined with the rapid mass-loss rate of the more transient Wolf-Rayet evolutionary stage, results in a fiducial range of $\sim 10^{49 - 51}$ ergs in a 5 Myr lifetime \citep{Leitherer1999, chu2005, smith2014, Ramachandran2019}. As a reference, the intense star-forming region traced by the supergiant shell within IC2574, a nearby dwarf galaxy, has an estimated kinetic energy input of the order of about $\sim 3\times 10^{53}$ ergs over a 1 Myr lifetime \citep{walter1998}. We can estimate the stellar mechanical wind energy input using a fiducial value for the intrinsic SFR for the \LPs\ of 1000 \Msun yr$^{-1}$ (see also \S \ref{meanprops}) and an expected cumulative fraction of 0.2\% by number for massive stars \citep[Kroupa IMF][]{kroupa2002}\footnote{Note, recent studies of high-\z\ star-forming galaxies suggest there may be a top-heavy IMF \citep{romano2017, zhang2018}.}. The approximate stellar mechanical wind energy input is of the order of $\sim 1\times 10^{52}$ ergs. SNe events occur at the end of the life-cycle of massive stars with initial stellar masses between $\sim 10 - 40$ \Msun \citep{heger2003}, each of which produces roughly $10^{51}$ ergs \citep{Jones1999}. If we assume that these massive stars consist of $\sim$ 7 \% of the total stellar mass fraction we derive a SNe rate of $\sim $6 SN yr$^{-1}$ for the \LPs, which is $\sim 300\times$ the value of the Milky Way \citep{Diehl2018}. For reference, the center of M82 has an estimated rate of $\sim 0.1$ SN yr$^{-1}$ \citep{kronberg1981, weiss1999}.\\

We can further estimate the energy input from SNe to be on the order of $10^{59}$ ergs using the reference time-frame of 70 Myr. Therefore this value is an upper limit, since we do not expect that the injection of energy from SNe will be a continuous process. Nonetheless, this value is comparable with our simplistic estimate of the possible stellar mechanical wind energy input from young massive stars over 70 Myr, which are expected to provide a steady stream of mechanical energy throughout their lifetimes via stellar winds with terminal velocities of the order of 1000 \kms \citep{conti1988, puls2008}. The direct impact of massive stars and SNe, however, may be biased towards their most immediate environments. It has also been shown theoretically that only a 1\% fraction, or less, of the SNe energy output is transformed into turbulent energy \citep{iffrig2015, martizzi2016}. Overall, the mixture of both SNe events occurring in parallel with the stellar evolution processes of the population of Wolf-Rayet stars and short-lived massive stars in young stellar associations can still potentially provide a large fraction of the necessary energy budget with respect to the value of E$_{\rm CO-70 Myr} \sim 10^{59}$ ergs. As well, the mechanical energy of $E$(massive stellar winds \& SNe) $\sim 10^{59 - 60}$ ergs,i.e. 4-5 orders of magnitude larger than E$_{\rm turb}$. \\

Despite the potentially strong energetic contributions from stellar evolution processes, massive stars may not be the only sources of such turbulent energy input. Gravity may play an equally important role in introducing a large amount of turbulent gas motion, which may sustain the relatively high turbulent velocity dispersions we derived for the \LPs. The right-hand side of Fig.~\ref{fig:SFE} shows the relationship between the IR luminosity surface density, $\Sigma_{\rm L_{\rm IR}}$, and the turbulent ram pressure for the \LPs. The equivalent turbulent gas ram pressure is connected to the vertical stabilizing force in a marginally stable gas disk. This pressure will increase as the SF activity increases, according to turbulence-regulated SF models \citep{krumholz2009b, bournaud2010, krumholz2018}. We use the mean values from the \textit{Turbulence} model to calculate the turbulent ram pressure to be $P_{\rm turb}$ = $\Delta_{\rm turb}^{2}\times n({\rm H_{\rm 2}} )$ = $10^{6.2 - 9.8 }$ (\kms )$^{2}$ \percc, which is significantly higher than the gas thermal pressure, $P_{\rm th} = P/k = $ n(${\rm H_{\rm 2}}$)$\times T_{\rm kin}$ = $10^{2.9 - 6.9}$ K \percc\ (with $k$ the Boltzmann constant). Many \LPs\ have large uncertainties due to our lack of constraints on the molecular gas density and our total errors, and future work may refine these values for individual \LPs\ using spatially resolved line measurements of the mean turbulent velocity dispersion. Both the Kendall's tau statistic, $\tau = 0.27$ and the Spearman's rank correlation, $r = 0.37$ indicate that there is only a mild positive correlation between the turbulent gas pressure across the range of $\Sigma_{\rm L_{\rm IR}}$ $\sim 10^{11 - 12}$ \Lsun kpc$^{-2}$ for the \LPs, if any. Since there is a large amount of turbulent energy present in the \LPs, it is inferred that the thermal pressure equilibrium of clouds is negligible overall in terms of regulating the SF activity. This result is consistent with theoretical studies \citep[e.g. ][]{ballesteros-Paredes1999}. It is important to measure the mean molecular gas thermal pressure, as it sets the background for thermal pressure balance throughout the ISM, as galaxies with higher than 50\% molecular gas to atomic gas fractions are subject to collapse \citep{krumholz2005}. \\

Our results indicate that the additional form of turbulent pressure is an important form of feedback to regulate the intense SF activity for these gas-rich \LPs. Using equation A7 from \citet{brucy2020} for the upper bound for possible turbulent power, $\Pi_{\rm SNe}$, injected by a SNe:

\begin{equation}
    \Pi_{\rm SNe} \sim 4\times10^{37} ( \Sigma_{\rm M_{\rm ISM}}/10 {[ \rm M_{\rm \odot} yr^{-1} ]})^{1.4} \, [{\rm ergs \,  s^{-1} }].
\end{equation}

This yields a much lower estimate of the total turbulent energy input from SNe over the 70 Myr of $\sim 10^{56}$ ergs, still $\sim$ 3 orders of magnitude less than $E_{\rm CO-70 Myr}$. This equation assumes the turbulence energy injection is proportional to the surface integrated SFR \citep{krumholz2018, brucy2020}. The high gas mass surface densities of the \LPs\ can therefore produce much more power from stellar feedback than the highest explored values for the SFR $\sim$ 100 \Msun yr$^{-1}$ in the work of \citet{brucy2020}, therefore these estimates may reflect lower limits. \citet{brucy2020} find that the large-scale turbulent energy injection has a much higher dependance on the gas mass surface densities,\textit{ by almost three orders of magnitude}. The general scenario is consistent with other studies, which find a more dominant turbulent energy contribution from large-scale motions instead of pure stellar feedback \citep{bournaud2010, renaud2012, krumholz2018, colling2018}. We can loosely estimate this large-scale turbulent energy input based on the theoretical model values of \citet{brucy2020} for a dense molecular medium, corresponding to $\sim 10^{40}$ ergs s$^{-1}$. We calculate, over 70 Myr, an estimate of $\sim10^{55}$ ergs. This is one order of magnitude higher than the fiducial $E_{\rm turb}$ $\sim 10^{54-55}$ ergs we estimate above for the \LPs. Therefore, a substantial amount of turbulence energy can also be supplied from large-scale disk motions.\\


Due to the conservation of angular momentum, any form of momentum injection into the molecular cloud surroundings will likely be radially transported towards the gravitational center of the galaxy. Indeed, mass accretion may play a primary role in increasing the level of turbulent energy within the ISM during these starburst episodes \citep{Schmidt2016}\footnote{This is in addition to the turbulent motions generated from hydrodynamic gas motions causing gravitational shear forces.}. The kpc-scale dynamics for these high-\z\ gas-rich star-forming galaxies cause gravitational instabilities, and the turbulent driving from these processes may be a primary source of the bulk kinetic energy density \citep{silk1997, schmidt2009, bournaud2010, krumholz2016, krumholz2018, colling2018, brucy2020}. These single-dish measurements offer a global view, or ``top-down'' perspective, of the molecular ISM conditions within the \LPs. The gas mass surface density drives these high SFR and IR luminosity surface densities in ``top-down'' processes \citep{krumholz2018}. The extreme gas mass surface densities may help to explain the extreme intrinsic IR luminosities exceeding 10$^{13}$ \Lsun. \\

We expect that such massive galaxy-wide SF will act to re-stabilize the forming disk over a timescale of $\sim 100$ Myr, meanwhile fresh gas is likely to be accreted, consumed or displaced within the ISM and/or in the form of massive galactic outflows \citep{quirk1973, cox1981, dopita1985, larson1987, ostriker2010, veilleux2020}. Altogether, such arguments for turbulence-regulated SF is consistent with evidence presented for both local (U)LIRGs and high-\z\ starbursts. The local (U)LIRGs, with similar CO line SED coverage, require high amounts of mechanical energy / turbulent activity to sustain the higher-J CO line emission \citep[e.g.][]{tacconi1999, lu2014, kamenetzky2016}, while the 10 kpc-scale turbulent molecular gas reservoirs are believed to extend the starburst phase of high-\z\ star-forming galaxies through the interplay of stellar/AGN feedback and intergalactic gas accretion \citep{falgarone2017}. \\


\section{Conclusions}
 
In this work we have studied the physical gas properties of the molecular and atomic ISM at high-\z. We have measured and compiled a compendium of roughly 200 CO and [CI] spectral lines measured in a sample of 24 strongly lensed star-forming systems at $z \sim 1-4$, selected from the all-sky sub/mm \Planck satellite (the \LPs). To yield deeper insight into the molecular ISM excitation conditions, we systematically measured the multi-J line excitation of CO and [CI] using spatially unresolved, single-dish observations (i.e. IRAM 30m ([CI] and CO(J$_{\rm up} = 3 - 11$)), GBT (CO(1-0)), and APEX ([CI] and CO(J$_{\rm up} = 4 - 12$))). \\

This work is the first major effort to simultaneously fit all of the available spectral line and dust continuum observations. The vast majority of previous high-\z\ studies have focused on non-LTE radiative transfer modelling of a single or double component model of the observed line emission, excluding the thermal background from the IR radiation field. In this work we perform two complementary modelling procedures to model all of the line/continuum data: i.) a two-component molecular medium model, which enabled us to highlight the dominant properties of the more diffuse/quiescent and denser/highly excited gas, and ii.) a more realistic description of a molecular ISM which is described by a, turbulence-driven, molecular gas density PDF. Our main results are summarized as follows:\\

$\bullet$ The broad [CI] and CO lines ($<FWZI> \, \sim 850$ \kms) are strikingly similar in line shape, therefore these emission lines trace comparable galactic dynamics across the spatially unresolved, kpc-scales.\\

$\bullet$ We have derived the mean CO line brightness temperature ratios for the \LPs\ out to the ratio of $L'_{CO(12-11)}/L'_{CO(1-0)}$, based on our best-fit, minimum-$\chi^{2}$ \textit{Turbulence} model. In addition, we have derived a set of median CO line brightness temperature ratios for a significant number of $z = 1 - 7 $ galaxies with CO line detections, including the compilations by \citet{carilli2013} and \citet{kirkpatrick2019}. Although the median values are in excellent agreement with the average best-fit, minimum-$\chi^{2}$ model derived values for the \LPs, the wide range in CO excitation observed in individual galaxies implies that the use of an average (or median-based) value used for scaling $L'$ measurements may be misleading when there is limited line/continuum data available.  \\

$\bullet$ There is a wide range in the observed intensities for the CO rotational ladder, with an order of magnitude dispersion tracing a range of gas excitation in these lensed IR-luminous, star-forming galaxies. We further explored this wide dynamic range in observed gas excitation following the methodology presented by \citet{rosenberg2015} for local IR luminous galaxies. We have thereby classified the CO excitation ladder with respect to the drop-off slope after the CO(5-4) transition by taking the ratio of the higher-J CO line luminosities to the mid-J CO line luminosities, and find the \LPs\ probe more than 4 orders of magnitude in CO excitation. This classification increases to higher excitation as the derived FIR luminosity increases.  \\

$\bullet$ There are 19 \LPs\ with a [CI] line detection, while sixteen have both [CI] lines detected. Our non-LTE radiative transfer modelling of these lines suggests the [CI] lines are indeed optically thin, which is important for reliable calibrations to carbon gas column density and total mass. The two ground-state fine-structure carbon lines are sub-thermally excited, however. We demonstrate, using 16 \LPs\ with both [CI] line detections, that the often-assumed LTE approximation to derive the carbon excitation temperature may under (or over) estimate the intrinsic carbon excitation temperature, depending on the gas excitation conditions of individual galaxies. We derived mean carbon gas excitation temperatures $T_{\rm exc} \sim$ 30 and 40 K for the \textit{Turbulence} model and \textit{2-component} model, respectively. In some of the \LPs\ we find values less than 20 K, and we would have misinterpreted the inferred total molecular gas mass if we had assumed the ideal, LTE prescription. We find, for the \textit{Turbulence} model, the sample mean for the \LPs\ of $< [{\rm [CI]}]/{\rm [H_{\rm 2} ]} > \sim 6.8  \times10^{-5}$, with a large dispersion.\\

$\bullet$ The average \textit{intrinsic} size of the modelled gas and dust emitting region for the \textit{Turbulence} model was derived to be $R_{\rm eff} (= 13.5 )/\sqrt{\mu_{\rm L} \sim 20.4} \sim 3 $ kpc. We have estimated the mean magnification using all of the available ranges derived for the \LPs\ which have lens models. For the \LPs\ without published magnification factors, we provide an estimate using the ``Tully-Fischer'' argument method presented by \citet{harris2012}, and our novel CO(1-0) line measurements. The intrinsic size for individual \LPs\, based on detailed lens modelling, agrees well with the expected intrinsic size derived in our modelling.\\ 

$\bullet$ We derived total molecular ISM masses in our modelling of the observed CO/[CI] lines and dust SED. Both of the modelling procedures are consistent in deriving $M_{\rm ISM}$, yet we find systematic offsets as the single-band 1-mm dust continuum method over-predicts the $M_{\rm ISM}$ derived using our robust modelling procedures. Our derived mean, mass-weighted, $T_{\rm d} \sim 40$ K for the sample of \LPs\ does not suggest the use of the recommended mass-weighted $T_{\rm d} = 25$ K value when using a $\sim$ 1 mm dust continuum observation to estimate the total molecular ISM mass. In fact, both of the modelling procedures indicate that the mass-weighted and luminosity-weighted $T_{\rm d}$ are close to identical, on average. \\

$\bullet$  We find a wide range in CO luminosity per mass, with a mean close to the Galactic value, i.e. $\alpha_{\rm CO} \sim 3.4$ \alphaunits, however there is a large dispersion. Each system has a unique value of $\alpha_{\rm CO}$, disfavoring the use of a single value common for active star-forming galaxies at high-\z\. Our modelling suggests the value of $\alpha_{\rm CO}$ increases with increasing gas mass surface density, as well as with gas volume density. The value of $\alpha_{\rm CO}$ decreases towards unity or less for increasing gas kinetic temperatures, specifically $T_{\rm kin}> $120 K.\\

$\bullet$  The more realistic description of the turbulent molecular gas offers a picture of the excitation conditions of the ISM in the \LPs. The large emitting regions are highly turbulent, as inferred by their mean turbulent velocity dispersion ( $\Delta V_{\rm turb} > 125 $\kms), and the gas kinetic temperature to dust temperature ratios  $T_{\rm kin}/T_{\rm d} > 2.5 $, on average, suggests the \LPs \, require a significant amount of mechanical activity on $>$kpc scales (the driving scale) in conjunction with their massive molecular gas reservoirs. Since the inferred gas depletion time, $M_{\rm ISM}$/SFR, is of the order of 70 Myr, there must be a significant amount of gas, likely supplied from the CGM over the lifetime of this $\sim 100 $Myr starburst episode. The $T_{\rm kin}/T_{\rm d}$ ratio also increases with the inferred SF efficiency (i.e. $L_{\rm IR}/M_{\rm ISM}$), which suggests the kinetic input from increased SNe and stellar winds may also play a role in characterizing the overall mechanical heating in the ISM.\\

\acknowledgments
We would like to sincerely thank the anonymous reviewer for their insightful and thoughtful comments which have influenced the quality of the manuscript. K.C.H would like to acknowledge the useful discussions to improve this manuscript with Nick Scoville, Alexandra Pope, Mark Krumholz, Stefanie M{\"u}hle, Tom Bakx and Chentao Yang. K.C.H would wholeheartedly like to thank all telescope station managers, operators and shift observers at all facilities: Carsten Kramer, Ignacio Ruiz, Manuel Ruiz, Frederic Damour, Joaquin Santiago, Kika, Victor Puela, Santiago Navarro, Salvador Sanchez, Martin Steinke, Jonathan Braine, Ivan Agudo, Maria Nuria Marcelino Lluch, Zsofia Nagy, Francesco Fontani, Elena Redaelli, Amber Bonsall, Amanda Jo, Toney Minter, Karen O'Neil, Rodrigo Parra, Mungo Jerry, Francisco Azagra, Felipe Mac-Auliffe, Paulina Vasquez, Eduoardo Gonzalez, Mauricio Martinez, Juan-Pablo Perez-Beapuis, Arnoud Belloche, Friedrich Wyrowski, Papito, Arshia Jacob, Parichay Mazumdar, Nina Brinkmann, and Hans Nguyen. This work is carried out within the Collaborative Research center 956, sub-project [A1, C4], funded by the Deutsche Forschungsgemeinschaft (DFG). T.K.D.L acknowledges funding from the European Research Council (ERC) under the European Union's Horizon 2020 research and innovation programme (grant agreement No. 694343). A.D.S acknowledges support from project PID2019-110614GB-C22 (MICINN). The research leading to these results has received funding from the European Union's Horizon 2020 research and innovation program under grant agreement No 730562 [RadioNet]. This publication is based on data acquired with the Atacama Pathfinder Experiment (APEX) Telescope. APEX is a collaboration between the Max-Planck-Institut fur Radioastronomie, the European Southern Observatory, and the Onsala Space Observatory. This publication also makes use of the Green Bank Observatory, a facility of the National Science Foundation operated under cooperative agreement by Associated Universities, Inc. This work is also based on observations carried out with the IRAM 30m telescope. IRAM is supported by INSU/CNRS (France), MPG (Germany) and IGN (Spain). This research has made use of NASA's Astrophysics Data System. This research has made use of adstex (\url{https://github.com/yymao/adstex}). \\

%

\vspace{5mm}
\facilities{GBT:100m, IRAM:30m, APEX:12m}




\clearpage

\appendix

\section{Notes on differential lensing}
\label{difflens}

The total size of the emitting region of the low-excitation and higher-excitation emission lines of CO may vary intrinsically in the source plane, therefore if these emitting regions are disproportionately distributed along the caustic, the observed fluxes could yield differential magnifications. That is, the more compact emission traced by higher-J CO transitions may be magnified by a factor that is significantly different than the diffuse, low-excitation gas \citep{blain1999, serjeant2012,hezaveh2012}, if it lies closer, on average, to the caustic. As AGN/QSOs have extreme luminosities centrally located within the host galaxy, these point-like objects may be subject to stronger differential lensing than starburst galaxies. This would lower the probability that such a distinction is made between the magnification factor for higher-J CO versus the low-J CO emitting regions. This is because the source of the higher-J emission in AGN/QSOs is likely confined to a concentrated region near the center of mass of the host galaxy, whereas the starburst galaxies have more extended reservoirs that may be well-mixed, and both low-J and high-J lines may be magnified differentially in a similar manner, on average. Differential lensing may be more pronounced when comparing low- and high-J CO line fluxes, however the bulk of this work is focused on global properties such as the total molecular gas mass -- which is most sensitive to the lowest-J CO line measurements.  High angular resolution imaging in the future is required to investigate differential lensing. We therefore focus on the observed quantities. \\

Due to the extreme starburst nature of the \LPs, the molecular ISM may have a large volume filling factor of gas, suggesting a smooth distribution of SF on galactic scales \citep{kennicutt1998,kennicutt2012}. An explicit accounting of this would require assigning a unique magnification factor to each measurement of the dust continuum and a magnification factor per velocity channel for the CO/[CI] lines to model the de-magnified integrated fluxes \citep[see e.g. ][]{leung2017}. Recent studies have shown a similar magnification factor for the low-J CO/adjacent dust continuum \citep{canameras2017b, canameras2018b}, while others show a non-negligible 20-40\% differential magnification of low-mid-J CO/dust continuum \citep{Yang:Gavazzi:2019}. It is clear that every lens configuration is a unique system, and therefore a coherent set of lens models is required for the lower-excitation versus the higher-excitation molecular/atomic gas at matching spatial resolution and S/N to diagnose these effects more systematically. However, in this study the effective source radius, described below, is directly connected to the apparent flux within the source solid angle, and is used as input to the model. For our two component modelling, we cannot explicitly determine the differential lens magnification factor, as the intrinsic ratio of the emitting radius for each component may be different from the modelled ratio. Thirteen out of the 24 \LPs \, presented here have lens models developed based on a wide range of high-angular resolution \HST near-IR data and/or ground-based optical/near-IR follow-up of the foreground and lensing environment \citep{diaz-sanchez2017, frye2019}. A small subset of the \LPs \, have a range of marginally resolved to highly resolved (down to beam sizes, $\theta \sim$0.1 - 0.2\arcsec) mm-radio interferometric dust continuum and/or single-line CO imaging to also aid lens modelling efforts \citep{geach2015, geach2018, bussmann2015, canameras2017a, canameras2017b, canameras2018a}. Overall, these systems have flux-weighted total magnification factors ranging between 10-40 (Table \ref{tab:summaryLPs}). \\

The lack of magnification factor estimates derived for each of the emission lines, in addition to multiple magnification factor estimates derived from sampling the rest-FIR thermal continuum emission, restricts our analyses to the magnified (apparent) quantities. \citet{harris2012} used a ``Tully-Fischer'' line luminosity/line-width relation to offer an empirical perspective on estimating the unknown magnification of the CO(1-0) line measured in 24 \Herschel-selected, strongly lensed galaxies. There may be an intrinsic dispersion among the \LPs \, along this empirical relation, and the inclination angle is unaccounted for, yet we can use equation 2 of \citet{harris2012} to estimate the lens magnification factor based on our GBT detected CO(1-0) line measurements. We derive the line full-width at half-maximum (FWHM) using a simple 1-D Gaussian model fit to the velocity line profiles as done by \citep{harrington2018}. In the case of LPs-J0305, LPs-J0226, LPs-J105353 and LPs-J112713, there is no available CO(1-0) line data. Therefore, we use the value of $L'_{\rm CO(1-0)}$ from the best-fit, minimum-$\chi^{2}$ \textit{Turbulence} model and the measured FWHM from the low-J CO line data. We report the derived magnification factors in Table \ref{tab:summaryLPs}. This magnification factor estimate, which is usually 1.5-3$\times$ higher than previously reported results, has systematic uncertainties in estimating the lens magnification estimate could be more than 50\% based on the intrinsic scatter within the calibration sample used in  \citet{harris2012}. \\

\section{\textit{Turbulence} model posterior distributions}

\begin{figure*}
 \includegraphics[scale = 0.35]{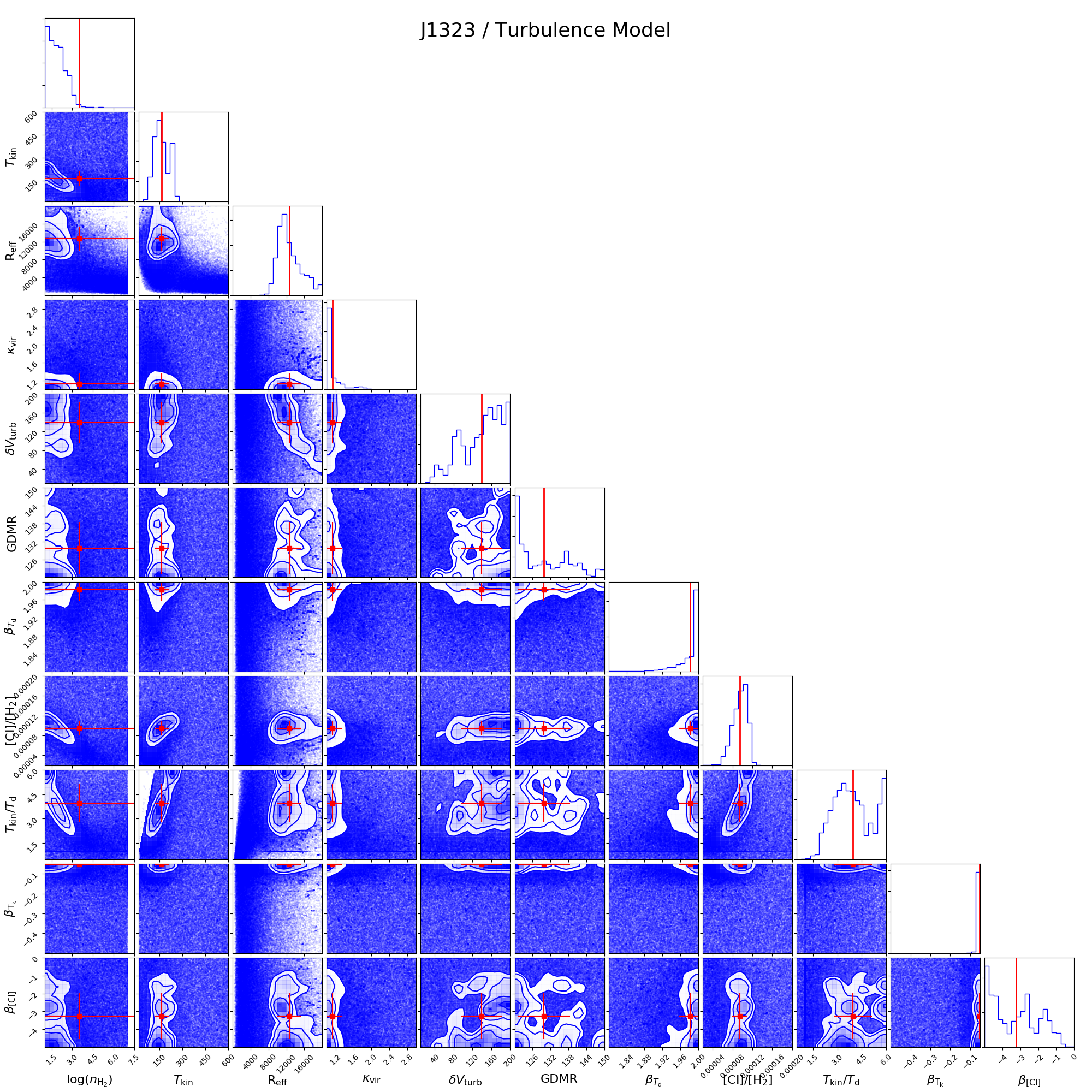}
 \caption{Posterior likelihood distributions within the explored solution space for the parameters used in the \textit{Turbulence} model. Denser regions indicate model calculations with higher likelihood. Note that the gas density is in log$_{\rm 10}$ units. The range of parameters correspond to the tightened parameter limits described in the text (see Tab. \ref{tab:paramspace}). Contours represent the 16th and 84th percentiles. The red points indicate the likelihood-weighted mean and standard deviation as based on $\sim$ 2 million model calculations. The general parameter space degeneracy observed here between parameters reflects many of the same features for the other \LPs; the corresponding posterior distributions can be found online.  }
 \label{fig:bfJ1323corner}
\end{figure*}

In \S \ref{sec:fit} we discuss our optimisation fitting procedure, for which we employ a Markov Chain process combined with the optimisation routine to sample the underlying probability distribution of each parameter \citep[see ][ and future work]{Pham2009,strandet2017}. In Fig. \ref{fig:bfJ1323corner} we show an example, for LPs-J1323, of the likelihood-weighted posterior distributions for the explored range of parameters (Tab. \ref{tab:paramspace}, corresponding to $\sim$ 2 million model calculations. As noted in \S \ref{sec:fit}, the average galaxy-integrated gas excitation conditions of the \LPs\ are inferred based on the mean, likelihood-weighted parameters (as determined by the $\chi^{2}$ value) and uncertainties, which encompass a wide variation in solution space (Fig. \ref{fig:bfJ1323corner}). The highest likelihood regions in Fig. \ref{fig:bfJ1323corner} correspond to the best-fit, minimum $\chi^{2}$ solutions used as input parameters to calculate the best-fit models to match the observed line/continuum SEDs, as seen in Fig.\ref{fig:bfJ1323}.\\

\section{Tabulated properties}
\label{sec:tables}

\begin{deluxetable}{ccccccccc}




\tablecaption{Individually derived mean values}

\tablenum{6}

\tablehead{\colhead{Source ID} & \colhead{log($ n_{   H_{ \rm 2}  } $)} & \colhead{Err, log($ n_{   H_{ \rm 2}  } $)} & \colhead{$T_{\rm k}$} & \colhead{Err, $T_{\rm k}$} & \colhead{$\mu_{\rm L}R_{\rm eff}$} & \colhead{Err, $\mu_{\rm L}R_{\rm eff}$} & \colhead{$\Delta V_{\rm turb}$} & \colhead{Err, $\Delta V_{\rm turb}$} \\ 
\colhead{(---)} & \colhead{(cm$^{-3}$)} & \colhead{(cm$^{-3}$)} & \colhead{(K)} & \colhead{(K)} & \colhead{(pc)} & \colhead{(pc)} & \colhead{(km s$^{-1}$)} & \colhead{(km s$^{-1}$)} } 

\startdata
LPsJ0116-24 & 4.92E+00 & 3.82E+00 & 6.97E+01 & 3.29E+01 & 1.46E+04 & 2.48E+03 & 1.48E+02 & 3.56E+01 \\
LPsJ0209+00 & 5.86E+00 & 6.17E-01 & 4.34E+01 & 3.54E+00 & 1.61E+04 & 2.96E+03 & 3.58E+01 & 3.31E+01 \\
LPsJ0226+23 & 4.67E+00 & 4.83E+00 & 2.93E+02 & 7.32E+01 & 1.68E+04 & 3.13E+03 & 1.69E+02 & 3.42E+01 \\
LPsJ0305-30 & 4.76E+00 & 4.21E+00 & 1.18E+02 & 5.17E+01 & 9.33E+03 & 2.10E+03 & 1.68E+02 & 3.00E+01 \\
LPsJ0748+59 & 4.89E+00 & 3.79E+00 & 7.60E+01 & 2.46E+01 & 1.56E+04 & 2.42E+03 & 1.35E+02 & 3.58E+01 \\
LPsJ0846+15 & 4.01E+00 & 1.01E+01 & 1.06E+02 & 2.87E+01 & 1.42E+04 & 2.77E+03 & 1.45E+02 & 4.31E+01 \\
LPsJ1053+60 & 5.27E+00 & 7.90E-01 & 4.52E+01 & 3.68E+00 & 1.49E+04 & 2.76E+03 & 1.56E+02 & 3.85E+01 \\
LPsJ1053+05 & 4.05E+00 & 4.52E+00 & 1.37E+02 & 5.25E+01 & 1.45E+04 & 2.81E+03 & 9.83E+01 & 5.56E+01 \\
LPsJ1127+42 & 4.59E+00 & 5.39E+00 & 6.74E+01 & 4.03E+01 & 1.15E+04 & 4.21E+03 & 1.22E+02 & 5.35E+01 \\
LPsJ1127+46 & 3.31E+00 & 2.14E+01 & 1.41E+02 & 4.90E+01 & 1.71E+04 & 2.41E+03 & 1.09E+02 & 4.45E+01 \\
LPsJ1138+32 & 4.45E+00 & 5.93E+00 & 2.97E+02 & 7.68E+01 & 7.17E+03 & 2.06E+03 & 1.55E+02 & 4.32E+01 \\
LPsJ1139+20 & 3.71E+00 & 4.11E+00 & 5.00E+01 & 8.92E+00 & 1.34E+04 & 2.20E+03 & 1.48E+02 & 3.52E+01 \\
LPsJ1202+53 & 3.27E+00 & 1.29E+01 & 8.97E+01 & 2.22E+01 & 1.56E+04 & 2.26E+03 & 1.56E+02 & 3.51E+01 \\
LPsJ1322+09 & 4.46E+00 & 6.57E+00 & 1.58E+02 & 6.47E+01 & 1.42E+04 & 3.94E+03 & 8.14E+01 & 5.00E+01 \\
LPsJ1323+55 & 3.50E+00 & 1.83E+01 & 1.64E+02 & 4.70E+01 & 1.29E+04 & 2.82E+03 & 1.39E+02 & 4.32E+01 \\
LPsJ1326+33 & 3.01E+00 & 1.89E+01 & 9.06E+01 & 2.27E+01 & 1.54E+04 & 2.96E+03 & 1.36E+02 & 4.82E+01 \\
LPsJ1329+22 & 4.52E+00 & 3.77E+00 & 2.43E+02 & 7.73E+01 & 1.47E+04 & 4.30E+03 & 8.60E+01 & 5.71E+01 \\
LPsJ1336+49 & 4.12E+00 & 8.36E+00 & 2.05E+02 & 6.47E+01 & 1.33E+04 & 2.22E+03 & 1.53E+02 & 3.71E+01 \\
LPsJ1428+35 & 5.17E+00 & 1.90E+00 & 4.85E+01 & 1.78E+01 & 6.65E+03 & 2.86E+03 & 6.68E+01 & 4.47E+01 \\
LPsJ1449+22 & 4.79E+00 & 3.50E+00 & 6.66E+01 & 3.95E+01 & 8.82E+03 & 4.22E+03 & 1.16E+02 & 5.66E+01 \\
LPsJ1544+50 & 2.64E+00 & 1.41E+01 & 4.18E+01 & 6.34E+00 & 1.70E+04 & 1.63E+03 & 1.70E+02 & 2.32E+01 \\
LPsJ1607+73 & 5.30E+00 & 2.03E+00 & 5.24E+01 & 2.22E+01 & 9.63E+03 & 3.37E+03 & 2.98E+01 & 2.82E+01 \\
LPsJ1609+45 & 2.82E+00 & 1.63E+01 & 1.86E+02 & 4.61E+01 & 1.80E+04 & 1.69E+03 & 1.48E+02 & 3.48E+01 \\
LPsJ2313+01 & 5.43E+00 & 1.82E+00 & 9.02E+01 & 4.40E+01 & 1.34E+04 & 2.53E+03 & 1.33E+02 & 4.15E+01 \\
\enddata


\tablecomments{Individual $\chi^{2}$ weighted mean and standard deviation of the \textit{Turbulence} model calculations of $\sim $2 million model.}


\end{deluxetable}

\begin{deluxetable}{ccccccccc}




\tablecaption{Individually derived mean values}

\tablenum{6}

\tablehead{\colhead{Source ID} & \colhead{$\kappa_{\rm vir}$} & \colhead{Err, $\kappa_{\rm vir}$} & \colhead{$\delta v / \delta r$} & \colhead{Err, $\delta v / \delta r$} & \colhead{$T_{\rm k}$ / $T_{\rm d}$} & \colhead{Err, $T_{\rm k}$ / $T_{\rm d}$} & \colhead{$T_{\rm d}$} & \colhead{Err,$T_{\rm d}$} \\ 
\colhead{(---)} & \colhead{(km s$^{-1}$ pc$^{-1}$ cm$^{3/2}$)} & \colhead{(km s$^{-1}$ pc$^{-1}$ cm$^{3/2}$)} & \colhead{(km s$^{-1}$ pc$^{-1}$)} & \colhead{(km s$^{-1}$ pc$^{-1}$)} & \colhead{(---)} & \colhead{(---)} & \colhead{(K)} & \colhead{(K)} } 

\startdata
LPsJ0116-24 & 1.17E+00 & 2.78E-01 & 4.58E+00 & 1.98E+01 & 1.71E+00 & 6.10E-01 & 4.06E+01 & 7.94E+00 \\
LPsJ0209+00 & 1.50E+00 & 4.66E-01 & 3.73E+01 & 3.46E+01 & 1.12E+00 & 8.11E-02 & 3.88E+01 & 1.81E+00 \\
LPsJ0226+23 & 1.79E+00 & 6.31E-01 & 2.38E+00 & 1.24E+01 & 5.29E+00 & 1.24E+00 & 5.55E+01 & 7.10E+00 \\
LPsJ0305-30 & 1.09E+00 & 2.01E-01 & 2.85E+00 & 1.40E+01 & 1.98E+00 & 8.94E-01 & 6.63E+01 & 5.02E+01 \\
LPsJ0748+59 & 1.30E+00 & 3.63E-01 & 3.80E+00 & 1.98E+01 & 1.63E+00 & 4.99E-01 & 4.66E+01 & 6.39E+00 \\
LPsJ0846+15 & 1.06E+00 & 1.56E-01 & 8.70E-01 & 6.56E+00 & 2.04E+00 & 5.53E-01 & 5.22E+01 & 4.12E+00 \\
LPsJ1053+60 & 1.17E+00 & 2.25E-01 & 1.39E+01 & 1.30E+01 & 1.14E+00 & 1.14E-01 & 3.99E+01 & 1.79E+00 \\
LPsJ1053+05 & 1.13E+00 & 2.42E-01 & 2.89E+00 & 4.41E+00 & 2.99E+00 & 1.22E+00 & 4.63E+01 & 3.09E+00 \\
LPsJ1127+42 & 1.47E+00 & 5.61E-01 & 2.99E+00 & 1.37E+01 & 2.31E+00 & 1.17E+00 & 2.82E+01 & 6.39E+00 \\
LPsJ1127+46 & 1.93E+00 & 5.11E-01 & 7.61E-01 & 2.83E+00 & 3.26E+00 & 1.19E+00 & 4.35E+01 & 1.99E+00 \\
LPsJ1138+32 & 2.26E+00 & 4.96E-01 & 3.04E+00 & 1.19E+01 & 5.04E+00 & 1.16E+00 & 5.93E+01 & 1.12E+01 \\
LPsJ1139+20 & 1.14E+00 & 1.92E-01 & 1.91E+00 & 2.61E+00 & 1.29E+00 & 2.67E-01 & 3.90E+01 & 2.13E+00 \\
LPsJ1202+53 & 1.09E+00 & 1.91E-01 & 7.50E-01 & 2.35E+00 & 2.43E+00 & 6.47E-01 & 3.72E+01 & 1.64E+00 \\
LPsJ1322+09 & 2.08E+00 & 5.49E-01 & 1.90E+00 & 1.21E+01 & 3.06E+00 & 1.24E+00 & 5.17E+01 & 7.26E+00 \\
LPsJ1323+55 & 1.13E+00 & 2.24E-01 & 5.16E-01 & 3.26E+00 & 3.95E+00 & 1.17E+00 & 4.19E+01 & 2.44E+00 \\
LPsJ1326+33 & 1.53E+00 & 4.17E-01 & 6.35E-01 & 2.12E+00 & 2.45E+00 & 6.51E-01 & 3.72E+01 & 1.39E+00 \\
LPsJ1329+22 & 2.12E+00 & 6.51E-01 & 8.67E+00 & 1.16E+01 & 3.98E+00 & 1.09E+00 & 6.15E+01 & 1.26E+01 \\
LPsJ1336+49 & 1.23E+00 & 3.28E-01 & 1.29E+00 & 7.74E+00 & 4.14E+00 & 1.36E+00 & 5.01E+01 & 4.85E+00 \\
LPsJ1428+35 & 1.50E+00 & 5.69E-01 & 1.49E+01 & 2.10E+01 & 1.12E+00 & 3.77E-01 & 4.30E+01 & 2.12E+00 \\
LPsJ1449+22 & 1.13E+00 & 2.53E-01 & 5.92E+00 & 1.40E+01 & 1.27E+00 & 5.69E-01 & 5.18E+01 & 8.01E+00 \\
LPsJ1544+50 & 1.13E+00 & 2.02E-01 & 4.40E-01 & 1.07E+00 & 1.36E+00 & 2.19E-01 & 3.07E+01 & 6.69E-01 \\
LPsJ1607+73 & 1.53E+00 & 6.26E-01 & 1.60E+01 & 2.75E+01 & 1.52E+00 & 5.16E-01 & 3.41E+01 & 2.45E+00 \\
LPsJ1609+45 & 1.60E+00 & 3.26E-01 & 7.10E-01 & 1.50E+00 & 4.52E+00 & 1.17E+00 & 4.12E+01 & 1.41E+00 \\
LPsJ2313+01 & 1.62E+00 & 5.07E-01 & 1.34E+01 & 3.65E+01 & 2.42E+00 & 1.00E+00 & 3.54E+01 & 9.16E+00 \\
\enddata


\tablecomments{Individual $\chi^{2}$ weighted mean and standard deviation based on the \textit{Turbulence} model calculations of $\sim $2 million model. }


\end{deluxetable}

\begin{deluxetable}{ccccccccc}
\label{tab:indmeansA}



\tablecaption{Individually derived mean values}

\tablenum{6}

\tablehead{\colhead{Source ID} & \colhead{CO/H$_{\rm 2}$} & \colhead{Err, CO/H$_{\rm 2}$} & \colhead{[CI]/H$_{\rm 2}$} & \colhead{Err, [CI]/H$_{\rm 2}$} & \colhead{$\mu_{\rm L}M_{\rm ISM}$} & \colhead{Err, $\mu_{\rm L}M_{\rm ISM}$} & \colhead{$L_{\rm FIR}$} & \colhead{Err,$L_{\rm FIR}$} \\ 
\colhead{(---)} & \colhead{(---)} & \colhead{(---)} & \colhead{(---)} & \colhead{(---)} & \colhead{(M$_{\rm \odot}$)} & \colhead{(M$_{\rm \odot}$)} & \colhead{(L$_{\rm \odot}$)} & \colhead{(L$_{\rm \odot}$)} } 

\startdata
LPsJ0116-24 & 1.34E-04 & 2.82E-05 & 5.46E-05 & 2.52E-05 & 4.73E+12 & 5.47E+12 & 9.23E+13 & 9.42E+13 \\
LPsJ0209+00 & 1.23E-04 & 2.42E-05 & 3.04E-05 & 2.68E-05 & 6.51E+12 & 5.49E+11 & 1.33E+14 & 4.30E+12 \\
LPsJ0226+23 & 1.27E-04 & 2.77E-05 & 5.00E-05 & 5.00E-05 & 2.27E+12 & 2.80E+12 & 4.66E+14 & 1.04E+14 \\
LPsJ0305-30 & 1.18E-04 & 1.85E-05 & 6.06E-05 & 1.94E-05 & 1.45E+12 & 3.46E+11 & 1.68E+14 & 6.43E+13 \\
LPsJ0748+59 & 1.44E-04 & 2.81E-05 & 8.29E-05 & 2.63E-05 & 2.94E+12 & 4.00E+12 & 2.30E+14 & 6.12E+13 \\
LPsJ0846+15 & 1.24E-04 & 1.59E-05 & 1.05E-04 & 1.99E-05 & 2.12E+12 & 1.71E+12 & 3.52E+14 & 5.80E+13 \\
LPsJ1053+60 & 1.40E-04 & 3.02E-05 & 2.30E-05 & 9.86E-06 & 1.55E+13 & 3.60E+12 & 1.83E+14 & 4.56E+12 \\
LPsJ1053+05 & 1.15E-04 & 1.86E-05 & 3.74E-05 & 1.36E-05 & 2.41E+12 & 2.30E+11 & 2.33E+14 & 1.14E+13 \\
LPsJ112714 & 1.38E-04 & 2.93E-05 & 3.86E-05 & 2.18E-05 & 1.25E+12 & 1.32E+12 & 5.62E+12 & 2.47E+12 \\
LPsJ112713 & 1.45E-04 & 2.82E-05 & 1.05E-04 & 1.91E-05 & 1.10E+12 & 1.21E+11 & 1.05E+14 & 8.38E+12 \\
LPsJ1138+32 & 1.68E-04 & 2.64E-05 & 9.00E-05 & 2.57E-05 & 3.63E+11 & 3.77E+11 & 7.00E+13 & 3.77E+13 \\
LPsJ1139+20 & 1.17E-04 & 1.94E-05 & 3.69E-05 & 1.57E-05 & 3.11E+12 & 5.44E+11 & 6.32E+13 & 2.77E+12 \\
LPsJ1202+53 & 1.17E-04 & 1.97E-05 & 3.71E-05 & 1.52E-05 & 2.76E+12 & 2.74E+11 & 8.11E+13 & 4.03E+12 \\
LPsJ1322+09 & 1.50E-04 & 2.40E-05 & 1.22E-04 & 2.71E-05 & 5.04E+11 & 9.59E+11 & 9.99E+13 & 2.72E+13 \\
LPsJ1323+55 & 1.24E-04 & 2.34E-05 & 9.46E-05 & 1.48E-05 & 1.36E+12 & 1.55E+11 & 9.32E+13 & 6.26E+12 \\
LPsJ1326+33 & 1.37E-04 & 2.69E-05 & 1.11E-04 & 2.02E-05 & 1.40E+12 & 1.97E+11 & 4.61E+13 & 2.40E+12 \\
LPsJ1329+22 & 1.45E-04 & 3.27E-05 & 6.44E-05 & 2.22E-05 & 1.62E+12 & 4.78E+11 & 3.90E+14 & 1.64E+14 \\
LPsJ1336+49 & 1.36E-04 & 2.92E-05 & 8.83E-05 & 2.12E-05 & 1.79E+12 & 5.38E+11 & 1.77E+14 & 3.00E+13 \\
LPsJ1428+35 & 1.32E-04 & 2.55E-05 & 5.00E-05 & 5.00E-05 & 6.24E+11 & 8.04E+10 & 1.83E+13 & 1.09E+12 \\
LPsJ1449+22 & 1.19E-04 & 2.02E-05 & 3.61E-05 & 1.68E-05 & 1.52E+12 & 1.14E+12 & 1.17E+14 & 3.25E+13 \\
LPsJ1544+50 & 1.19E-04 & 2.03E-05 & 6.23E-05 & 1.91E-05 & 3.26E+12 & 3.32E+11 & 3.57E+13 & 6.12E+11 \\
LPsJ1607+73 & 1.30E-04 & 2.49E-05 & 5.00E-05 & 5.00E-05 & 5.89E+11 & 9.51E+10 & 7.78E+12 & 9.10E+11 \\
LPsJ1609+45 & 1.47E-04 & 2.77E-05 & 1.07E-04 & 2.02E-05 & 2.23E+12 & 2.09E+11 & 1.56E+14 & 6.78E+12 \\
LPsJ2313+01 & 1.46E-04 & 2.89E-05 & 1.00E-04 & 4.18E-05 & 2.99E+12 & 5.24E+12 & 4.11E+13 & 1.54E+13 \\
\enddata


\tablecomments{Individual $\chi^{2}$ weighted mean and standard deviation of the \textit{Turbulence} model calculations of $\sim $2 million model. The FIR luminosity calculated by integrating the dust SED between 40-120${\rm \mu m}$.}


\end{deluxetable}


\subsection{Observations and Line Measurements}

\begin{deluxetable}{cccccccccc}
\label{tab:allOBS}



\tablecaption{Summary of observations}

\tablenum{7}

\tablehead{\colhead{Source ID} & \colhead{Line ID} & \colhead{Receiver} & \colhead{Date of Observation} & \colhead{Mean T$_{\rm sys}$} & \colhead{Int. Time} & \colhead{$\nu_{\rm obs}$} & \colhead{Error $\nu_{\rm obs}$} & \colhead{Gain} & \colhead{RMS} \\ 
\colhead{(---)} & \colhead{(---)} & \colhead{(---)} & \colhead{(d:m:y)} & \colhead{(K)} & \colhead{(min)} & \colhead{(GHz)} & \colhead{(GHz)} & \colhead{(Jy K$^{-1}$)} & \colhead{(Jy km s$^{-1}$)} } 

\startdata
LPsJ0116-24 & CO(1-0) & Ka Band & 20.10.17 & 170 & 60 & 36.88243 & 0.00184 & 1.60 & 0.041 \\
 & CO(6-5) & PI230 & 30.05.18 & 91 & 128 & 221.31994 & 0.01107 & 41.50 & 0.984 \\
 & [CI](2-1) & PI230 & 28.05.18 & 123 & 355 & 259.02220 & 0.01295 & 43.00 & 0.615 \\
\enddata


\tablecomments{The root mean square (RMS) sensitivity is calculated as the integrated flux (Jy \kms) within $\Delta V \approx$ 500-1000 \kms (channel width $\sim$ 80-110 \kms). Integration time corresponds to the averaged scans used for analysis. This table is published in its entirety in the machine-readable format. A portion is shown here for guidance regarding its form and content. Also note, a few detected lines in this work were previously reported in the literature, \citep{canameras2018, nesvadba2019}, but we present deeper (5-10$\times$ better S/N) observations for those $\sim$ 20 CO/[C{\sc i}] lines. Duplicate line measurements include LPs-J105353 (CO(5-4), CO(6-5), CO(8-7), CO(9-8)), LPs-J112714 (CO(4-3), CO(7-6), [C{\sc i}](2-1)), LPs-J1202 (CO(4-3), CO(5-4), CO(7-6), [C{\sc i}](1-0), [C{\sc i}](2-1) ), LPs-J1323 (CO(4-3), CO(5-4), CO(7-6), [C{\sc i}](1-0), [C{\sc i}](2-1)) and LPs-J1609 (CO(5-4), CO(6-5), CO(8-7), CO(9-8) ) \citep{canameras2018b,nesvadba2019}.}


\end{deluxetable}

\begin{deluxetable}{ccccccccccccc}



\rotate

\tablecaption{Measured line properties}

\tablenum{8}

\tablehead{\colhead{ID	} & \colhead{J$_{\rm up}$} & \colhead{Line ID} & \colhead{Redshift} & \colhead{ErrRedshift} & \colhead{$S_{\rm \nu} \Delta V$} & \colhead{Err$S_{\rm \nu} \Delta V$} & \colhead{$L' $} & \colhead{Err$L' $ } & \colhead{L$_{\rm lin}$} & \colhead{ErrL$_{\rm lin}$} & \colhead{Lower Bound} & \colhead{Upper Bound} \\ 
\colhead{(---)} & \colhead{(---)} & \colhead{(---)} & \colhead{(---)} & \colhead{(---)} & \colhead{(Jy km s$^{-1}$)} & \colhead{(Jy km s$^{-1}$)} & \colhead{($\times10^{10}$ K km/s pc$^{2}$)} & \colhead{($\times10^{10}$ K km/s pc$^{2}$)} & \colhead{($\times10^{8}$ L$_{\rm \odot}$)} & \colhead{($\times10^{8}$ L$_{\rm \odot}$)} & \colhead{(km s$^{-1}$)} & \colhead{(km s$^{-1}$)} } 

\startdata
LPsJ0116 & 1 & CO(1-0) & 2.12537 & 1.34E-04 & 4.68 & 1.64 & 1.06E+02 & 3.72E+01 & 5.22E-01 & 1.83E-01 & -300 & 500 \\
LPsJ0116 & 3 & CO(3-2) & 2.12490 & 1.34E-04 & 51.10 & 10.22 & 1.29E+02 & 2.58E+01 & 1.71E+01 & 3.42E+00 & -500 & 500 \\
LPsJ0116 & 6 & CO(6-5) & 2.12431 & 1.34E-04 & 55.20 & 16.56 & 3.49E+01 & 1.05E+01 & 3.69E+01 & 1.11E+01 & -200 & 650 \\
LPsJ0116 & 7 & CO(7-6) & 2.12443 & 1.34E-04 & 33.25 & 9.98 & 1.54E+01 & 4.63E+00 & 2.59E+01 & 7.78E+00 & -300 & 500 \\
LPsJ0116 & 8 & CO(8-7) & 2.12398 & 1.34E-04 & 30.34 & 10.62 & 1.08E+01 & 3.77E+00 & 2.70E+01 & 9.46E+00 & -250 & 500 \\
LPsJ0116 & 9 & CO(9-8) & 2.12417 & 1.34E-04 & 19.72 & 6.90 & 5.54E+00 & 1.94E+00 & 1.97E+01 & 6.91E+00 & -250 & 200 \\
LPsJ0116 & 2 & [CI](2-1) & 2.12443 & 1.34E-04 & 31.34 & 9.40 & 1.44E+01 & 4.33E+00 & 2.45E+01 & 7.35E+00 & -1200 & -550 \\
\enddata


\tablecomments{All observations reported in Tab. \ref{tab:allOBS} were used to derive the line-integrated measurements (Full-Width at Zero Intensity; FWZI) within the integral regions marked above. This table is published in its entirety in the machine-readable format. A portion is shown here for guidance regarding its form and content.}

\label{tab:linemeasurements}
\end{deluxetable}


\begin{deluxetable}{cccccc}




\tablecaption{Best-fit Turbulence model line properties}

\tablenum{9}

\tablehead{\colhead{Source ID} & \colhead{J$_{\rm up}$} & \colhead{$S_{\rm \nu} \Delta V$} & \colhead{$L'_{\rm CO}$} & \colhead{$L_{\rm CO}$} & \colhead{R(J$_{\rm up }$,1)} \\ 
\colhead{(---)} & \colhead{(---)} & \colhead{(Jy km s$^{-1}$)} & \colhead{(10$^{10}$  K km s$^{-1}$ pc$^{2}$)} & \colhead{(10$^{8}$ L$_{\rm \odot}$ )} & \colhead{(---)} } 

\startdata
LPsJ0116 & 1 & 6.37E+00 & 1.45E+02 & 7.10E-01 & 1.00E+00 \\
 & 2 & 2.18E+01 & 1.24E+02 & 4.85E+00 & 8.55E-01 \\
 & 3 & 3.80E+01 & 9.61E+01 & 1.27E+01 & 6.64E-01 \\
 & 4 & 4.89E+01 & 6.95E+01 & 2.18E+01 & 4.80E-01 \\
 & 5 & 5.22E+01 & 4.75E+01 & 2.91E+01 & 3.28E-01 \\
 & 6 & 4.84E+01 & 3.06E+01 & 3.24E+01 & 2.11E-01 \\
 & 7 & 4.00E+01 & 1.86E+01 & 3.12E+01 & 1.28E-01 \\
 & 8 & 2.94E+01 & 1.05E+01 & 2.62E+01 & 7.23E-02 \\
 & 9 & 1.89E+01 & 5.32E+00 & 1.90E+01 & 3.68E-02 \\
 & 10 & 1.03E+01 & 2.35E+00 & 1.15E+01 & 1.62E-02 \\
 & 11 & 4.60E+00 & 8.66E-01 & 5.64E+00 & 5.98E-03 \\
 & 12 & 1.50E+00 & 2.37E-01 & 2.00E+00 & 1.64E-03 \\
 & 13 & 3.09E-01 & 4.16E-02 & 4.47E-01 & 2.87E-04 \\
 & 14 & 4.87E-02 & 5.66E-03 & 7.58E-02 & 3.91E-05 \\
 & 15 & 6.08E-03 & 6.16E-04 & 1.02E-02 & 4.26E-06 \\
\enddata


\tablecomments{Best-fit CO excitation ladders,as determined from the best model solution in the top 1\% of the best $\chi^{2}$ solutions. The model error for each value is of order 5\% based on the dispersion of best-fit values within the top solutions. This table is published in its entirety in the machine-readable format. A portion is shown here for guidance regarding its form and content.}

\label{tab:modelfluxesratios}
\end{deluxetable}

\begin{deluxetable}{ccccc}




\tablecaption{Ancillary dust photometry}

\tablenum{10}

\tablehead{\colhead{Source ID} & \colhead{Observed Frequency} & \colhead{Flux density} & \colhead{Err, Flux density} & \colhead{Telescope} \\ 
\colhead{(---)} & \colhead{(GHz)} & \colhead{(mJy)} & \colhead{(mJy)} & \colhead{(---)} } 

\startdata
LPsJ0116 & 857 & 513 & 462 & Planck \\
 & 545 & 555 & 336 & Planck \\
 & 273 & 66 & 10 & ALMA/Band 6 \\
\enddata


\tablecomments{Ancillary dust photometry for the \LPs. This table is published in its entirety in the machine-readable format. A portion is shown here for guidance regarding its form and content.}

\label{tab:continuumdata}
\end{deluxetable}


\bibliography{H2020refsNEW}{}
\bibliographystyle{aasjournal}



\end{document}